\documentclass[12pt]{article}
\usepackage{titlesec}
\usepackage{lineno}
\usepackage{lipsum}
\usepackage{changepage}
\usepackage{graphicx}
\usepackage{amsmath}
\usepackage{caption}
\usepackage{setspace}
\usepackage{color}
\usepackage{booktabs}
\usepackage{enumerate}

\usepackage{ntheorem}
\newtheorem*{proof}{Proof}

\usepackage[backend=biber, style=numeric-comp, giveninits=true, sorting=none, doi=false, url=false, isbn=false, maxnames=3, minnames=3, date=short,giveninits=true]{biblatex}
\DeclareNameAlias{default}{family-given}

\DeclareDelimFormat{multinamedelim}{\addcomma\space}
\DeclareDelimFormat{finalnamedelim}{\addspace and\space}

\AtBeginBibliography{
  
}
\DefineBibliographyStrings{english}{
  andothers = {\mkbibemph{et\addabbrvspace al\adddot}},
}

\renewbibmacro{in:}{}

\newcommand*{\addsemicolon}{\addpunct{\semicolon}}

\DeclareFieldFormat[article]{title}{#1\isdot}
\DeclareFieldFormat[article]{journaltitle}{\mkbibitalic{#1}\isdot}
\DeclareFieldFormat[article]{volume}{#1}
\DeclareFieldFormat[article]{number}{#1}
\DeclareFieldFormat[article]{pages}{#1}
\DeclareFieldFormat{year}{#1}

\DeclareFieldFormat[article]{volume}{\mkbibbold{#1}}
\DeclareBibliographyDriver{article}{
  \printnames{author}
  \setunit{\addperiod\space}
  \printfield{title}
  \setunit{\addperiod\space}
  \printfield{journaltitle}
  \setunit{\space}
  \printfield{year}
  \setunit{\addsemicolon\space}
  \printfield{volume} 
  \setunit{\addcolon\space}
  \printfield{pages}
  \finentry
}

\DeclareBibliographyDriver{book}{
 \printnames{author}
 \setunit{\addperiod\space}
 \printfield{title}
 \setunit{\addperiod\space} 
\printlist{location}
\setunit{\addcolon\space}
 \printlist{publisher}
 \setunit{\addcomma\space}
 \printfield{year}
 \finentry}

\DeclareFieldFormat[incollection]{title}{#1\isdot}
\DeclareFieldFormat[incollection]{pages}{#1}
\DeclareBibliographyDriver{incollection}{
  \printnames{author}
  \setunit{\addperiod\space}
  \printfield{title}
  \setunit{\addspace}
  \printtext{In:\addspace}
  \printnames{editor}
  \ifnameundef{editor}
    {}
    {\setunit{\addspace}\printtext{\mkbibparens{\usebibmacro{editorstrg}}}}
  \setunit{\addperiod\space}
  \printfield{booktitle}
  \setunit{\addperiod\addspace}
  \printlist{location}
  \setunit{\addcolon\space}
  \printlist{publisher}
  \setunit{\addcomma\space}
  \printfield{year}
  \setunit{\addcomma\space}
  \printfield{pages}
  \finentry}

\DeclareFieldFormat[preprint]{title}{#1\isdot}
\DeclareBibliographyDriver{preprint}{
 \printnames{author}
 \setunit{\addperiod\space}
 \printfield{title}
 \setunit{\addperiod\space} 
 \printfield{note}
 \finentry}

\usepackage[
 left=3cm,  
 right=2.6cm,  
 top=2.1cm,   
 bottom=2.1cm, 
]{geometry}

\usepackage{newtxtext,newtxmath} 
\usepackage{helvet}

\DeclareCaptionFormat{boldnumberfirstsentence}{
 \textbf{#1#2}#3\par
}

\DeclareCaptionLabelSeparator{colonspace}{. }
\newcommand*\patchAmsMathEnvironmentForLineno[1]{
 \expandafter\let\csname old#1\expandafter\endcsname\csname #1\endcsname
 \expandafter\let\csname oldend#1\expandafter\endcsname\csname end#1\endcsname
 \renewenvironment{#1}
   {\linenomath\csname old#1\endcsname}
   {\csname oldend#1\endcsname\endlinenomath}}
\newcommand*\patchBothAmsMathEnvironmentsForLineno[1]{
 \patchAmsMathEnvironmentForLineno{#1}
 \patchAmsMathEnvironmentForLineno{#1*}}
\AtBeginDocument{
\patchBothAmsMathEnvironmentsForLineno{equation}
\patchBothAmsMathEnvironmentsForLineno{align}
\patchBothAmsMathEnvironmentsForLineno{flalign}
\patchBothAmsMathEnvironmentsForLineno{alignat}
\patchBothAmsMathEnvironmentsForLineno{gather}
\patchBothAmsMathEnvironmentsForLineno{multline}
}

\title{\LARGE\bfseries 
Behavioral alignment in social networks
}
\date{}
\author{\parbox[c]{16cm}{
\onehalfspacing \centering ~\\[-0.4cm] 
Yu Xia$^{1,2,3}$ \quad
Alex McAvoy$^{4,5}$ \quad
Qi Su$^{1,2,3}\footnote{Correspondence: qisu@sjtu.edu.cn}$ \\ 
		\footnotesize
		$^{1}$School of Automation and Intelligent Sensing, Shanghai Jiao Tong University, Shanghai, 200240, China \\
		  $^{2}$Key Laboratory of System Control and Information Processing, Ministry of Education of China, Shanghai, 200240, China \\
        $^{3}$Shanghai Key Laboratory of Perception and Control in Industrial Network Systems, Shanghai, 200240, China\\
        $^{4}$ School of Data Science and Society, University of North Carolina at Chapel Hill, Chapel Hill, NC 27599, USA \\
        $^{5}$ Department of Mathematics, University of North Carolina at Chapel Hill, Chapel Hill, NC 27599, USA}
        }

\begin{document}
\maketitle

\begingroup
\titleformat*{\section}{\normalfont\large\bfseries\centering}
\section*{ABSTRACT}
\begin{adjustwidth}{1cm}{1cm} 
\small 
The orderly behaviors observed in large-scale groups, such as fish schooling and the organized movement of crowds, are both ubiquitous and essential for the survival and stability of these systems. Understanding how such complex collective behaviors emerge from simple local interactions and behavioral adjustments is a significant scientific challenge. Historically, research has predominantly focused on imitation and social learning, where individuals adopt the strategies of more successful peers to refine their behavior. However, in recent years, an alternative learning approach based on self-exploration and introspective learning has garnered increasing attention. In this paradigm, individuals assess their own circumstances and select strategies that best align with their specific conditions. Two examples are coordination and anti-coordination, where individuals align with and diverge from the local majority, respectively. In this study, we analyze networked systems of coordinating and anti-coordinating individuals, exploring the combined effects of system dynamics, network structure, and behavioral patterns. We address several practical questions, including the number of equilibria, their characteristics, the equilibrium time, and the resilience of the system. We find that the number of equilibrium states can be extremely large, even increasing exponentially with minor alterations to the network structure. Moreover, the network structure has a significant impact on the average equilibrium time. Despite the complexity of these findings, we find that variations can be captured by a single, simple network characteristic (the average path length), which we illustrate in both synthetic and empirical networks. 
\end{adjustwidth}
\endgroup

\begin{adjustwidth}{1cm}{1cm} 
\textbf{Keywords}:  game theory, complex network, best-response, evolutionary dynamics
\end{adjustwidth}

\section*{INTRODUCTION}
The nature of individual decision-making processes can profoundly influence evolutionary outcomes in large-scale systems, including many populations found in social, biological, economic, and ecological contexts \cite{edwards1954theory, wilson2000sociobiology, RobertMay_ecosystem, VonNeumann_gameTheory}.
A long-standing question across these domains is how simple local interactions, collectively give rise to organized, system-level behavior, such as consensus, polarization, or innovation diffusion \cite{Couzin2005,Shirado2017}. 
Addressing this question requires understanding both the behavioral rules individuals follow and how these rules interact with the underlying network structures.
Imitation is an important and well-studied social learning mechanism, in which individuals copy the behavior of those who are more successful or authoritative \cite{nowak2006book, Pingle1995ImitationVersusRationality}.
When combined with social network structure, imitation can drive the evolution of prosocial traits like cooperation \cite{ohtsuki2006nature}, behaviors which otherwise would not be chosen by perfectly rational agents. A drawback of imitation is that it requires individuals to have extensive information, including the behaviors of their direct neighbors as well as the neighbors of those neighbors, both to assess the behaviors and imitate those they evaluate as being successful \cite{Apesteguia2007imitation_experiment}.

Conversely, coordination and anti-coordination as decision-making approaches allow individuals to independently seek the most beneficial strategy based on analyzing information from interaction partners.
These forms of decision-making involve the ability to self-assess, enabling a deeper level of autonomy and robustness not achievable through imitation alone.
Coordination, where individuals conform to the majority, and anti-coordination, where individuals diverge from the majority, can also be found in many real-life scenarios.
Both are fundamental behavioral mechanisms consistently observed across diverse empirical systems.
Examples of coordination include the formation of driving conventions, where people drive on the same side of the road, and the choice of language within multilingual communities, where individuals tend to use the language adopted by most people \cite{Lewis1969Convention,Giles1977Language}. On the other hand, instances of anti-coordination include auction strategies, where some bidders choose less popular items in order to avoid competition, and problems of resource allocation, where variation in resource consumption and use can mitigate over-exploitation \cite{Cassady1967Auction,Ostrom2000Resource}.

It is known that systems following either coordinating or anti-coordinating patterns must converge to an equilibrium state under asynchronous activation \cite{ramazi2016PNAS_coordinationandAnticoordination}. In fact, convergence to an equilibrium in such systems occurs in a finite number of strategy updates \cite{zhu2023Automatica_equilibriumAnalysis}.
However, convergence properties of the system are sensitive to initial strategy composition and behavioral switching threshold \cite{Ramazi2018TAC_asynchronous, RAMAZI2020Automatica_heterogeneousThreshold,ramazi2022lowerConvergence,zhu2023payoffbasedControl}. These studies focus mainly on convergence itself, leaving some critical aspects yet to be explored. Dynamical systems often exhibit a multitude of equilibrium states, a phenomenon observed in epidemiology, network synchronization, and the evolution of ecosystems \cite{Arenas2008Synchronization, PRE2001Epidemic, May1977eco, Huntington2020eco}.
The variety of equilibrium states can have profound implications for a system. For example, shallow lake systems have two equilibrium states: clear and turbid.
The clear state is rich in submerged vegetation, while in the turbid state, animal diversity is lost and algae is reduced \cite{Scheffer1993lake}.
It is natural, then, to be concerned with a system's equilibrium probability distribution and equilibration time.
For instance, in evolutionary game theory, fixation probabilities and times are important indicators of whether cooperation is favored \cite{Slatkin1981Fixatimeandprob, Whitlock2003Fixatimeandprob, allen2017anystructure, McAvoy2021FixationProb, wang2023reproductive}.

The physical structure of a system profoundly influences the outcomes of evolution \cite{nowak1992_SpatialChaos, Li2020,su2022asymmetric, su2022multilayer, su2022contextualized, Sheng2024HighOrder}. 
Degree distribution heterogeneity and the clustering coefficient represent two important measures of heterogeneity in networked systems \cite{Watts1998smallworld,barabasi1999emergence}. In many networks, there exist highly-connected nodes, called hubs, and nodes tend to form tightly-knit groups in networks with high clustering coefficients.
Both of these properties greatly influence the dynamical characteristics of networked systems.
For example, heterogeneity in the interaction frequency between pairs of individuals can facilitate the evolution of cooperation \cite{Wu2021_Chaos}.
In disease transmission models, heterogeneous networks might see rapid outbreaks of epidemics because hubs can accelerate disease spread, and networks with high clustering coefficients may lead to intense local outbreaks, as infections spread rapidly within closely interconnected groups \cite{Romualdo2001Disease_scalefree,Newman2002Disease_cluster,Xia2012}. 
In quite a different setting, heterogeneous networks are strongly resilient to random network attacks but remain vulnerable to targeted strategies, and networks with high clustering coefficients possess a higher percolation threshold, which increases the cost of attacks \cite{Holme2002NetworkAttack,Iyer2013NetworkAttack}.
Hence, thoroughly analyzing the impact of network structure is essential for advancing our understanding of network dynamics.

In this paper, we investigate the dynamics of coordination and anti-coordination systems, including key properties such as the number of equilibrium states, equilibration time, probability distributions, and system resilience.
Our analysis reveals that structural features of the underlying network, such as the degree distribution and clustering coefficient, strongly affect the number of equilibrium states. Notably, even small changes in network topology can lead to orders-of-magnitude variations in the number of equilibria. We further show that both the number of equilibrium states and the time required to reach equilibrium are governed by a simple topological measure: the average path length. As the average path length increases, coordination systems tend to equilibrate more slowly, while anti-coordination systems tend to equilibrate more rapidly.
Furthermore, we validate these theoretical findings using twenty empirical networks spanning diverse domains, sizes, and densities. These include the karate club friendship network in a university setting \cite{Zachary1977KarateClub}, interaction patterns among Mexican political elites \cite{1996MexicanPoliticalElites}, a Facebook friendship network within a computer company \cite{magnani2013combinatorial}, and social associations among wild Grévy’s zebras \cite{Sundaresan2007ZebraNetwork}. 

\section*{RESULTS}
\subsection*{Model}
We consider a population of $N$ individuals, denoted by $\mathcal{N} = \left\{1,2,\dots, N\right\}$. The spatial structure of the population is represented by a network with $N$ nodes, in which each individual occupies exactly one node, and edges represent connections between individuals. For nodes $i$ and $j$, we denote by $k_{ij}$ the edge weight between $i$ and $j$, which is $1$ if there is an edge between $i$ and $j$ and is $0$ otherwise. Every individual has one of two strategies, $A$ or $B$, and plays a game with each neighbor.
When both players adopt strategy $A$, they both get a payoff $a$.
Similarly, when both players employ strategy $B$, they both receive a payoff $d$.
If one player chooses strategy $A$ and the other uses strategy $B$, the $A$-individual gets $b$ and the $B$-individual gets $c$.

Two classes of games are of immediate interest. The first is a coordination game, in which both players prefer to use the same strategy rather than different strategies. This class of games is characterized by the inequalities $a>c$ and $d>b$. There are three Nash equilibria in a coordination game: two pure Nash equilibria at $\left(A,A\right)$ and $\left(B,B\right)$, as well as a mixed equilibrium in which both players use $A$ with probability $\tau =\left(d-b\right) /\left(a-b-c+d\right)$. The second is an anti-coordination game, in which both players would prefer to use opposite strategies. This class of games is characterized by $c>a$ and $b>d$. Here, both $\left(A,B\right)$ and $\left(B,A\right)$ are Nash equilibria, as is the strategy in which both players use $A$ with probability $\tau$. In the main text, we focus our attention on \emph{pure} coordination and anti-coordination games, which means $a=d$ for the former and $b=c$ for the latter. Thus, for each game, both pure-strategy Nash equilibria are Pareto efficient and the players are concerned only with coordinating or not. (In Supplementary Information, we consider examples where this assumption does not necessarily hold.)

In both games, every player is assigned an aggregate score obtained by accumulating the payoffs from all pairwise interactions. Let $s_{i}\left(t\right)$ be individual $i$'s strategy at time $t$, where $s_{i}\left(t\right) =1$ represents strategy $A$ and $s_{i}\left(t\right) =0$ represents strategy $B$. The accumulated payoffs to individual $i$ when $i$ uses strategies $A$ and $B$, respectively, at time $t$ are
\begin{subequations}
\begin{align}
u_{i}^{A}\left(t\right) &= \sum_{j \in \mathcal{N}} k_{ij} \Big(a s_{j}\left(t\right) + b \left(1-s_{j}\left(t\right)\right)\Big) ; \\
u_{i}^{B}\left(t\right) &= \sum_{j \in \mathcal{N}} k_{ij}\Big( cs_{j}\left(t\right) + d \left(1-s_{j}\left(t\right)\right)\Big) .
\end{align}\label{eq:payoff}
\end{subequations}

\begin{figure}[t]
  \centering
  \includegraphics[width=\linewidth]{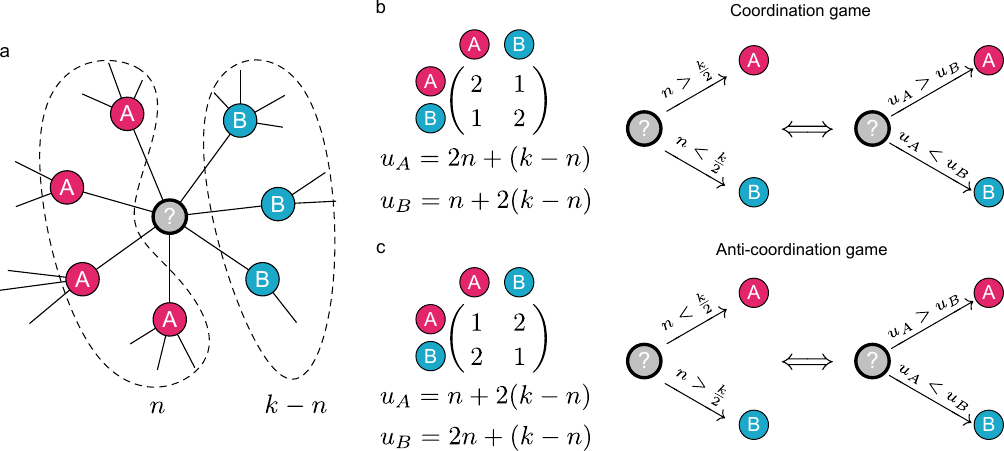}
  \caption{
  Strategy evolution under coordination and anti-coordination decision-making patterns. (a) The structure is described by a network, where each individual (node) adopts either strategy $A$ or $B$ to interact with neighbors. An individual (marked by ``?'') is chosen to update its strategy to maximize payoff, interacting with $n$ individuals using strategy $A$ and $k-n$ individuals using strategy $B$. (b) An example of a coordination game, where the chosen individual receives a higher payoff when adopting the strategy used by the majority of his neighbors. For this payoff matrix, taking the strategy $A$ used by the majority brings a higher payoff than taking $B$ whenever $n>k/2$. The individual opts for strategy $B$ when $n<k/2$. The individual maintains its current strategy if $n=k/2$, which is equivalent to $u^{A} = u^{B}$. (c) An example of an anti-coordination game. When $n > k/2$, adopting strategy $B$ yields a higher payoff. If $n < k/2$, strategy $A$ leads to a higher payoff. The payoffs are the same when $n = k/2$.}
  \label{fig1}
\end{figure}

In each step, a random individual, $i$, is chosen uniformly from the system to update its strategy. Under best-response dynamics \cite{hopkins1999noteonBR}, this individual evaluates strategies $A$ and $B$, adopting the one yielding the greater total payoff (or maintaining the current strategy if both yield the same payoff).
If $i$ is chosen at time $t$ for a strategy revision, then, from Eq.~\ref{eq:payoff}, we have
\begin{align}
    u_{i}^{A}-u_{i}^{B} 
    &= \sum_{j \in \mathcal{N}} k_{ij}\Big( a s_j + b \left(1-s_j\right) - c s_j - d \left(1-s_j\right) \Big) \nonumber \\
    &= \left(a-b-c+d\right) n_i - \left(d-b\right) k_i, \label{eq:payoff_difference}
\end{align}
where $n_i = \sum_{j \in \mathcal{N}} k_{ij} s_j$ denotes the number of $i$'s neighbors using $A$, and $k_i = \sum_{j \in \mathcal{N}} k_{ij}$ is the total number of neighbors of $i$ (i.e., the degree of $i$).

In a coordination game, $a-b-c+d>0$. Thus, under best-response dynamics, Eq.~\ref{eq:payoff_difference} gives
\begin{align}
s_i \left(t+1\right) &= 
\begin{cases}
  1 & n_i > \tau k_i , \\
  s_i \left(t\right) & n_i = \tau k_i, \\
    0 & n_i < \tau k_i ,
\end{cases}\label{eq:coor}
\end{align}
where, again, $\tau =\left(d-b\right) /\left(a-b-c+d\right)$, which coincides with the probability of playing $A$ in the mixed-strategy Nash equilibrium. On the other hand, $a-b-c+d<0$ in an anti-coordination game, which yields a next-round action of
\begin{align}
  s_i \left(t+1\right) &=
  \begin{cases}
    1 & n_i < \tau k_i, \\
    s_i \left(t\right) & n_i = \tau k_i, \\
    0 & n_i > \tau k_i.
  \end{cases}\label{eq:anti}
\end{align}
Figure~\ref{fig1} shows examples of coordination and anti-coordination games, as well as switching thresholds under best-response dynamics.

It is worth noting what best-response dynamics would lead to for other classes of games, such as social dilemmas. In a prisoner's dilemma, which is defined by $c>a>d>b$, we have $d-b>0$, but $a-b-c+d$ can be positive, negative, or zero. However, in all three cases, the payoff difference in Eq.~\ref{eq:payoff_difference} is negative, which means that a player's best response is always $B$. This finding is not surprising, given that $B$ is a dominant action in a prisoner's dilemma and will always be used by a rational agent. However, there are also weaker social dilemmas such as snowdrift games, which are defined by $c>a>b>d$. This, in fact, is an example of an anti-coordination game (just not a \emph{pure} anti-coordination game). For a snowdrift game, we have $\tau\in\left(0,1\right)$; in particular, unlike in a prisoner's dilemma, the best response is non-trivial (Eq.~\ref{eq:anti}).

A configuration of $A$ and $B$ is defined as an equilibrium state if, under the update rule in Eq.~\ref{eq:coor} or Eq.~\ref{eq:anti}, each individual’s prescribed next action matches their current action.
A state $\mathbf{s}^{\ast}=\left(s_1^{\ast},\dots,s_N^{\ast}\right)$ is an equilibrium if 
$u_i\left(\mathbf{s}^{\ast}\right)=\max\left\{u_i^A\left(\mathbf{s}^{\ast}\right) ,u_i^B\left(\mathbf{s}^{\ast}\right)\right\}$ for all $i\in\mathcal{N}$,
where $u_i\left(\mathbf{s}^{\ast}\right)$ denotes player~$i$'s payoff in state $\mathbf{s}^{\ast}$, and $u_i^X\left(\mathbf{s}^{\ast}\right)$ ($X\in\left\{A,B\right\}$) is player~$i$'s payoff when the other players use $s_j^{\ast}$ $\left(j\neq i\right)$ and $i$ adopts strategy~$X$.
Consequently, no individual has the incentive to change strategy in an equilibrium state and $s_i^{\ast}\left(t+1\right) = s_i^{\ast}\left(t\right)$ for all $i \in \mathcal{N}$.
We first explore the number of equilibrium states on representative networks and conduct a quantitative analysis of a class of networks to provide intuition about the interplay between network structure and evolutionary dynamics.
Next, we investigate the absorption capacity of the equilibrium states and the average time required to reach them. 
Here, we define the absorption capacity of a given equilibrium state as the probability that the system, starting from a random initial configuration, eventually reaches that state.
Finally, we examine the robustness of various network structures.

\subsection*{The number of equilibrium states}
\begin{figure}[!t]
  \centering
  \includegraphics[width=\linewidth]{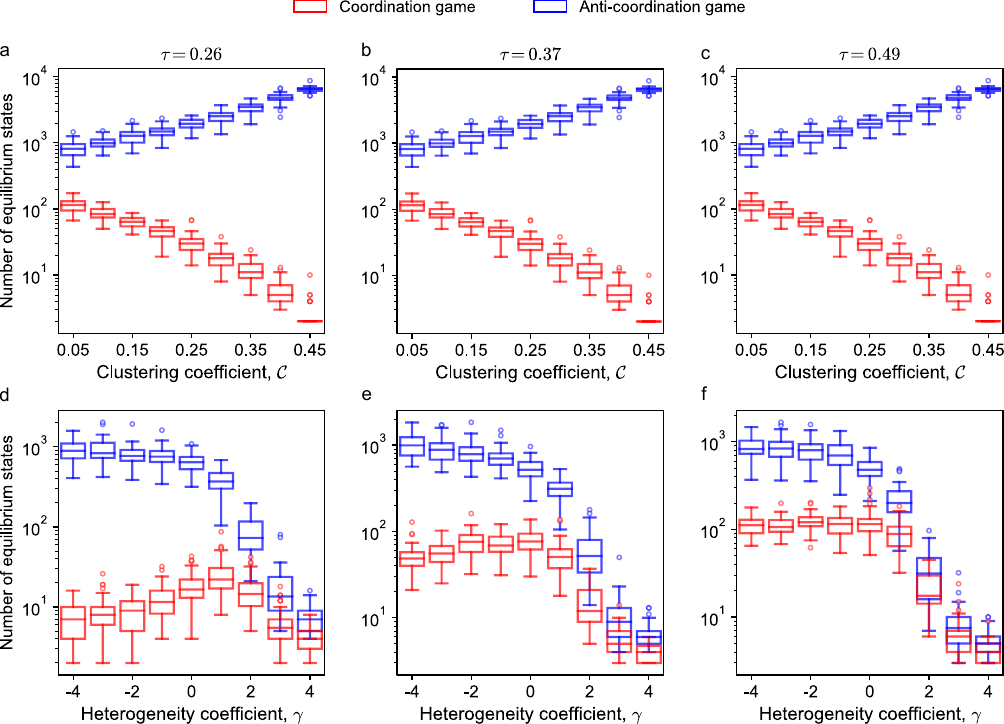}
  \caption{
  A minor change in network structure can dramatically change the number of equilibrium states. 
  We consider two important network characteristics: closed triads (measured by the clustering coefficient, $\mathcal{C}$) and node degree heterogeneity distribution (measured by the distribution heterogeneity coefficient, $\gamma$). 
  We generate networks with varying clustering coefficients by rewiring edges in a degree-preserving manner, starting with regular ring lattice networks \cite{Rao1996Degree_preserving}. Networks with different heterogeneity coefficients are generated based on the connection kernel and the preferential attachment mechanism \cite{barabasi1999emergence, PRL2000_GrowingRandomNetworks}. 
  The number of equilibrium states is presented for a broad range of clustering coefficients, $\mathcal{C}$, (a-c) and heterogeneity coefficients, $\gamma$, (d-f), for three thresholds, $\tau = 0.26, 0.37, 0.49$, and two types of games, coordination and anti-coordination. 
  For each kind of network, we sample 50 networks and present the median, quartiles, and outliers. 
  Modifications to network structures significantly change the number of equilibrium states. 
  Parameters: $N=30$ and average degree $\Bar{k}=4$.}
  \label{fig2}
\end{figure}
We first investigate the impact of closed triads, a widely studied network property, on evolutionary outcomes. Closed triads, commonly found in empirical networks \cite{Feld1981cluster, FRIEDKIN1984cluster, Louch2000cluster}, indicate that two individuals connected to a common third party are likely to connect to each other as well. The prevalence of closed triads is quantified by the (global) clustering coefficient,
\begin{equation}
  \mathcal{C} = \frac{\sum_{i,j,k \in \mathcal{N}} k_{ij} k_{jk} k_{ki}}{\sum_{i,j,k \in \mathcal{N}} k_{ij} k_{jk}}.
\end{equation}
Here, we perform an exhaustive search of the entire state space to identify and count all possible equilibrium states.
In Fig.~2a-c, we show the number of equilibrium states as a function of the clustering coefficient.
As the clustering coefficient increases, there is a decrease in the number of equilibrium states in coordination games, while this number increases in anti-coordination games.
For coordination games in networks with high clustering coefficients, if interconnected nodes choose the same strategy, then it becomes highly probable for their common neighbors to adopt the strategy, eventually resulting in the strategy being employed by the vast majority or even all nodes.
Therefore, the number of equilibrium states in networks with high clustering coefficients is typically lower than that in networks with low clustering coefficients.
Consequently, the number of equilibrium states in coordination games is low because the entire network essentially forms a large cluster.
In contrast, for anti-coordination games in networks with high clustering coefficients, individuals choose strategies opposite to most of their neighbors.
Therefore, if the strategy proportions within the clusters remain at the same level, more individuals can choose between two strategies, culminating in a greater number of equilibrium states (see Fig.~S2 in the Supplementary Information for an example of the effect of clustering coefficients under different games). Thus, clusters promote the spread of strategies in coordination games but act as a barrier in anti-coordination games. In other words, clusters are usually taken over by one strategy in coordination games, while they always include two opposing strategies in anti-coordination games.

We also investigate how degree heterogeneity, measured by the heterogeneity coefficient $\gamma$, affects the number of equilibrium states. This coefficient parametrizes a connection kernel for preferential attachment networks, with a new node being connected to a degree-$k$ node with probability proportional to $k^{\gamma}$ \cite{PRL2000_GrowingRandomNetworks}.
Heterogeneity in the degree distribution grows with $\gamma$. 
Notably, at $\gamma =1$, networks exhibit scale-free characteristics. 
Meanwhile, networks with $\gamma <0$ are akin to random networks, and those with $\gamma >1$ are more closely related to star networks (see Fig.~S3 in the Supplementary Information for the degree distribution of these networks).
Figure~2d-f illustrates that the number of equilibrium states decreases with more degree heterogeneity in both coordination and anti-coordination games.
In contrast to homogeneous networks, heterogeneous networks typically have hubs whose degree greatly exceeds the average. 
In these networks, individuals tend to adopt the same (resp. opposite) strategy as that of the majority of hubs in coordination (resp. anti-coordination) games.
At the same time, as degree heterogeneity intensifies, the number of hubs decreases due to the unchanged total edge count, which means that the number of equilibrium states is determined by the number of hubs at high $\gamma$ values.
Therefore, we observe that in heterogeneous networks, the number of equilibrium states is markedly low in both coordination and anti-coordination games compared with homogeneous networks, with the count of equilibrium states being very close under both types of games in such networks (see Fig.~S4 in the Supplementary Information for an example).
These results highlight the dominance of hubs in both coordination and anti-coordination games.

\subsection*{A simple network measure capturing evolutionary outcomes}
\begin{figure}[!t]
  \centering
  \includegraphics[width=\linewidth]{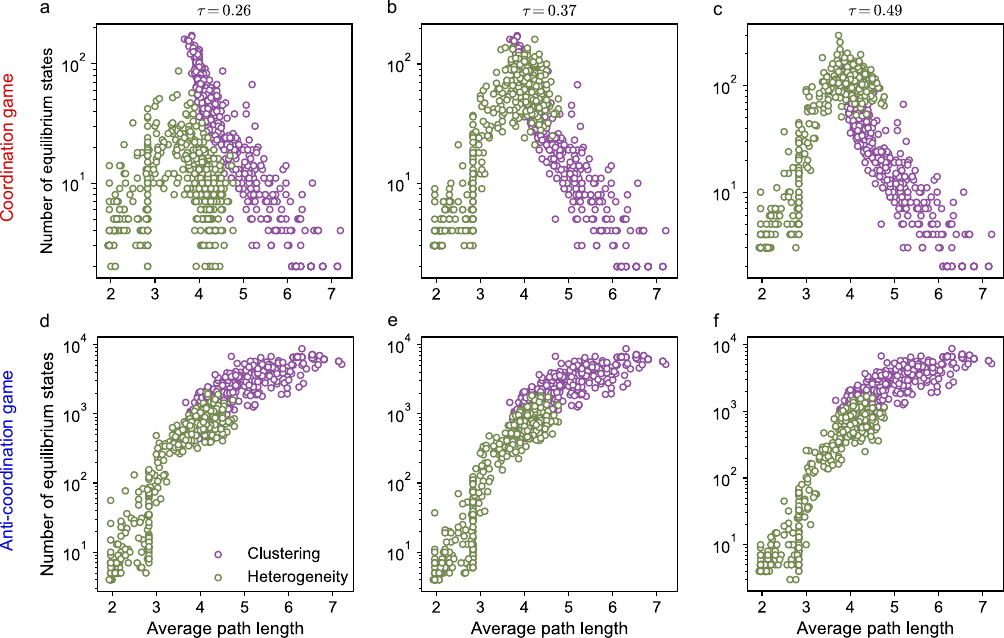}
  \caption{The number of equilibrium states as a function of average path length. Panels ad, be, and cf correspond to $\tau$ values of $0.26$, $0.37$, and $0.49$, respectively. We investigate networks with different structures, including those with variations in clustering coefficients (corresponding to Fig.~2a-c) and in degree distribution heterogeneity (corresponding to Fig.~2d-f). Every dot is the number of equilibrium states of a network. For a given type of game, the average path length describes the number of equilibrium states: a moderate average path length leads to a higher number of equilibrium states in coordination games (a-c), whereas in anti-coordination games, a greater average path length results in a higher number of equilibrium states (d-f). Parameter values: network size $N=30$ and average degree $\Bar{k}=4$.}
  \label{fig3}
\end{figure}
Both the clustering coefficient and degree distribution are fundamental properties of complex networks, and can have strong impacts on evolutionary dynamics.
An important quantity in the graphs shown in Fig.~2 is the average path length, which represents the mean number of edges that nodes in the network have to pass through to reach the most distant node. 
By reorganizing the data presented in Fig.~2 and modifying the horizontal axis to represent the average path length, we obtain Fig.~3. 
Here, the clustering coefficient is positively correlated with the average path length, and the heterogeneity coefficient is negatively correlated with it (see details in Fig.~S24).
Figure~3 shows that as the average path length increases, the number of equilibrium states increases, peaks, and then decreases in coordination games (see Fig.~3a-c). On the other hand, anti-coordination games see continual growth in the number of equilibrium states as the average path length of the network grows (see Fig.~3d-f).

To better understand this observation, we can analyze a sequential root-leaf structure, a representative network on which it is possible to more explicitly characterize the number of equilibrium states. 
This network is parameterized by $n$, the number of hubs, and $m$, the number of leaves per (interior) hub (the terminal nodes have $m+1$ leaves (see Fig.~4a).
For a fixed network size, increasing the number of hubs leads to a longer average path length (but requires a decrease in the number of leaves per hub to maintain constant size). 
In both coordination and anti-coordination games, we provide an explicit formula for the number of equilibrium states, as a function of the number of root nodes, $n$, leaf nodes, $m$, and the behavioral switching threshold, $\tau$. The basic structure of the analysis is given in Supplementary Information, \S{2}, where the number of equilibrium states is characterized in terms of the eigenvalues of a $4\times 4$ binary matrix.

\begin{figure}[!t]
  \centering
  \includegraphics[width=.97\linewidth]{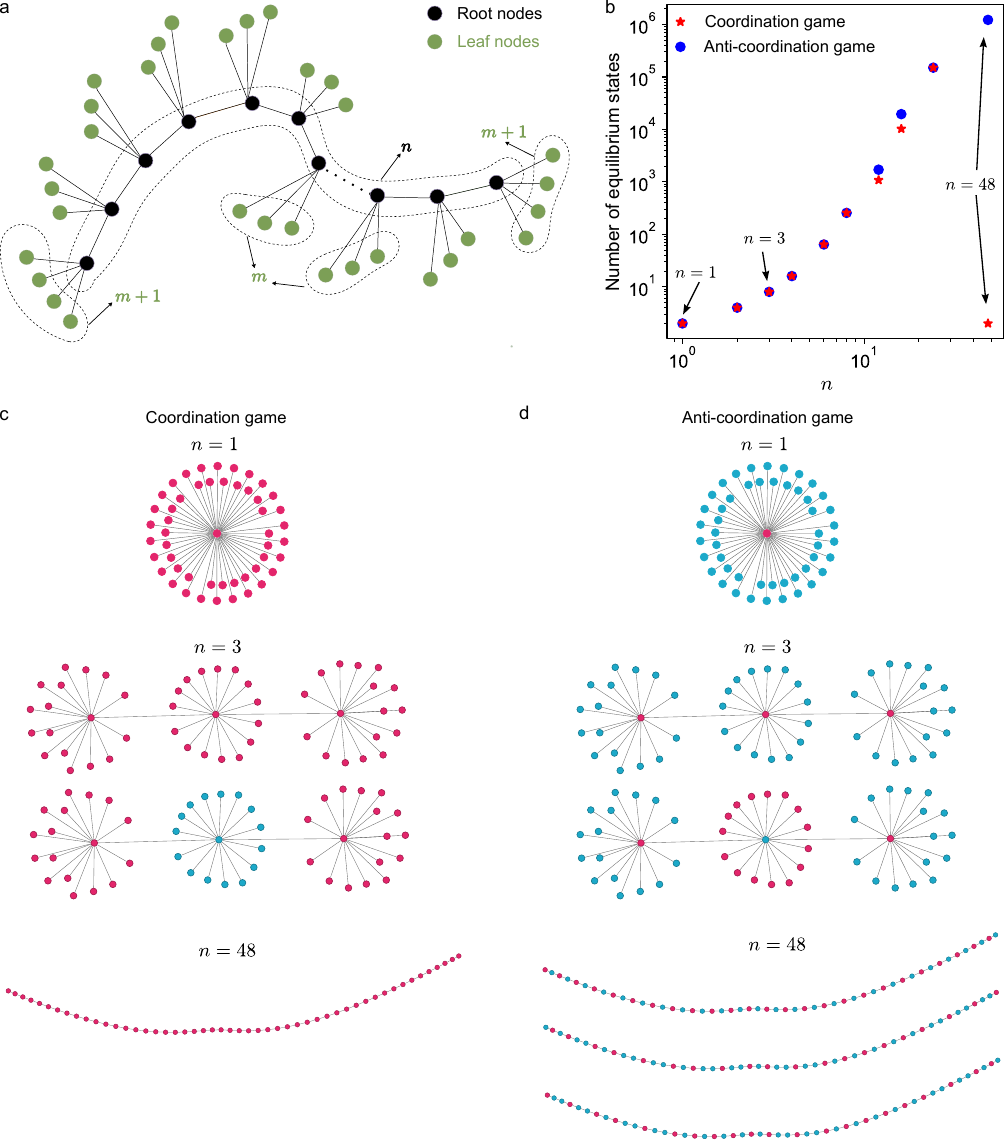}
  \caption{A sequential root-leaf structure. (a) In this idealized structure, $n$ root nodes are sequentially connected. Each root node is linked to $m$ leaf nodes, with the exception that the terminal root nodes at both ends each connect to $m+1$ leaf nodes. (b) In coordination games, there is an initial increase followed by a decrease in the number of equilibrium states as $n$ grows. Conversely, in anti-coordination games, the quantity of equilibrium states grows with $n$. (c) Typical equilibrium states for the sequential root-leaf structure with different root nodes in coordination games. (d) Typical equilibrium states for the sequential root-leaf structure with different root nodes in anti-coordination games. Red (blue) nodes in cd denote individuals of type $A$ ($B$). Parameters: network size $N = 50$ and behavioral switching threshold $\tau=0.37$.}
  \label{fig:cal}
\end{figure}

In a coordination game, when $m\geq2$ and $2/\left(m+2\right)\leq \tau \leq m/\left(m+2\right)$, the number of equilibrium states is $2^{n}$. On the other end of the spectrum, when $m=0$ and $\tau \neq 0.5$, the number of equilibrium states is simply $2$, owing to the fact that the network is just a linear structure in which all-$A$ and all-$B$ are the only equilibria (Fig.~4c). For other parameter ranges of $m$ and $\tau$, we can characterize the growth of the number of equilibrium states (see SI), but the important property is that this structure captures the behavior observed in Fig.~3a-c, in which the number of equilibrium states is maximized at an intermediate average path length (see Fig.~4b).

In anti-coordination games, the number of equilibrium states is determined by hubs as well. 
We note an upward trend in the equilibrium state count with low and increasing values of $n$. 
But unlike the coordination game, more equilibrium states exist when $n$ is significantly large, corresponding to the cases when $m=0$ and $\tau \neq 1/2$ (see Fig.~4b). For example, when $n=48$, $m=0$, and $\tau =0.37$, there are $1{,}221{,}537$ equilibrium states.
In fact, hubs vanish in networks with long path lengths, individuals must adopt strategies opposite most of their neighbors (Fig.~4d), thereby introducing more potential equilibria.

\subsection*{Distribution of equilibrium states}
\begin{figure}[!t]
  \centering
  \includegraphics[width=\linewidth]{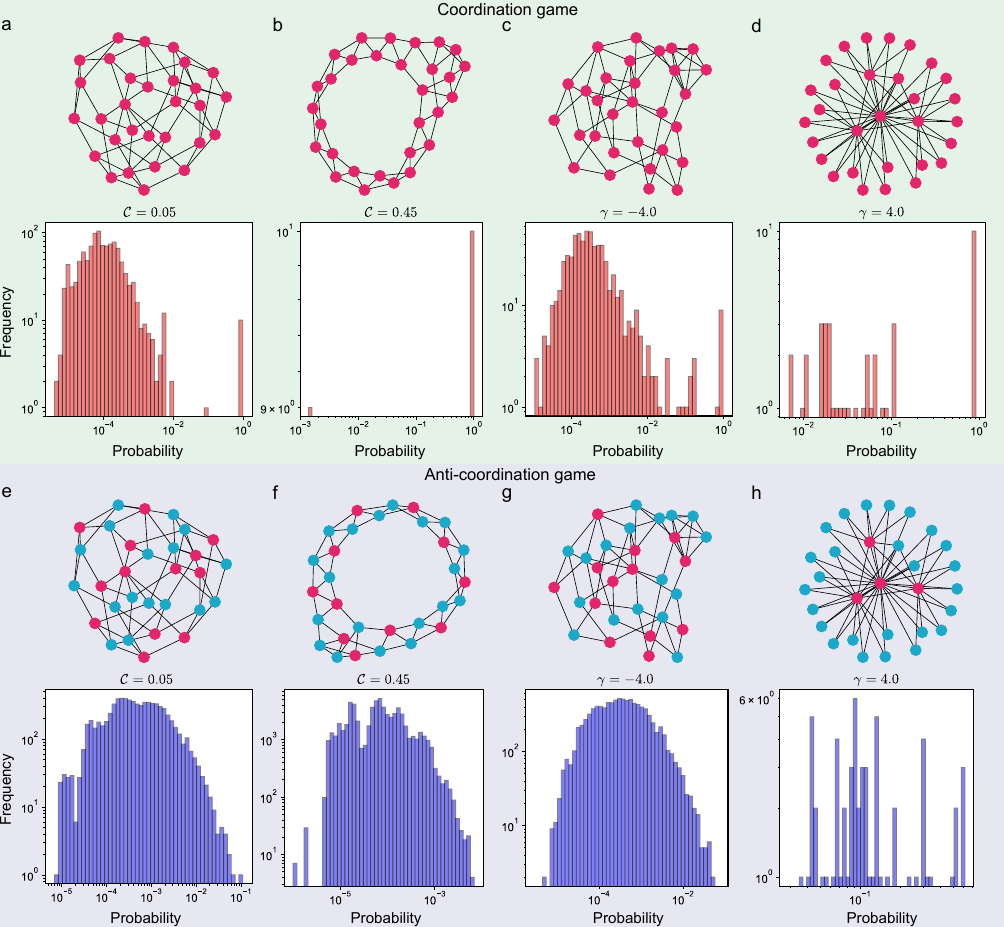}
  \caption{Preferred decision-making patterns in coordination and anti-coordination games. We analyzed the arrival probability of each equilibrium state for both coordination games (a-d) and anti-coordination games (e-f). In each histogram of panel (a-h), the horizontal axis represents the arrival probability, while the vertical axis indicates the number of equilibrium states whose probabilities fall within the corresponding interval. The equilibrium state with the strongest absorption capacity corresponding to each parameter is shown in each panel, where the red (blue) nodes represent individuals with strategy $A$ ($B$). Each histogram displays the results from simulations conducted $3\times 10^7$ times on each of $10$ distinct networks. Parameter values: network size $N=30$, average degree $\Bar{k}=4$, and behavioral switching threshold $\tau = 0.37$.}
  \label{fig5}
\end{figure}
The set of equilibrium states is vast, so it is natural to ask what the differences are among them and which ones are more likely to be reached.
In this section, we explore the equilibrium probability distribution of various networks, a classic topic in network dynamics. 
We begin with a random initial setup in which each individual adopts strategy $A$ with probability $1/2$ (and $B$ otherwise).
Best-response dynamics then unfolds on the network until an equilibrium is reached. This process is repeated many times, and we record the frequency of each equilibrium appearing in the final state as a proxy for the probability of observing various equilibria.

Figure 5 illustrates the degree distribution when the clustering coefficient is $0.05$ and $0.45$ and the heterogeneity coefficient is $-4$ and $4$. (See Fig.~S5 in Supplementary Information for the distribution under other parameter values.)
In the coordination game, there exist equilibrium states with strong absorption capacity in any network.
The behavioral switching threshold determines if one strategy is favored over the other: when $\tau<0.5$, strategy $A$ is favored in coordination games and disfavored in anti-coordination games.
From the data, we find that after half of the individuals in the system adopt the favored strategy, the system tends to evolve into a state where all individuals adopt the same strategy (see Fig.~S6 in Supplementary Information).

This phenomenon can be explained via the observation that since each individual adopts strategy $A$ with the probability equal to $0.5$ and the behavioral switching threshold is $\tau<0.5$, the inequality $n_i > \tau k_i$ is likely to hold.
Therefore, the probability of reaching the state where all individuals adopt strategy $A$ is high.
However, in anti-coordination games, only heterogeneous networks have equilibrium states with strong absorption capacity.
This occurs because, in heterogeneous networks, hubs exist, and once the strategy of these hubs is determined, the strategies of many of the remaining nodes are determined. 
Conversely, in homogeneous networks, hubs do not exist. 
Individuals simply need to be in the minority among their neighbors, indicating that the system lacks a specific driving force to reach a predetermined equilibrium state.

In Fig.~5, we show representative equilibrium states of networks under different parameters (that is, equilibrium states with the highest probability of being reached).
We find the preferred equilibrium states are those in which all individuals adopt the same strategy (the favored strategy) in coordination games (see Fig.~5a-d).
In contrast, the hubs emerging in heterogeneous networks play a significant role in anti-coordination games.
Here, hubs adopt the same strategy and non-hub nodes choose the opposite strategy (see Fig.~5h).
In homogeneous networks, strategies emerge layer by layer, where ``layer'' refers to a group of nodes that have few or no direct connections among themselves.
Therefore, individuals on nodes of one layer select one strategy, and individuals on an adjacent layer choose the opposing strategy.
This phenomenon is particularly evident in networks with high clustering coefficients (see Fig.~5f).

\subsection*{Equilibrium time}
\begin{figure}[!t]
  \centering
  \includegraphics[width=.87\linewidth]{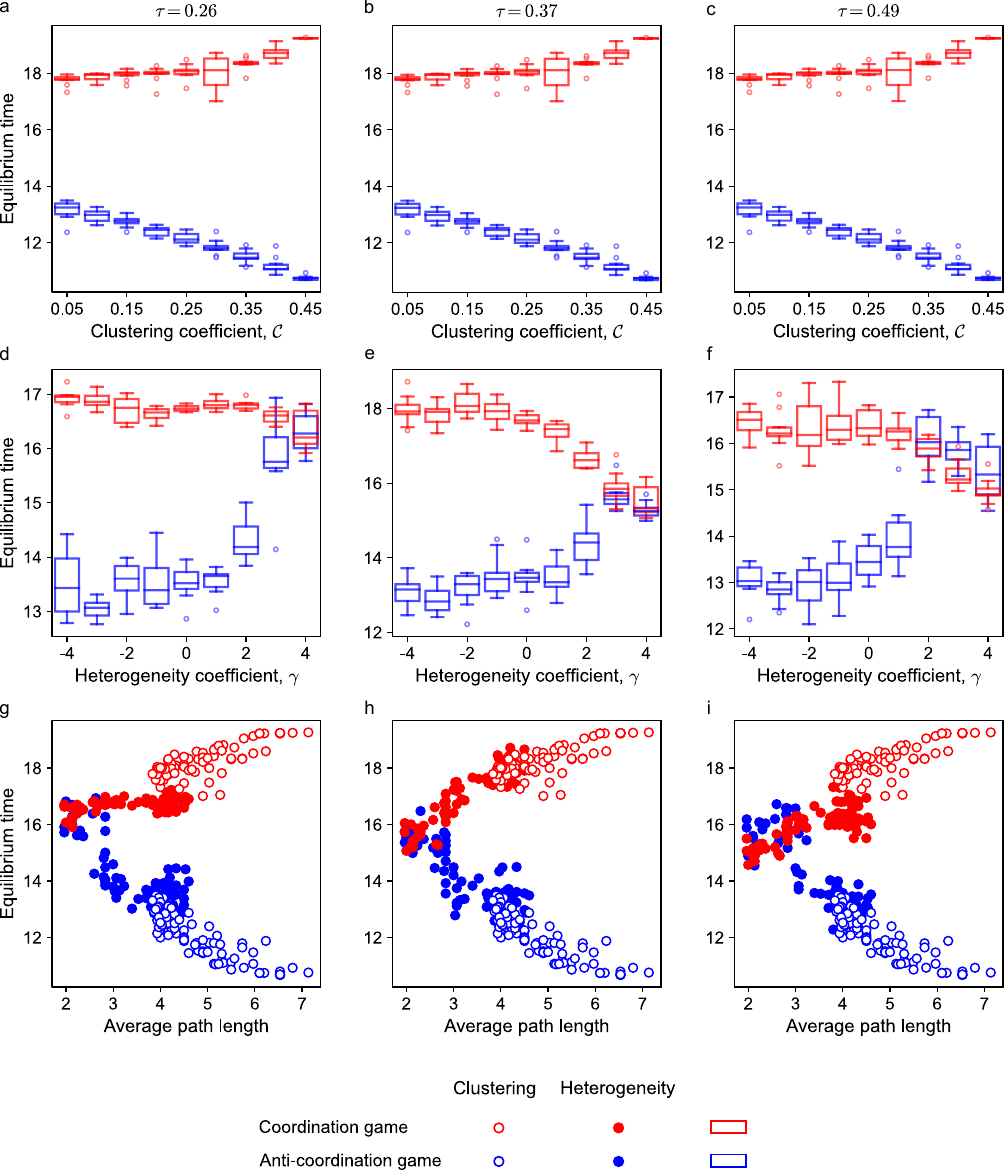}
  \caption{Equilibrium time as a function of clustering coefficient, heterogeneity, and average path length. The average equilibrium time of different networks is shown for coordination games and for anti-coordination games. Each box in panels (a-f) represents the average equilibrium time for $50$ networks, showing the median, quartiles, and outliers. Open and solid dots in (g-i) correspond to the data in (a-c) and (d-f), respectively. Panel (adg), (beh), and (cfi) correspond to $\tau$ values of $0.26$, $0.37$, and $0.49$, respectively. The average equilibrium time increases (decreases) in coordination (anti-coordination) games as the clustering coefficient increases, while it decreases (increases) in coordination (anti-coordination) games as the heterogeneity coefficient increases. (g-i), The equilibrium time increases in coordination games and decreases in anti-coordination games as the average path length increases. The simulation is repeated $3\times 10^7$ times for each network. Parameter values: network size $N=30$ and average degree $\Bar{k}=4$.}
  \label{fig4}
\end{figure}
In addition to which states the system can reach, an important quantity is the time it takes to reach these states. In this section, we consider the equilibrium time, which is measured by the total number of strategy changes in the system prior to reaching an equilibrium state.

Fig.~6a-c demonstrates that, in coordination games, the average time to reach an equilibrium increases as the clustering coefficient increases.
We know already that strategies diffuse in a clustered pattern in coordination games. 
A high clustering coefficient enables the favored strategy to dominate the entire network, while one strategy spreading throughout the network requires the adjustment of strategies in a larger number of individuals, thereby increasing the average equilibrium time. 
On the other hand, increasing degree heterogeneity leads to a decrease in the average equilibrium time in coordination games (see Fig.~6d-f).
In homogeneous networks, these games have equilibrium states with strong absorption capacity, and such equilibria can be reached from very different initial states, increasing the equilibrium time. In heterogeneous networks, hubs dominate the system, which leads to faster equilibration times.

Conversely, we observe coexistence of strategies $A$ and $B$ in anti-coordination games.
Thus, to keep the strategy proportions stable among the clusters, only relatively few nodes need to switch strategies, leading to a decrease in the average time to achieve equilibrium in networks with high clustering coefficients (see Fig.~6a-c).
We also find that greater degree heterogeneity results in a longer equilibrium time in anti-coordination games (see Fig.~6d-f). Other nodes must choose the opposite strategy to the hubs, resulting in more strategy changes.

We can derive intuition about these outcomes by examining the number of equilibrium states. More equilibrium states lead (in principle) to more evolutionary outcomes. Consider two extreme cases: all states are equilibria and only one state is an equilibrium.
If all the states are equilibria, then no matter what the initial state is, the system reaches the equilibrium state directly without any strategy changes. 
Conversely, if there is only one equilibrium state and the initial states are uniformly distributed, many more strategy changes are required (on average).
Therefore, at least informally, systems with fewer equilibrium states need more strategy changes to reach an equilibrium state.
In coordination games, networks with high clustering coefficients have fewer equilibrium states than those with lower clustering coefficients (see Fig.~2a-c), thus reaching equilibrium slower. Similar logic applies to anti-coordination games.

Once again, the effects of network structure on equilibrium time can be understood via the basic network metric of average path length.
Figure 6g-i depicts the variation in average equilibrium time as the average path length changes.
We analyze all the states of the star (see Fig.~S7 and Fig.~S8 in Supplementary Information) and the chain network (see Fig.~S9 and Fig.~S11 in Supplementary Information) of size six.
We find the average equilibrium time grows (shrinks) as the average path length increases under coordination (anti-coordination) games.

Networks with short path lengths tend to have hubs, and the strategies of all other nodes are restricted by these hubs.
Thus, networked systems with short path lengths generally converge rapidly to equilibrium states in both coordination and anti-coordination games.
The star graph of size six illustrates this conclusion (see Fig.~S7 and Fig.~S8 in Supplementary Information for examples).
In contrast, hub nodes tend not to exist in networks with long path lengths, and some nodes switch strategies twice before reaching equilibrium.
For instance, consider the linear segment $ABABA$. Before eventually reaching the all-$A$ state, it may first transition through $ABBBA$ and then become $AAAAA$; the middle $A$ individual switches its strategy twice.
In the chain network of size six, the equilibrium states where all individuals choose strategy $A$ have strong absorption capacity in coordination games (see Fig.~S9 in Supplementary Information). 
Before all individuals adopt strategy $A$, there must be two interconnected nodes (individuals) employing strategy $A$, and the $A$-node (individual) without an $A$-neighbor might first switch to strategy $B$ (see Fig.~S10 in Supplementary Information for an example).

In anti-coordination games, however, due to the repulsion between strategies of the same type, individuals adjust their strategies to become the minority among those of their connections. For instance, in the linear segment $AAAAA$, after two of the individuals switch to strategy $B$, yielding $ABABA$, all individuals tend to keep their strategy unchanged.
In this case, the strategy $B$ spreads through the network in a leapfrog manner, meaning it bypasses immediate neighbors and instead spreads to the neighbors of neighbors.
Therefore, in networks with long average path lengths, only a few individuals need to adjust their strategies to ensure all connected individuals' strategies are the best response to their neighbors, expediting the transition to an equilibrium state (see Fig.~S12 in Supplementary Information for an example).

\subsection*{Robustness}
\begin{figure}[!t]
  \centering
  \includegraphics[width=.85\linewidth]{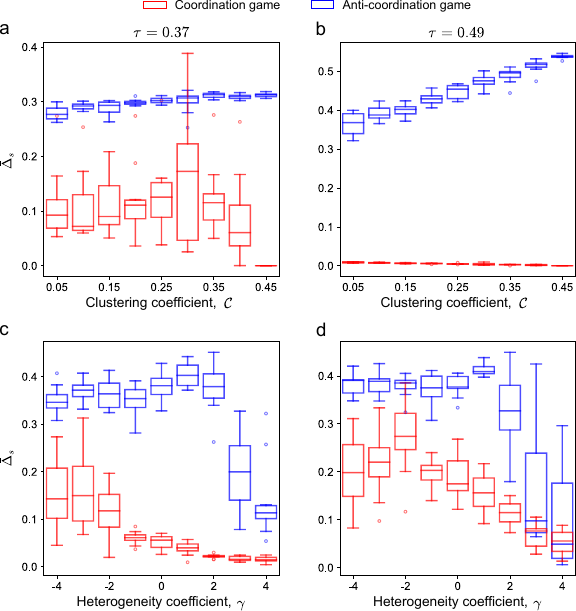}
  \caption{Coordination games are more robust than anti-coordination games. We assess the robustness of the system by randomly adding a new edge to the equilibrium state and measuring the resulting distance (i.e., the number of strategy changes required to reach the final equilibrium state from the initial equilibrium state.) The red (blue) boxes represent the average number of strategy changes in coordination (anti-coordination) games, and each of them is the average over 20 networks. (ab), The average number of strategy switches varies with changes in clustering coefficients. (cd), The average number of strategy switches varies with changes in degree heterogeneity. We used a network size of $N=30$ and an average degree of $\Bar{k}=4$.}
  \label{fig6}
\end{figure}
The robustness of dynamical systems is a topic that has long been a concern in many fields \cite{Carlson2002Robustness, Kitano2004RobustnessBiological, Gao2011Robustness}.
Here, we explore the robustness of different networks by starting a network in an equilibrium state, perturbing it, and then measuring how many strategy changes are required to return to an equilibrium. The initial equilibrium state is selected based on the distribution of equilibrium states, implying that an equilibrium state reached with greater frequency is more likely to be chosen as the initial state. For a given initial state, we perturb the network by adding a single, randomly-chosen edge. If $s_{i}\left(0\right)$ and $s_{i}^{\ast}$ denote the state of node $i$ prior to the perturbation and after equilibrium has been restored following the perturbation, respectively, then we measure robustness of the network via the average number of strategy changes,
\begin{equation}
\Delta_{s} = \sum_{i \in \mathcal{N}} \left|s_{i}^{\ast} - s_{i}\left(0\right)\right| .
\end{equation}

In most cases, networks with coordination games exhibit greater robustness than those with anti-coordination games (see Fig.~7).
For networks with coordination games, in the equilibrium state with the strongest absorption capacity, all individuals adopt the same strategy (see Fig.~5), thus adding new edges does not have an effect according to Eq.~\ref{eq:coor}.
For networks with anti-coordination games, there exist no equilibrium states with strong absorption capacity except in heterogeneous networks.
Thus, in most cases, adding edges causes more individuals to change strategies in anti-coordination games, resulting in larger values of $\Delta_{s}$.

However, in some equilibrium states for coordination games in which two strategies coexist, adding a new edge can drive the system to a monomorphic state (see Fig.~S13 in Supplementary Information for an example).
Thus, adding new edges to networks in coordination games may result in several individuals switching strategies when their neighbors with $A$ strategy increase and the value of $\lfloor \tau k \rfloor$ remains constant.
This is why there are particularly sensitive networks when the clustering coefficient equals $0.30$ in Fig.~7a.
Meanwhile, in heterogeneous networks, fewer individuals tend to switch their strategy than they do in homogeneous networks. Hubs, which are often present in heterogeneous networks, tend to maintain their current strategy under small perturbations, and the strategy changes among other nodes cannot influence the hubs, yielding greater robustness.

Our work focuses mainly on populations in which all individuals have the same behavioral switching threshold.
In Supplementary Information, we consider a complete bipartite graph with nodes divided into two disjoint subsets: one ``left'' set containing $n_l$ nodes and threshold $\tau_l$, and the other ``right'' set containing $n_r$ nodes and threshold $\tau_r$ (see Supplementary Information, \S4, for details). When $\tau_l > n_r/ \left(n_r+1\right)$ and $\tau_r <1/n_l$, and initially the left set uses $B$ and the right set uses $A$, adding an edge to connect two nodes in the left subset leads to an inversion in which the left set uses $A$ and the right set uses $B$.
Thus, even a small perturbation can result in dramatic changes in systems with non-uniform thresholds.
We also investigate examples of adding more than one edge to the system. The results are similar to those shown here (see Fig.~S14 in Supplementary Information).

\subsection*{Real-world networks}
\begin{figure}[!t]
  \centering
  \includegraphics[width=.5\linewidth]{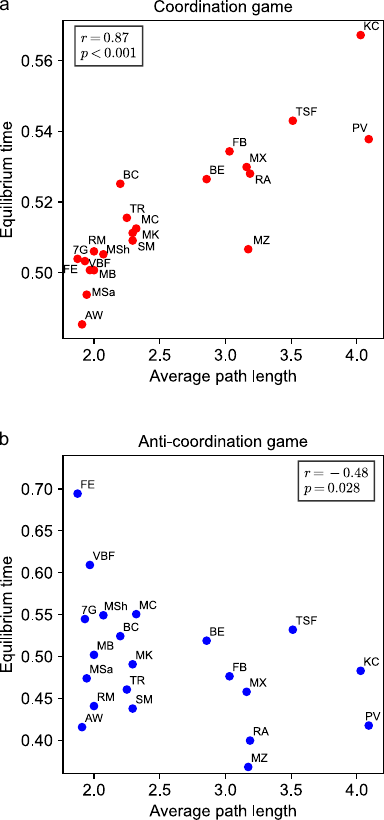}
  \caption{Average path length determines the equilibrium time in empirical networks. 
  The equilibrium time of 20 real-world networks for both coordination games (a) and anti-coordination games (b) is shown as a function of the average path length.
  The equilibrium time of each network is normalized by $N$.
  The equilibrium time increases in coordination games and decreases in anti-coordination games as the average path length of empirical networks increases.
  }
  \label{fig_empirical}
\end{figure}
Here we analyze the system dynamics across 20 empirical networks (see Supplementary Table~S4 for details) that span a wide range of real-world contexts, including physical and online social relationships (e.g., school friendships and Facebook ties), workplace and organizational structures (e.g., CEO clubs and employee interactions), small-scale communities (e.g., villages and tribal societies), and animal social systems (e.g., bison and sheep). 
These networks vary substantially in size (from $N=16$ to $N=39$) and average degree (from $\bar{k}=3.55$ to $\bar{k}=22.25$), reflecting the structural diversity of natural and social systems.
In each empirical network, we simulate strategy updates from random initial conditions and record the average number of time steps required to reach equilibrium, normalized by the network size $N$.  
Figure~\ref{fig_empirical} demonstrates how average path length systematically shapes equilibrium time in empirical networks. 
We find that equilibrium time increases with average path length in coordination games (Fig.~\ref{fig_empirical}a), whereas it decreases in anti-coordination games (Fig.~\ref{fig_empirical}b).  
We calculate the Pearson correlation coefficient $r$, which measures the strength and direction of the linear relationship between two variables.
Specifically, $r>0$ means that larger values of one variable tend to be associated with larger values of the other, $r<0$ means that larger values of one tend to be associated with smaller values of the other, and $|r|$ closer to 1 indicates a stronger relationship. 
To check whether the observed correlation could simply be due to random chance rather than a real link between the variables, we use a permutation test to calculate the $p$-value to assess statistical significance.
The resulting trends reveal a significant positive correlation between average path length and equilibrium time in coordination games ($r = 0.87,\ p < 0.001$), and a significant negative correlation in anti-coordination games ($r = -0.48,\ p = 0.028$), corroborating the patterns previously identified in synthetic network models such as random-regular and scale-free networks.
Therefore, such patterns are not artifacts of any individual network but reflect a robust principle generalizing across diverse empirical systems.  
These findings highlight the predictive power of average path length and support its theoretical importance in shaping system dynamics in strategic interactions, both in controlled models and in complex, real-world environments.

\section*{DISCUSSION}
In this paper, we analyze the interplay between network structure and best-response dynamics, emphasizing the behavioral patterns arising in both coordination and anti-coordination games.
We find that coordination games typically exhibit fewer equilibrium states, greater equilibration times, and more robustness than anti-coordination games.
The comparison between coordination games and anti-coordination games on a common network reveals fundamental differences between the two behavioral patterns, even though the underlying logic of decision-making at an individual level is to maximize payoff.
In addition, we systematically study the influence of spatial structure, investigating how closed triads and hubs (two important components in complex networks) affect system dynamics. We find a simple measure, the average path length in the network, that serves as a good indicator of the resulting dynamics.

Our analysis of network properties, including the clustering coefficient, heterogeneity coefficient, and the average path length, applies for general complex networks (e.g., Erd\H{o}s-R\'enyi, scale-free, small-world, and empirical networks).
However, in highly structured special cases, exceptions to our findings may arise.
In the complete bipartite graph, for example, there are typically only two equilibrium states in both coordination and anti-coordination games (see Section 4 of Supplementary Information). 
This finding lies in contrast to our previous finding that in networks with low clustering coefficients, there are more equilibrium states in coordination games.
The number of equilibrium states can fluctuate dramatically when the behavioral switching threshold changes, especially if its product with the degree of the highest frequency in the network is an integer (see Fig.~S15 and Fig.~S16 in Supplementary Information for examples).
An integer product makes it possible for equality to hold in Eq.~\ref{eq:coor} and Eq.~\ref{eq:anti}, and such individuals are free to adopt either strategy $A$ or $B$ in an equilibrium, resulting in a greater quantity of equilibrium states overall.

Unlike reinforcement learning, our model is grounded in introspection-based dynamics, where individuals choose strategies based on neighbors’ behaviors in the immediately preceding round and aim only to improve the current payoff. 
Compared to link-prediction studies, which aim to infer missing edges from existing network structures and may treat circle-like or tree-like topologies as prediction targets \cite{kekeShang2019tree,kekeShang2022circle}, our study adopts a different perspective. 
Here, the network is used as a fixed experimental condition to examine how structural properties, particularly the average path length, influence the evolutionary dynamics and equilibrium outcomes of coordination and anti-coordination systems.
Unlike prior data-driven simulations of multi-dimensional opinion dissemination for specific events \cite{XU2024}, our study provides a rigorous theoretical framework for coordination and anti-coordination dynamics on graphs, and finds average path length can serve as a general structural metric, which hold across a variety of networked systems.

Imitation, along with coordination and anti-coordination, are all significant heuristic decision-making patter
ns.
Imitation has received considerable attention due to the potential of imitation systems to complete complex tasks under simple rules \cite{Pingle1995ImitationVersusRationality, Byrne1998imitation, Heyes2001imitation, imitation2018review_computer}.
However, the advantages of coordination and anti-coordination games in this regard are more substantial than those of imitation. 
In imitation systems, decision-makers need not discern only the behavioral information of the individuals they interact with but also the payoffs resulting from these behaviors. 
On the other hand, in coordinating and anti-coordinating systems, decision-makers can make decisions based only on the observed behaviors of others. 
Compared with aspiration dynamics \cite{Zhou2021}, which also relies on local information, our model uses observed behaviors of others while aspiration dynamics uses internal benchmarks.
They also differ in structural sensitivity: aspiration dynamics are robust to network variations, whereas our model exhibits a systematic dependence on network topology, governed by the average path length.
The distinction between fixation time in evolutionary game theory \cite{Taylor2006time, Hadjichrysanthou2011StarTime, Tkadlec2021time} and equilibrium time discussed here lies in their respective focuses: fixation time quantifies the time for a single mutant trait to become fixed in the system, whereas our notion of equilibrium time captures the time required to reach an equilibrium state from a randomly initialized state. 
In other words, fixation time is a specific measure of a particular property, while equilibrium time assesses the system's overall relaxation time.

Best-response dynamics, an introspective learning pattern, is quite different from social learning patterns, such as death-birth or imitation rules that have long been studied in the field of evolutionary dynamics.
Some researchers study this topic under probabilistic decision-making in which individuals choose the strategy yielding the highest payoff or imitate the most successful neighbors according to a given probability distribution, and they reveal the distinction between the two learning patterns \cite{su2022contextualized}.
Here, we analyze this problem under deterministic decision-making. The idea is that, with sufficient rationality, decision-makers can find the optimal solution as the best response to the environment \cite{HSimon1979, HSimon1991}.
Best-response dynamics may be viewed as an extreme version of logit-response dynamics \cite{alosferrer:GEB:2010}, which are parameterized by a rationality term. As this rationality level approaches zero, an individual adopts actions uniformly at random. As it approaches infinity, an individual adopts actions based on payoff maximization, yielding best-response dynamics in the limit. In between, rationality is bounded \cite{simon1982models}, and although better-performing actions are favored, there is the potential for suboptimal play. Humans do not always behave perfectly rationally, and as a result it could be the case that some states that appear to be equilibria are technically not equilibria, but still there should be some exponential escape time from a neighborhood of these states, making them quasi-equilibria.

Looking ahead, it is crucial to develop and refine computational methods aimed at calculating (or approximating) the equilibrium distribution and times, as these aspects have implications for many real-life scenarios.
For example, voting processes can be modeled as coordinating systems \cite{Parhami1994vote, vote2019linear}, and understanding both the transient dynamics and equilibria can inform strategies for targeted information delivery. 
In non-human systems such as power grids, interpreting strategies in the network in terms of resource distribution allows one to study efficient allocation of resources (this would typically be an anti-coordinating system). 
Of course, in these systems and others, networks can be highly heterogeneous and individuals may not always have the same behavioral switching thresholds as we have assumed here.
Stochasticity and multi-player interactions are common in collective systems. 
For instance, stochasticity may result from random decision errors, while multi-player interactions can arise when multiple individuals contribute to a shared resource.
Recent studies have shown that such factors can fundamentally alter the underlying dynamics \cite{CHEN2024114565, Duong2016, Duong20162, Duong2025Random}, leading to qualitatively different equilibrium structures and evolutionary outcomes. 
In the context of networked systems with best-response decision rules, exploring how stochasticity and multi-player interactions jointly shape system dynamics represents a challenging yet highly meaningful direction for future research.

\section*{FUNDING}
This work is supported by the National Natural Science Foundation of China (62473252) and Shanghai Pujiang Program (23PJ1405500).

\section*{AUTHOR CONTRIBUTIONS}
All authors developed the model, performed the research and wrote the paper. 

\sloppy
\printbibliography
\fussy

\clearpage
\setcounter{page}{1} 
\setcounter{figure}{0} 
\renewcommand{\thefigure}{S\arabic{figure}} 

{
\centering
\Large Supplementary Information for \\
\bfseries Behavioral alignment in social networks

\vspace{2cm} 
}

\section{Model}
The system is described by a network consisting of $N$ nodes, which are denoted by $\mathcal{N}=\left\{1,2,\dots,N\right\}$. The edges are tracked by an adjacency matrix $\left\{k_{ij}\right\}_{i,j \in \mathcal{N}} \in \{0,1\}$, where $k_{ij} = k_{ji}$ because the interaction is symmetric.
Here $k_{ij} = 1$ means individual $i$ and $j$ interact with each other, while $k_{ij} = 0$ indicates that they do not. 
If $i=j$, then $k_{ij} = 0$ (that is, there are no-self loops).
The total degree of node $i$ is $k_{i} := \sum_{j \in \mathcal{N}} k_{ij}$.

Individuals choose their strategy from the set $\mathcal{S}= \left\{A,B\right\}$. 
When individual $i$ selects strategy $A$ and one of its neighbors uses strategy $A$ (or $B$), it receives a payoff of $a_i$ (or $b_i$).
Similarly, if individual $i$ chooses strategy $B$ and one of its neighbors adopts strategy $A$ (or $B$), it obtains a payoff of $c_i$ (or $d_i$).
Let $s_i\left(t\right)$ denote individual $i$'s strategy at time $t$, where $s_i\left(t\right) =1$ represents strategy $A$ and $s_i\left(t\right) =0$ represents strategy $B$. The total payoff to $i$ is then
\begin{equation}
    u_i\left(t\right) = \sum_{j \in \mathcal{N}} k_{ij}\left(s_i\left(t\right)\left(a_i s_j\left(t\right) + b_i \left(1-s_j\left(t\right)\right)\right) + \left(1-s_i\left(t\right)\right)\left(c_i s_j\left(t\right) + d_i \left(1-s_j\left(t\right)\right)\right)\right) .
    \label{eq:u_i}
\end{equation}
At each time step, a random individual is selected to update its strategy.
Under best-response dynamics, the chosen individual tries to maximize $u_i\left(t\right)$, adopting the strategy that leads to a higher payoff.

The number of $i$'s neighboring individuals adopting strategy $A$ ($B$) at time $t$ is denoted by $n_i\left(t\right)$ ($k_i - n_i\left(t\right)$).
Define the behavioral switching threshold of individual $i$ as $\tau_i = (d_i-b_i)/(a_i+d_i-b_i-c_i)$.
By comparing the payoff generated by strategy $A$ and $B$, we get the best-response dynamics of individual $i$.
In a coordination game, which satisfies $a_{i}>c_{i}$ and $d_{i}>b_{i}$ we find that
\begin{equation}
    s_i\left(t+1\right) =
    \begin{cases}
        1 & \quad n_i\left(t\right) > \tau_i k_i, \\
        s_i\left(t\right) & \quad n_i\left(t\right) = \tau_i k_i, \\
        0 & \quad n_i\left(t\right) < \tau_i k_i .
    \end{cases}
    \label{eq:coor}
\end{equation}

In anti-coordination games, $c_{i}>a_{i}$ and $b_{i}>d_{i}$, which gives
\begin{equation}
    s_i\left(t+1\right) =
    \begin{cases}
        1 & \quad n_i\left(t\right) < \tau_i k_i, \\
        s_i\left(t\right) & \quad n_i\left(t\right) = \tau_i k_i, \\
        0 & \quad n_i\left(t\right) > \tau_i k_i .
    \end{cases}
    \label{eq:anti}
\end{equation}

\section{Sequential root-leaf structure}
\subsection{Analysis of the number of equilibrium states}
As shown in Fig. 4\textbf{a} of the main text, in the sequential root-leaf structure the root nodes are connected sequentially, and each root node is connected to $m$ leaf nodes ($m+1$ for the terminal root nodes).
For a fixed network size, increasing the number of hubs leads to a longer average path length (but requires a decrease in the number of leaves per hub to maintain constant size). 
The strategy of node $i \in \mathcal{N}$ at time $t$ is denoted by $s_i\left(t\right)\in\left\{0,1\right\}$, where $s_i\left(t\right) =1$ and $s_i\left(t\right) =0$ correspond to strategies $A$ and $B$, respectively. Here, we consider the scenario in which all individuals have the same behavioral switching threshold, meaning $\tau_i=\tau_j=\tau$ for all $i,j \in \mathcal{N}$. Denote the strategy of node $i$ in the equilibrium state as $s_i^*$ and the number of equilibrium states as $f\left(n,m,\tau\right)$, which is a function of the number of root nodes, $n$; the number of leaf nodes, $m$; and the behavioral switching threshold, $\tau$.

In an equilibrium state, a leaf node's strategy must either match that of the corresponding root (in a coordination game) or be of the opposite type (in an anti-coordination game). Therefore, we can define a reduced, elementary (and totalistic) graph automaton with the same number of fixed points as the full system, which can then be used to study the number of equilibrium states. In what follows, we let $a_{n}$, $b_{n}$, $c_{n}$, and $d_{n}$ denote the number of equilibrium states in the relevant elementary automaton of length $n$, when the leftmost states are $AA$, $AB$, $BA$, and $BB$, respectively. For $X,Y,Z\in\left\{A,B\right\}$, let $S\left(\boxed{X}Y\right)$ and $S\left(X\boxed{Y}Z\right)$ be indicators for whether $X$ in the endpoint $XY$ and $Y$ in the interior triple $XYZ$ are stable, respectively. (Note that, since the automaton is totalistic, we have $S\left(X\boxed{Y}Z\right) =S\left(Z\boxed{Y}X\right)$ for all $X,Y,Z\in\left\{A,B\right\}$.)

For $n>2$, we have
\begin{subequations}
\begin{align}
a_{n} &= S\left(A\boxed{A}A\right) a_{n-1} + S\left(A\boxed{A}B\right) b_{n-1} ; \\
b_{n} &= S\left(A\boxed{B}A\right) c_{n-1} + S\left(A\boxed{B}B\right) d_{n-1} ; \\
c_{n} &= S\left(A\boxed{A}B\right) a_{n-1} + S\left(B\boxed{A}B\right) b_{n-1} ; \\
d_{n} &= S\left(A\boxed{B}B\right) c_{n-1} + S\left(B\boxed{B}B\right) d_{n-1} .
\end{align}
\end{subequations}
The boundary conditions are
\begin{subequations}
\begin{align}
a_{2} &= S\left(\boxed{A}A\right) ; \\
b_{2} &= S\left(\boxed{A}B\right) ; \\
c_{2} &= S\left(\boxed{B}A\right) ; \\
d_{2} &= S\left(\boxed{B}B\right) .
\end{align}
\end{subequations}
More compactly, we can write
\begin{align}
\begin{pmatrix}
a_{n} \\ b_{n} \\ c_{n} \\ d_{n}
\end{pmatrix}
&= 
\begin{pmatrix}
S\left(A\boxed{A}A\right) & S\left(A\boxed{A}B\right) & 0 & 0 \\
0 & 0 & S\left(A\boxed{B}A\right) & S\left(A\boxed{B}B\right) \\
S\left(A\boxed{A}B\right) & S\left(B\boxed{A}B\right) & 0 & 0 \\
0 & 0 & S\left(A\boxed{B}B\right) & S\left(B\boxed{B}B\right)
\end{pmatrix}
\begin{pmatrix}
a_{n-1} \\ b_{n-1} \\ c_{n-1} \\ d_{n-1}
\end{pmatrix} \nonumber \\
&= \begin{pmatrix}
S\left(A\boxed{A}A\right) & S\left(A\boxed{A}B\right) & 0 & 0 \\
0 & 0 & S\left(A\boxed{B}A\right) & S\left(A\boxed{B}B\right) \\
S\left(A\boxed{A}B\right) & S\left(B\boxed{A}B\right) & 0 & 0 \\
0 & 0 & S\left(A\boxed{B}B\right) & S\left(B\boxed{B}B\right)
\end{pmatrix}^{n-2}
\begin{pmatrix}
a_{2} \\ b_{2} \\ c_{2} \\ d_{2}
\end{pmatrix} .\label{eq:matrix_recurrence}
\end{align}
(For later use, we denote by $M$ the matrix in Eq.~\ref{eq:matrix_recurrence}.) In total, the number of fixed points is
\begin{align}
f\left(m,n,\tau\right) &= S\left(\boxed{A}A\right) a_{n} + S\left(\boxed{A}B\right) b_{n} + S\left(\boxed{B}A\right) c_{n} + S\left(\boxed{B}B\right) d_{n} = z^{\intercal}M^{n-2}z ,
\end{align}
where $z^{\intercal}=\begin{pmatrix}a_{2} & b_{2} & c_{2} & d_{2}\end{pmatrix}$. For the values that $M$ can take (see below), it must be true that either $M^{3}=M^{2}$ or $M$ is diagonalizable. In the former case, $M^{n}=M^{2}$ for all $n\geqslant 2$. In the latter case, we let $v_{1},v_{2},v_{3},v_{4}\in\mathbb{C}^{4}$ be an eigenbasis for $M$, corresponding to eigenvalues $\lambda_{1},\lambda_{2},\lambda_{3},\lambda_{4}\in\mathbb{C}$, respectively. If $D$ is the diagonal matrix of eigenvalues and $V$ is the matrix whose columns are the eigenvectors, then
\begin{align}
f\left(m,n,\tau\right) &= z^{\intercal}M^{n-2}z = z^{\intercal}VD^{n-2}V^{-1}z = \sum_{i=1}^{4} \lambda_{i}^{n-2}\left(z^{\intercal}V\right)_{i}\left(V^{-1}z\right)_{i} .
\end{align}
This expression can be evaluated simply by computing $z$ and the eigenvalues and eigenvectors of $M$, and the result will be an integer even when some of these constituent terms are complex.

All the same, it can be useful to write this expression slightly differently when some of these terms have non-zero imaginary parts. For the matrices $M$ appearing below, either zero or two of the eigenvalues are strictly complex. In the latter case, we let $\lambda_{1}$ and $\lambda_{2}$ be complex conjugates (because $M$ is real), while $\lambda_{3}$ and $\lambda_{4}$ are real. Let $\lambda_{1} =\lambda_{1}^{r}+i\lambda_{1}^{i}$ with corresponding eigenvector $v_{1}=v_{1}^{r}+iv_{1}^{i}$. Since
\begin{align}
Mv_{1} &= Mv_{1}^{r} +iMv_{1}^{i} = \lambda_{1}^{r}v_{1}^{r}-\lambda_{1}^{i}v_{1}^{i} + i\left(\lambda_{1}^{i}v_{1}^{r}+\lambda_{1}^{r}v_{1}^{i}\right) ,
\end{align}
we have $Mv_{1}^{r} =\lambda_{1}^{r}v_{1}^{r}-\lambda_{1}^{i}v_{1}^{i}$ and $Mv_{1}^{i} =\lambda_{1}^{i}v_{1}^{r}+\lambda_{1}^{r}v_{1}^{i}$. Thus,
\begin{align}
M\underbrace{\begin{pmatrix}v_{1}^{r} & -v_{1}^{i} & v_{3} & v_{4}\end{pmatrix}}_{\widetilde{V}} &= \underbrace{\begin{pmatrix}v_{1}^{r} & -v_{1}^{i} & v_{3} & v_{4}\end{pmatrix}}_{\widetilde{V}}\begin{pmatrix}\lambda_{1}^{r} & -\lambda_{1}^{i} & 0 & 0 \\ \lambda_{1}^{i} & \lambda_{1}^{r} & 0 & 0 \\ 0 & 0 & \lambda_{3} & 0 \\ 0 & 0 & 0 & \lambda_{4}\end{pmatrix} .
\end{align}
If $\theta_{1}$ is the phase of $\lambda_{1}$ and $\left|\lambda_{1}\right| =\sqrt{\left(\lambda_{1}^{r}\right)^{2}+\left(\lambda_{1}^{i}\right)^{2}}$ is its modulus, then we obtain
\begin{align}
f\left(m,n,\tau\right) &= z^{\intercal}\widetilde{V}\begin{pmatrix}\left|\lambda_{1}\right|^{n-2}\cos\left(n-2\right)\theta & -\left|\lambda_{1}\right|^{n-2}\sin\left(n-2\right)\theta & 0 & 0 \\ \left|\lambda_{1}\right|^{n-2}\sin\left(n-2\right)\theta & \left|\lambda_{1}\right|^{n-2}\cos\left(n-2\right)\theta & 0 & 0 \\ 0 & 0 & \lambda_{3}^{n-2} & 0 \\ 0 & 0 & 0 & \lambda_{4}^{n-2}\end{pmatrix}\widetilde{V}^{-1}z ,
\end{align}
which expresses $f$ in terms of only real constituent terms.

\subsubsection{Coordination game}
Consider the elementary automaton defined as follows. For interior root nodes,
\begin{align}
\widetilde{s}_{i}\left(t+1\right) &= 
\begin{cases}
0 & m + \widetilde{s}_{i-1}\left(t\right) + \widetilde{s}_{i+1}\left(t\right) < \left(m+2\right)\tau , \\
1 & \widetilde{s}_{i-1}\left(t\right) + \widetilde{s}_{i+1}\left(t\right) > \left(m+2\right)\tau , \\
\widetilde{s}_{i}\left(t\right) & \textrm{otherwise} .
\end{cases}
\end{align}
For the terminal root nodes,
\begin{align}
\widetilde{s}_{i}\left(t+1\right) &= 
\begin{cases}
0 & m+1 + \widetilde{s}_{i,\textrm{neighbor}}\left(t\right) < \left(m+2\right)\tau , \\
1 & \widetilde{s}_{i,\textrm{neighbor}}\left(t\right) > \left(m+2\right)\tau , \\
\widetilde{s}_{i}\left(t\right) & \textrm{otherwise} .
\end{cases}
\end{align}
The fixed points of this automaton are in one-to-one correspondence with the equilibrium points of the sequential root-leaf structure.

\begin{table}[h]
    \centering
\begin{tabular}{c c c}
\toprule 
    Component & Condition  \\
    \midrule
        \boxed{A}$A$ & $\tau \leq 1$ \\
        \boxed{A}$B$ & $\tau \leq \frac{m+1}{m+2}$ \\
        \boxed{B}$A$ & $\tau \geq \frac{1}{m+2}$ \\
        \boxed{B}$B$ & $\tau \geq 0$ \\
        $A$\boxed{A}$A$ & $\tau \leq 1$ \\
        $A$\boxed{A}$B$ & $\tau \leq \frac{m+1}{m+2}$ \\
        $B$\boxed{A}$B$ & $\tau \leq \frac{m}{m+2}$ \\
        $A$\boxed{B}$A$ & $\tau \geq \frac{2}{m+2}$ \\
        $A$\boxed{B}$B$ & $\tau \geq \frac{1}{m+2}$ \\
        $B$\boxed{B}$B$ & $\tau \geq 0$ \\
    \bottomrule
\end{tabular}
    \caption{\textbf{Components of the sequential root-leaf structure in the coordination game.}\label{tab:coor_srls}}
\end{table}

\subsubsection{Anti-coordination game}
Consider the elementary automaton defined as follows. For interior root nodes,
\begin{align}
\widetilde{s}_{i}\left(t+1\right) &= 
\begin{cases}
1 & m + \widetilde{s}_{i-1}\left(t\right) + \widetilde{s}_{i+1}\left(t\right) < \left(m+2\right)\tau , \\
0 & \widetilde{s}_{i-1}\left(t\right) + \widetilde{s}_{i+1}\left(t\right) > \left(m+2\right)\tau , \\
\widetilde{s}_{i}\left(t\right) & \textrm{otherwise} .
\end{cases}
\end{align}
For the terminal root nodes,
\begin{align}
\widetilde{s}_{i}\left(t+1\right) &= 
\begin{cases}
1 & m+1 + \widetilde{s}_{i,\textrm{neighbor}}\left(t\right) < \left(m+2\right)\tau , \\
0 & \widetilde{s}_{i,\textrm{neighbor}}\left(t\right) > \left(m+2\right)\tau , \\
\widetilde{s}_{i}\left(t\right) & \textrm{otherwise} .
\end{cases}
\end{align}
The fixed points of this automaton are in one-to-one correspondence with the equilibrium points of the sequential root-leaf structure.

\begin{table}[h]
    \centering
\begin{tabular}{c c c}
\toprule 
    Component & Condition  \\
    \midrule
        \boxed{A}$A$ & $\tau \geq \frac{1}{m+2}$ \\
        \boxed{A}$B$ & $\tau \geq 0$ \\
        \boxed{B}$A$ & $\tau \leq 1$ \\
        \boxed{B}$B$ & $\tau \leq \frac{m+1}{m+2}$ \\
        $A$\boxed{A}$A$ & $\tau \geq \frac{2}{m+2}$ \\        
        $A$\boxed{A}$B$ & $\tau \geq \frac{1}{m+2}$ \\
        $B$\boxed{A}$B$ & $\tau \geq 0$ \\
        $A$\boxed{B}$A$ & $\tau \leq 1$ \\
        $A$\boxed{B}$B$ & $\tau \leq \frac{m+1}{m+2}$ \\
        $B$\boxed{B}$B$ & $\tau \leq \frac{m}{m+2}$ \\
    \bottomrule
\end{tabular}
    \caption{\textbf{Components of the sequential root-leaf structure in an anti-coordination game.}}
    \label{tab:anti_srls}
\end{table}

\subsubsection{Parameters of interest}
Based on Tables~\ref{tab:coor_srls}~and~\ref{tab:anti_srls}, there are several relevant subdivisions of $\left[0,1\right]$ for $\tau$. When $m=0$, we are interested in the subdivisions $\left\{0\right\}$, $\left(0,\frac{1}{2}\right)$, $\left\{\frac{1}{2}\right\}$, $\left(\frac{1}{2},1\right)$, $\left\{1\right\}$. When $m=1$, we are interested in the subdivisions $\left[0,\frac{1}{3}\right)$, $\left\{\frac{1}{3}\right\}$, $\left(\frac{1}{3},\frac{2}{3}\right)$, $\left\{\frac{2}{3}\right\}$, and $\left(\frac{2}{3},1\right]$. Finally, when $m\geqslant 2$, we are interested in the subdivisions $\left[0,\frac{1}{m+2}\right)$, $\left[\frac{1}{m+2},\frac{2}{m+2}\right)$, $\left[\frac{2}{m+2},\frac{m}{m+2}\right]$, $\left(\frac{m}{m+2},\frac{m+1}{m+2}\right]$, and $\left(\frac{m+1}{m+2},1\right]$. The specific eigenvalue calculations for each interval and each kind of game may be found in the attached notebook.

\section{Star}
\subsection{Analysis of the number of equilibrium states}
Define $(A_r, A_l)$ as the state of the star graph, where $A_r\in \{0,1\}$ and $A_l \in \{0,1,\cdots,N-1\}$ correspond to the number of individual choosing $A$ at the root node and leaf nodes respectively.
There are only two equilibrium states for the star graph, that is, $(0,0)$, $(1,N-1)$ in the coordination game and $(0,N-1)$ and $(1,0)$ in the anti-coordination game, which means all individuals at leaf nodes must adopt the same(opposite) strategy in the coordinating(anti-coordinating) game.

Take the coordination game as an example.
Suppose there exists another equilibrium state except $(0,0)$ and $(1, N-1)$.
Thus, individuals adopting strategy $A$ and $B$ coexist at leaf nodes.
If the individual at the root node uses strategy $A$, then for the individual $i$ who adopts strategy $B$ at leaf nodes, we obtain

\begin{equation}
    n_i(t) = 1 > \tau_i k_i = \tau_i,
\end{equation}

which means individual $i$ tend to switch to strategy $A$.
If the individual at root node uses strategy $B$, then for the individual $i$ who adopts strategy $A$ at leaf nodes, we have

\begin{equation}
    n_i(t) = 0 < \tau_i k_i = \tau_i,
\end{equation}

which means individual $i$ tend to switch to strategy $B$.
Therefore the current state of the system is not an equilibrium state and there are only two equilibrium states $(0,0)$ and $(1,N-1)$ in the coordination game.

\subsection{Analysis of the equilibrium time}
Let $p_{i \rightarrow j}^{h \rightarrow l}$ represent the probability of moving from state $(h,i)$ to state $(l,j)$. The average time until absorption, starting from state $(h,i)$, is denoted by $T_i^h$. Then for the coordination game, we obtain

\begin{equation}
    p_{i \rightarrow i-1}^{0 \rightarrow 0} = \frac{i}{N},
\end{equation}
\begin{equation}
    p_{i \rightarrow i+1}^{1 \rightarrow 1} = \frac{N-i-1}{N},
\end{equation}
\begin{equation}
    p_{i \rightarrow i}^{0 \rightarrow 1} = \frac{1-H(\tau(N-1)-i)}{N},
\end{equation}
\begin{equation}
    p_{i \rightarrow i}^{1 \rightarrow 0} = \frac{1-H(i-\tau(N-1))}{N},
\end{equation}

where $H(\cdot)$ is the Heaviside step function and $N$ is the number of nodes on the star. Meanwhile, we have

\begin{equation}
    T_{i}^0 = p_{i \rightarrow i-1}^{0 \rightarrow 0} T_{i-1}^0 + p_{i \rightarrow i}^{0 \rightarrow 1} T_i^1 + (1- p_{i \rightarrow i-1}^{0 \rightarrow 0}- p_{i \rightarrow i}^{0 \rightarrow 1})T_i^0+1,
\end{equation}
\begin{equation}
    T_{i}^1 = p_{i \rightarrow i+1}^{1 \rightarrow 1} T_{i+1}^1 +  p_{i \rightarrow i}^{1 \rightarrow 0} T_i^0 +(1- p_{i \rightarrow i+1}^{1 \rightarrow 1}-  p_{i \rightarrow i}^{1 \rightarrow 0})T_{i}^1 +1. 
\end{equation}

The boundary conditions are set as $T_0^0 = 0$ and $T_{N-1}^1 = 0$. Rearranging the equations, we obtain

\begin{equation}
    T_{i}^0 = \frac{ p_{i \rightarrow i-1}^{0 \rightarrow 0}}{p_{i \rightarrow i-1}^{0 \rightarrow 0}+ p_{i \rightarrow i}^{0 \rightarrow 1}} T_{i-1}^0 + \frac{p_{i \rightarrow i}^{0 \rightarrow 1}}{p_{i \rightarrow i-1}^{0 \rightarrow 0}+ p_{i \rightarrow i}^{0 \rightarrow 1}} T_i^1 + \frac{1}{p_{i \rightarrow i-1}^{0 \rightarrow 0}+ p_{i \rightarrow i}^{0 \rightarrow 1}},
\end{equation}
\begin{equation}
    T_i^1 = \frac{p_{i-1 \rightarrow i}^{1 \rightarrow 1}+ p_{i-1 \rightarrow i-1}^{1 \rightarrow 0}}{p_{i-1 \rightarrow i}^{1 \rightarrow 1}} T_{i-1}^1 - \frac{ p_{i-1 \rightarrow i-1}^{1 \rightarrow 0}}{p_{i-1 \rightarrow i}^{1 \rightarrow 1}}T^0_{i-1} - \frac{1}{p_{i-1 \rightarrow i}^{1 \rightarrow 1}}.
\end{equation}

Let $\pi_{i \rightarrow j}^{ h \rightarrow l}$ represent the probability of transition, given that the system does not stay in its current state, namely

\begin{equation}
    \pi_{i \rightarrow i+1}^{1 \rightarrow 1} = 1-\pi_{i \rightarrow i}^{1 \rightarrow 0} = \frac{p_{i \rightarrow i+1}^{1 \rightarrow 1}}{p_{i \rightarrow i+1}^{1 \rightarrow 1} + p_{i \rightarrow i}^{1 \rightarrow 0}},
\end{equation}
\begin{equation}
    \pi_{i \rightarrow i}^{0 \rightarrow 1} = 1 - \pi_{i \rightarrow i-1}^{0 \rightarrow 0} = \frac{p_{i \rightarrow i}^{0 \rightarrow 1}}{p_{i \rightarrow i}^{0 \rightarrow 1} + p_{i \rightarrow i-1}^{0 \rightarrow 0}}.
\end{equation}

Thus we obtain

\begin{equation}
    T_i^0 = \pi_{i \rightarrow i-1}^{0 \rightarrow 0} T_{i-1}^{0} + \pi_{i \rightarrow i}^{0 \rightarrow 1} T_{i}^{1} + \frac{1}{p_{i \rightarrow i-1}^{0 \rightarrow 0}+ p_{i \rightarrow i}^{0 \rightarrow 1}},
\end{equation}
\begin{equation}
    T_i^1 =\frac{1}{\pi_{i-1 \rightarrow i}^{1 \rightarrow 1}} T_{i-1}^1 -\frac{\pi_{i-1 \rightarrow i-1}^{1 \rightarrow 0}}{\pi_{i-1 \rightarrow i}^{1 \rightarrow 1}}  T^0_{i-1} - \frac{1}{\pi_{i-1 \rightarrow i}^{1 \rightarrow 1} (p_{i-1 \rightarrow i}^{1 \rightarrow 1}+ p_{i-1 \rightarrow i-1}^{1 \rightarrow 0})}.
\end{equation}

Solving inductively, we get

\begin{equation}
    T_{i}^{1} = A(1,i)T_1^1 - \sum_{l =2}^{i} A(l,i) B(l),
\end{equation}

where

\begin{equation}
    A(l,m) = 1+ \sum_{j = 1}^{m-1} \pi_{j \rightarrow j}^{1 \rightarrow 0} \prod_{k = l}^{j} \frac{\pi_{k \rightarrow k-1}^{0 \rightarrow 0}}{\pi_{k \rightarrow k+1}^{1 \rightarrow 1}},
\end{equation}
\begin{equation}
    B(l) = \frac{\pi_{l-1 \rightarrow l-1}^{1 \rightarrow 0}}{\pi_{l-1 \rightarrow l}^{1 \rightarrow 1}} \sum_{j =1}^{l-1} (\frac{\prod_{k =j+1}^{l-1} \pi_{k \rightarrow k-1}^{0 \rightarrow 0}}{p_{j \rightarrow j-1}^{0 \rightarrow 0} + p_{j \rightarrow j}^{0 \rightarrow 1}}) + \frac{1}{p_{l-1 \rightarrow l}^{1 \rightarrow 1}}.
\end{equation}

And for the anti-coordination game, we have

\begin{equation}
    p_{i \rightarrow i+1}^{0 \rightarrow 0} = \frac{N-i-1}{N},
\end{equation}
\begin{equation}
    p_{i \rightarrow i-1}^{1 \rightarrow 1} = \frac{i}{N},
\end{equation}
\begin{equation}
    p_{i \rightarrow i}^{1 \rightarrow 0} = \frac{1-H(\tau(N-1)-i)}{N},
\end{equation}
\begin{equation}
    p_{i \rightarrow i}^{0 \rightarrow 1} = \frac{1-H(i-\tau(N-1))}{N},
\end{equation}

where $H(\cdot)$ is the Heaviside step function and $N$ is the number of nodes on the star. Similarly, we have

\begin{equation}
    T_{i}^0 = p_{i \rightarrow i+1}^{0 \rightarrow 0} T_{i+1}^0 + p_{i \rightarrow i}^{0 \rightarrow 1} T_i^1 + (1- p_{i \rightarrow i+1}^{0 \rightarrow 0}- p_{i \rightarrow i}^{0 \rightarrow 1})T_i^0+1,
\end{equation}
\begin{equation}
    T_{i}^1 = p_{i \rightarrow i-1}^{1 \rightarrow 1} T_{i-1}^1 +  p_{i \rightarrow i}^{1 \rightarrow 0} T_i^0 +(1- p_{i \rightarrow i-1}^{1 \rightarrow 1}-  p_{i \rightarrow i}^{1 \rightarrow 0})T_{i}^1 +1. 
\end{equation}

with boundary conditions $T_0^1 = 0$ and $T_{N-1}^0 = 0$. And the conditional transition probability are

\begin{equation}
    \pi_{i \rightarrow i+1}^{0 \rightarrow 0} = 1-\pi_{i \rightarrow i}^{0 \rightarrow 1} = \frac{p_{i \rightarrow i+1}^{0 \rightarrow 0}}{p_{i \rightarrow i+1}^{0 \rightarrow 0} + p_{i \rightarrow i}^{0 \rightarrow 1}},
\end{equation}
\begin{equation}
    \pi_{i \rightarrow i}^{1 \rightarrow 0} = 1 - \pi_{i \rightarrow i-1}^{1 \rightarrow 1} = \frac{p_{i \rightarrow i}^{1 \rightarrow 0}}{p_{i \rightarrow i}^{1 \rightarrow 0} + p_{i \rightarrow i-1}^{1 \rightarrow 1}}.
\end{equation}

Rearranging the equations, we obtain

\begin{equation}
    T_{i}^1 = \pi_{i \rightarrow i-1}^{1 \rightarrow 1} T_{i-1}^1 +  \pi_{i \rightarrow i}^{1 \rightarrow 0} T_i^0 + \frac{1}{p_{i \rightarrow i}^{1 \rightarrow 0} + p_{i \rightarrow i-1}^{1 \rightarrow 1}},
\end{equation}
\begin{equation}
    T_{i}^0 = \frac{1}{\pi_{i-1 \rightarrow i}^{0 \rightarrow 0}} T_{i-1}^0 - \frac{\pi_{i-1 \rightarrow i-1}^{0 \rightarrow 1}}{\pi_{i-1 \rightarrow i}^{0 \rightarrow 0}} T_{i-1}^1 - \frac{1}{\pi_{i-1 \rightarrow i}^{0 \rightarrow 0} (p_{i-1 \rightarrow i}^{0 \rightarrow 0} + p_{i-1 \rightarrow i-1}^{0 \rightarrow 1})}
\end{equation}

Similarly, by solving inductively, we have

\begin{equation}
    T_{i}^{0} = C(1,i)T_1^0 - \sum_{l =2}^{i} C(l,i) D(l),
\end{equation}

where

\begin{equation}
    C(l,m) = 1+ \sum_{j=1}^{m-1} \pi_{j \rightarrow j}^{0 \rightarrow 1} \prod_{k =l}^{j} \frac{\pi^{1 \rightarrow 1}_{k \rightarrow k-1}}{\pi_{k \rightarrow k+1}^{0 \rightarrow 0}},
\end{equation}
\begin{equation}
    D(l) = \frac{\pi_{l-1 \rightarrow l-1}^{0 \rightarrow 1}}{\pi_{l-1 \rightarrow l}^{0 \rightarrow 0}} \sum_{j =1}^{l-1} (\frac{\prod_{k =j+1}^{l-1} \pi_{k \rightarrow k-1}^{1 \rightarrow 1}}{p_{j \rightarrow j-1}^{1 \rightarrow 1} + p_{j \rightarrow j}^{1 \rightarrow 0}}) + \frac{1}{p_{l-1 \rightarrow l}^{0 \rightarrow 0}}.
\end{equation}
Note that the equilibrium time calculated in this section may not equal the strategy switch number because an individual may not change his strategy after being activated.
\section{Bipartite graph}
Here we consider the complete bipartite graph where every node at the first(right) subset is connected to all the nodes at the second(left) subset.
In our complete bipartite graph model, each node (individual) at the left subset has the same threshold, denoted by $\tau_l$, and the count of these nodes is $n_l$. Conversely, the threshold of nodes at the right subset is $\tau_r$, with the number of nodes denoted as $n_r$.
Define the pair $(A_l, A_r)$ to represent the state of a bipartite graph, where $A_l$, taking values from ${0,1,\ldots, n_l}$, denotes the number of $A$-individuals at the left subset, and $A_r$, within the range ${0,1, \ldots, n_r}$, indicates the number of $A$-individuals at the right subset. 
For individual $i$ at the left subset, $n_i$ equals $A_r$ and $k_i = n_r$. 
Similarly, for individual $j$ at the right subset, $n_j=A_l$ and $k_j = n_l$. 
The total number of nodes in the graph, which equates to the total number of individuals, is given by $N = n_l + n_r$.

\subsection{Analysis of the number of equilibrium states}
In the coordination game, we find two(or three) equilibrium states:$(0, 0)$, $(n_l, n_r)$ and $(\tau_r n_l, \tau_l n_r)$ for the system, where the third equilibrium state $(\tau_r n_l, \tau_l n_r)$ is valid only if $\tau_r n_l$ and $\tau_l n_r$ are integers.
The equilibrium states $(n_l,0)$ and $(0,n_r)$ are readily verifiable. 
Taking $(n_l,0)$ as an example, the left individuals, satisfying $n_i=0 < \tau_l k_i=\tau_l n_r$, choose strategy $A$. 
In contrast, the right individuals, with $n_j=n_l > \tau_r k_j=\tau_r n_l$, adopt strategy $B$. 
Thus, all individuals on the left select strategy $A$, and those on the right select strategy $B$, ensuring the equilibrium of the system. 
The case of $(0,n_r)$ follows similarly.
In addition, if $\tau_l n_r$ and $\tau_r n_l$ are both integers, then $(\tau_r n_l, \tau_l n_r)$ is also an equilibrium state of the system.
Because according to the best-response dynamics, for both the left and right individuals, $n_i=\tau_l k_i=\tau_l n_r$ and $n_j=\tau_r k_j=\tau_r n_l$ holds, all individuals keep the current strategy unchanged.
It is important to note that the equilibrium state $(\tau_r n_l, \tau_l n_r)$ is unstable. Specifically, it cannot be reached from any adjacent state. This implies that any small perturbation can drive the system completely away from this equilibrium.

Here we explain why the equilibrium state $(\tau_r n_l, \tau_l n_r)$ is unstable.
It is easy to verify that starting from the four equilibrium states adjacent to it, the system never reaches the equilibrium state $(\tau_r n_l, \tau_l n_r)$, i.e

\begin{equation*}
    (\tau_r n_l -1,\tau_l n_r) \rightarrow (\tau_r n_l -1,\tau_l n_r+1),
\end{equation*}
\begin{equation*}
    (\tau_r n_l +1,\tau_l n_r) \rightarrow (\tau_r n_l +1,\tau_l n_r-1),
\end{equation*}
\begin{equation*}
    (\tau_r n_l ,\tau_l n_r -1) \rightarrow (\tau_r n_l +1,\tau_l n_r-1),
\end{equation*}
\begin{equation*}
    (\tau_r n_l ,\tau_l n_r +1) \rightarrow (\tau_r n_l -1,\tau_l n_r+1).
\end{equation*}

Concisely, in the state $(\tau_r n_l-1, \tau_l n_r)$, left individuals, for whom $n_i=A_r=\tau_l n_r=\tau_l k_i$, tend to remain their current strategy. 
In contrast, right individuals with $n_i=A_l=\tau_r n_l-1 < \tau_r n_l$ are likely to adopt strategy $A$ when activated. 
This invariably shifts the system from $(\tau_r n_l-1, \tau_l n_r)$ to $(\tau_r n_l-1, \tau_l n_r+1)$. 
Hence, under the assumption that a maximum of one individual updates its strategy per time step, the equilibrium state $(\tau_r n_l, \tau_l n_r)$ can never be reached unless it is the initial state of the system.
The mechanism is similar in the remaining cases.

Similarly, in the anti-coordination game, the system achieves equilibrium at the states $(n_l, 0)$ and $(0, n_r)$, representing scenarios where all individuals on one side adopt one strategy while individuals on the opposite side adopt the other. 
Additionally, a third potential equilibrium state, $(\tau_r n_l, \tau_l n_r)$, exists only when both $\tau_r n_l$ and $\tau_l n_r$ result in integer values.
\subsection{Analysis of the equilibrium probability}
Figure S18 illustrates the relationship between the number of $A$-individuals on the left and right sides and the equilibrium state in a bipartite graph in the anti-coordination game. 
The horizontal axis represents the number of $A$-individuals on the left and the vertical axis represents the number of $A$-individuals on the right, where the green(or blue) dots indicate that starting from this state, the system eventually reaches the state $(0,n_r)$ (or $(n_l,0)$).

Let $\alpha_A = \lceil \tau_r n_l \rceil$ and $\beta_A = \lceil \tau_l n_r \rceil$. 
We use the notation $x(A_l, A_r)$ to represent the probability that starting from state $(A_l, A_r)$, the system ultimately reaches the state $(n_l,0)$. 
Thus we obtain

\begin{equation}
\begin{aligned}
x(A_l,A_r)&=x(A_l+1,A_r)\cdot P^{l+}(A_l,A_r)+x(A_l-1,A_r)\cdot P^{l-}(A_l,A_r)\\
&+x(A_l,A_r+1)\cdot P^{r+}(A_l,A_r)+x(A_l,A_r-1)\cdot P^{r-}(A_l,A_r)\\
&+x(A_l,A_r)\cdot \{1- P^{l+}(A_l,A_r)-P^{l-}(A_l,A_r)-P^{r+}(A_l,A_r)-P^{r-}(A_l,A_r)
\},
\end{aligned}
\end{equation}

where $P^{l+}(A_l,A_r)$($P^{l-}(A_l,A_r)$) represents the probability that the number of left individuals choosing strategy $A$ increases(decreases) by one in the next time step from the state $(A_l, A_r)$.
Similarly, $P^{l+}(A_l,A_r)$ and $P^{l-}(A_l,A_r)$ represent the probabilities for the right individuals under the same conditions.

Consequently, the probability of reaching the state $(n_l,0)$ from any initial condition can be computed using the recurrence method

\begin{equation}
x(\alpha_A-m,\beta_A-k)=\sum_{t=1}^{k}\frac{\frac{(t+m-2)!}{(t-1)!}\cdot \frac{(n_l-\alpha_A+m)!}{(n_l-\alpha_A)!}\cdot \frac{(n_r-\beta_A+k)!}{(n_r-\beta_A+1+k-t)!}}{(m-1)!\cdot \frac{(N-\alpha_A-\beta_A+k+m)!}{(N-\alpha_A-\beta_A+k+1-t)!}}.
\end{equation}

Similarly, define $y(B_l,B_r)$ as the probability that the system starts from the state where $B_l $individuals on the left choose strategy $B$ and $B_r$ individuals on the right choose strategy $B$, and finally reaches a state where all the individuals on the left choose strategy $B$ and all the individuals on the right choose strategy $A$. 
Let $a=\lfloor \tau_r n_l \rfloor$, $b=\lfloor \tau_l n_r \rfloor$.
According to the symmetry, we obtain

\begin{equation}
y(\alpha_B-m,\beta_B-k)=\sum_{t=1}^{k}\frac{\frac{(t+m-2)!}{(t-1)!}\cdot \frac{(n_l-\alpha_B+m)!}{(n_l-a)!}\cdot \frac{(n_r-\beta_B+k)!}{(n_r-\beta_B+1+k-t)!}}{(m-1)!\cdot \frac{(N-\alpha_B-\beta_B+k+m)!}{(N-\alpha_B-\beta_B+k+1-t)!}}.
\end{equation}

Thus,

\begin{equation}
\begin{aligned}
    x(n_l-a+m,n_r-b+k) =& 1-y(\alpha_B-m,\beta_B-k)\\
    =& 1-\sum_{t=1}^{k}\frac{\frac{(t+m-2)!}{(t-1)!}\cdot \frac{(n_l-\alpha_B+m)!}{(n_l-a)!}\cdot \frac{(n_r-\beta_B+k)!}{(n_r-\beta_B+1+k-t)!}}{(m-1)!\cdot \frac{(N-\alpha_B-\beta_B+k+m)!}{(N-\alpha_B-\beta_B+k+1-t)!}}.
\end{aligned}
\end{equation}

Consequently, the probability of arriving at different equilibrium states from any initial condition in the anti-coordination game can be computed. 
This method is analogous to the one applied in the coordination game.
\subsection{Analysis of the robustness}
In this section, we analyze the robustness of the complete bipartite graph by adding new edges in equilibrium states.
According to the properties of the complete bipartite graph, the new edges can be only added between two nodes at the same subset.
\subsubsection{Coordination game}
In equilibrium states $(0,0)$ and $(n_l,n_r)$, all individuals adopts the same strategy.
Consequently, the addition of new edges does not result in any changes to their strategies.
Intuitively, the addition of new edges in states $(0,0)$ and $(n_l,n_r)$ has no effect on the proportion of individuals' neighbors employing the same strategy.
Actually, the proportion for any individual in states $(0,0)$ and $(n_l,n_r)$ is $1$.
Take state $(n_l,n_r)$ for an example, where all individuals adopt strategy $A$.
After adding new edges, for any individual $i$, $n_i > \tau_i k_i$ holds due to $n_i = k_i$ and $\tau_i < 1$.
Therefore, individual $i$ does not change its strategy Because all of its neighbors adopt strategy $A$.

However, the addition of new edges exerts a considerable effect on the equilibrium state $(\tau_r n_l, \tau_l n_r)$.
Here we assume $\tau_r n_l$ and $ \tau_l n_r$ are both integers to ensure the existence of equilibrium state $(\tau_r n_l, \tau_l n_r)$.
Randomly select two nodes to form a new edge. There are six cases (left side $A-A$, left side $B-B$, left side $A-B$, right side $A-A$, right side $B-B$, right side $A-B$). 
Here, the first three cases are discussed; the other three are similar.

\begin{enumerate}[I.]
\item \textbf{Left side $A-A$}

In this case, for the two $A$ nodes with new edges,
\begin{equation}
\left.
\begin{aligned}
n_i &= \tau_l n_r + 1 \\
k_i &= n_r + 1 \\
\end{aligned}
\right\}
\Rightarrow n_i > \tau k_i
\end{equation}
Thus, both $A$ nodes tend to choose the $A$ strategy, keeping their strategy unchanged; the neighbor count and state of other nodes remain unchanged, so the system's state does not change.

\item \textbf{Left side $B-B$}

Similarly, for the $B$ nodes with new edge,
\begin{equation}
\left.
\begin{aligned}
n_i &= \tau_l n_r \\
k_i &= n_r + 1 \\
\end{aligned}
\right\}
\Rightarrow n_i < \tau k_i
\end{equation}
Thus, both $B$ nodes tend to choose the $B$ strategy, keeping their strategy unchanged; the neighbor count and state of other nodes remain unchanged, so the system's state does not change.

\item \textbf{Left side $A-B$}

In this case, one $A$ node and one $B$ node are chosen to be connected, labeling the $A$ node as node one and the $B$ node as node two. Then,
\begin{enumerate}[1.]
    \item For node one
\begin{equation}
\left.
\begin{aligned}
n_i &= \tau_l n_r \\
k_i &= n_r + 1 \\
\end{aligned}
\right\}
\Rightarrow n_i < \tau k_i
\end{equation}
If node one is activated in the next time step, it switches its strategy to $B$.
    \item For node two
\begin{equation}
\left.
\begin{aligned}
n_i &= \tau_l n_r + 1 \\
k_i &= n_r + 1 \\
\end{aligned}
\right\}
\Rightarrow n_i > \tau k_i
\end{equation}  
If node two is activated in the next time step, it switches its strategy to $A$.
\end{enumerate}
Moreover, for individual one and individual two in this scenario, $k_i = n_r + 1$, the situation $n_i = \tau k_i$ never occurs. 

In this case, if the $A$ node (node one) is activated first, the system definitely reaches the state $(0,0)$
\begin{proof}
    Assuming node one is activated first, according to the best response dynamics, individual one switches its strategy to $B$. The state of the system becomes $(\tau_r n_l - 1, \tau_l n_r)$.
    
    At this point, for all other nodes on the left side, except for individual one and individual two, the equation $n_i = \tau_l n_r$ holds.
    For individual one and individual two, the inequality $n_i = \tau_l n_r < \tau_l k_i = \tau_l (n_r + 1)$ holds, and these two nodes currently adopt strategy $B$, so all nodes on the left side keep their current state unchanged.

    For the nodes on the right side, the inequality $n_j = \tau_r n_l - 1 < \tau_r n_l$ holds. Thus, according to best response dynamics, all nodes on the right side tend to switch their strategy to $B$. When any right-side node is activated, if its strategy is $B$, it remains unchanged; if its strategy is $A$, it switches to $B$.

    Therefore, starting from the state $(\tau_r n_l - 1, \tau_l n_r)$, the only reachable state is $(\tau_r n_l - 1, \tau_l n_r - 1)$

    Now, for individual one and individual two, the inequality $n_i = \tau_l n_r - 1 < \tau_l k_i = \tau_l (n_r + 1)$ holds: for the other nodes on the left side, except for individual one and individual two, the inequality $n_i = \tau_l n_r - 1 < \tau_l k_i = \tau_l n_r$ holds; for the nodes on the right side, the inequality $n_i = \tau_r n_l - 1 < \tau_r n_l$ holds. 
    Thus, at this time and all subsequent times, for any node, we have
    
    \begin{equation}
        n_i < \tau k_i.
        \label{coor_bestresponse1}
    \end{equation}
    Thus, all nodes tend to adopt strategy $B$. 
    Therefore, only when the system's state reaches $(0,0)$ does it stabilize.

\end{proof}
Similarly, if the $B$ node (individual two) is first activated, the system definitely reaches the state $(n_l, n_r)$.
\end{enumerate}
Additionally, it is straightforward to calculate the probability that the system reaches the state $(0,0)$ after adding a random edge:

\begin{equation}
p=\frac{C^1_{\tau_r n_l} C^1_{n_l - \tau_r n_l}+C^1_{\tau_l n_r} C^1_{n_r-\tau_l n_r}}{C^2_{n_r}+C^2_{n_l}} \times \frac{1}{2},
\end{equation}

and the probability that the system reaches the state $(n_l, n_r)$ after adding a random edge:

\begin{equation}
    p=\frac{C^1_{\tau_r n_l} C^1_{n_l - \tau_r n_l}+C^1_{\tau_l n_r} C^1_{n_r-\tau_l n_r}}{C^2_{n_r}+C^2_{n_l}} \times \frac{1}{2}.
\end{equation}

\subsubsection{Anti-coordination game}
First, we analyze the threshold of adding new edges for the strategy switching in equilibrium states. 
Consider that the initial state is $(0,n_r)$, adding an edge on the left influences only the connected nodes' strategies.
Suppose a single edge addition triggers a strategy switch, then we obtain

\begin{equation}
    n_i=n_r< \tau_l k_i= \tau_l (n_r+1) \quad \Rightarrow \quad \tau_l> \frac{n_r}{n_r+1}.
\end{equation}
If a minimum of two edges must be added to the same node for a strategy switch, the following conditions must hold

\begin{equation}
\left.
\begin{aligned}
n_i =n_r  &\geq \tau_l (n_r+1) \\
n_i =n_r  &< \tau_l k_i= \tau_l (n_r+2)\\
\end{aligned}
\right\} \quad
\Rightarrow \quad \frac{n_r}{n_r+1} \geq \tau_l > \frac{n_r}{n_r+2}.
\end{equation}
In general, if $K$ edges are required on the left for a switch

\begin{equation}
     \frac{n_r}{n_r+K-1} \geq \tau_l > \frac{n_r}{n_r+K}.
\end{equation}
And if $K$ edges are required on the right

\begin{equation}
     \frac{K-1}{n_l+K-1} \leq \tau_l < \frac{K}{n_l+K}.
\end{equation}

Next, we discuss the evolution of the system from state $(0,n_r)$ after the strategy switch occurs. 
We initially consider the cases that adding a single edge on the left side can trigger a strategy switch(that is, $\tau_l> n_r/(n_r+1)$ ). 
After activation of one vertex of the new edge, the individual at this node switches its strategy to $A$. 
For another vertex of the new edge, it follows that

\begin{equation}
n_i = n_r+1 > \tau_l k_i = \tau_l (n_r+1).
\end{equation}

For other nodes on the left, we obtain

\begin{equation}
n_i = n_r > \tau_l k_i = \tau_l n_r.
\end{equation}
Clearly, when they are activated, they all remain their current strategy.

We now classify based on the value of $\tau_r$, to discuss the evolution of the system.

\begin{enumerate}[I.]
    \item $\tau_r \geq \frac{1}{n_l}$

    In this case, for the right-side nodes,
    
    \begin{equation}
    n_j = 1 \leq \tau_r k_j = \tau_r n_l.
    \end{equation}
    According to the best response dynamics, the right-side nodes tend to maintain their current state(when $n_j = \tau_r n_l$) or choose the $A$ strategy(when $n_j < \tau_r n_l$). 
    Given that the initial state of the right-side nodes is all $A$, combined with the analysis above, the overall system state remains unchanged. 
    That is, after one individual at the vertex of the new edge updates its strategy to $A$, the system reaches a new equilibrium.

    \item $\tau_r < \frac{1}{n_l}$

    In this case, for the right-side nodes, it follows that
    
    \begin{equation}
    n_j = 1 > \tau_r k_j = \tau_r n_l.
    \end{equation}
    After activation, a node on the right switches its strategy from $A$ to $B$.

    Obviously, when the population on both sides is sufficiently large, both conditions $\tau_l> \frac{n_r}{n_r+1}$ and $n_j = 1 > \tau_r k_j = \tau_r n_l$ are rather stringent.
    Thus, for most systems, the equilibrium state is not easily disrupted, or even if disrupted, does not result in large-scale changes.

    Next, we discuss the final state to which the system evolves. 
    Starting from the state $(1,n_r-1)$, the system eventually reach equilibrium state $(n_l,0)$.

    First, we prove that $(n_l,0)$ is the equilibrium state of the system after adding a new edge.
    \begin{proof}
        For nodes on the right side, the inequality $n_i = n_l > \tau_r n_l$ clearly holds, so individuals on the right-side choose the $B$ strategy, maintaining their state unchanged.

        For nodes on the left side with no addition of the new edge, $n_i = 0 < \tau_l n_r$ clearly holds, so they adopt the $A$ strategy, also remaining unchanged.

        For nodes on the left side with the addition of the new edge, $n_i = 1 < \tau_l (n_r+1)$ also holds, so they choose the $A$ strategy, remaining unchanged as well.

        Thus, when the system is in state $(n_l,0)$, no matter which node is activated, they do not change their strategy. Therefore, state $(n_l,0)$ is one equilibrium state of the system.
    \end{proof}

    Next, we prove that starting from $(1,n_r-1)$, the system eventually reach the equilibrium state $(n_l,0)$.

    \begin{proof}
    We explain this from two perspectives:
    \begin{itemize}
        \item First, when $A_r \neq n_r$, we conclude that the individuals at left-side nodes all tend to choose strategy $A$.

        For left-side nodes with the addition of the new edge, we have
        
        \begin{equation}
        \left.
        \begin{aligned}
            n_i &= A_r + a, \quad a \in \{ 0,1\} \\
            \tau_l k_i &= \tau_l (n_r+1) > n_r\\
            A_r &\leq n_r -1
        \end{aligned}
        \right \}
        \Rightarrow \tau_l k_i > n_i.
        \end{equation}

        For other nodes on the left, we get
        
        \begin{equation}
        \left.
        \begin{aligned}
            n_i &= A_r \\
            \tau_l k_i &= \tau_l n_r > \frac{n_r^2}{n_r+1} > n_r -1\\
            A_r &\leq n_r -1
        \end{aligned}
        \right \}
        \Rightarrow \tau_l k_i > n_i.
        \end{equation}

        Thus, all individuals at left-side nodes tend to choose the $A$ strategy.
        \item Second, when $A_l \neq 0$, we say that all individuals at right-side nodes tend to choose the strategy $B$.

        For any right-side individuals, we have
        
        \begin{equation}
        \left.
        \begin{aligned}
            n_i &= A_l \\
            \tau_r k_i &= \tau_r n_l < 1\\
            A_l &\geq 1
        \end{aligned}
        \right \}
        \Rightarrow \tau_r k_i < n_i
        \end{equation}
        Therefore, all right-side individuals tend to choose the $B$ strategy.

        \item In summary, starting from $(1,n_r-1)$, $A_r \neq n_r$ and $A_l \neq 0$ are satisfied.
        According to the above statement, when a left-side $A$ individual is activated, $A_l$ remains unchanged, and when a left-side $B$ individual is activated, $A_l$ increases; when a right-side $A$ individual is activated, $A_r$ decreases, and when a left-side $B$ individual is activated, $A_r$ remains unchanged, thus $A_r \neq n_r$ and $A_l \neq 0$ are always met. 
        Therefore, the system reaches equilibrium only when there are no $B$ individuals on the left side and no $A$ individuals on the right side, which results in the system's state remaining unchanged whichever individual is activated.

        Therefore, starting from $(1,n_r-1)$, the system eventually reaches state $(n_l,0)$.
    \end{itemize}
    \end{proof}
\end{enumerate}

Furthermore, we analyze the scenario where the initial state is $(0, n_r)$, and adding at least $f$ edges to a node on the left side is required to change the system state.
At this time, the following inequality must hold

\begin{equation}
\frac{n_r}{n_r+f-1} \geq \tau_l > \frac{n_r}{n_r+f}.
\end{equation}

Consider the case where exactly $f$ edges are added to a particular node on the left side, while the number of new edges added to other nodes on the left is less than $f$. 
When the node with $f$ new edges is activated, it switches its strategy to $A$, changing the system's state to $(1, n_r)$.

For the other nodes on the left side, we obtain

\begin{equation}
        \left.
        \begin{aligned}
            n_i &= n_r + 1 \\
            k_i &= n_r + g_i\\
            0 \leq &g_i < f\\
        \end{aligned}
        \right \}
        \Rightarrow n_i > \tau_l (n_r+g_i).
\end{equation}
Therefore, other nodes on the left adopt strategy $B$, which means they maintain their strategy unchanged.

For nodes on the right side, where $n_i = 1$, whether the strategy is updated or not depends on their threshold. 
We focus on the case when $\tau_r < \frac{1}{n_l}$ otherwise no individual at right-side node changes its strategy. 
In this case, when a node on the right is activated, it switches to strategy $B$, hence the system's state becomes $(1, n_r - 1)$. 
From this state, the subsequent evolution process of the system is discussed in the following.

For the right-side nodes, we obtain

\begin{equation}
\tau_r n_l < 1,
\end{equation}
implying that as long as not all nodes on the left adopt strategy $B$, the right-side nodes tend to choose strategy $B$, continuously reducing $A_r$.

For the node on the left with $f$ new edges, we have

\begin{equation}
    n_i = A_r + a, \quad 0 \leq a \leq f, 
    \end{equation}
    \begin{equation}
    \tau_l  (n_r + f) > n_r.
\end{equation}

For a node on the left with $g_i$ new edges added:
\begin{equation}
n_i = A_r + a, \quad 0 \leq a \leq g_i,
    \end{equation}
    \begin{equation}
\tau_l  (n_r + g_i) > n_r \frac{n_r+g_i}{n_r+f} = n_r - n_r \frac{f-g_i}{n_r+f} > n_r - f + g_i.
\end{equation}

It can be observed that if $n_r \geq f$ is guaranteed, then adding $h$ edges to the left side, some of which have $f$ edges on certain nodes while others have fewer than $f$, the bipartite graph evolve from the state $(0, n_r)$ to $(n_l, 0)$.(\textbf{Necessary and sufficient condition})

\begin{proof}

\begin{enumerate}[I.]
    \item Sufficiency

We prove the sufficiency by contradiction. 
Assume if $n_r \geq f$, starting from $(0, n_r)$, the system cannot eventually reach $(n_l, 0)$, then one of the following conditions must be true in the equilibrium state: $A_r \neq 0$ or $A_l \neq n_l$.

\begin{itemize}
    \item If $A_r \neq 0$, according to the best response dynamics, this implies $A_l = 0$ in the equilibrium state, otherwise the state is not an equilibrium state. 
    For the node on the left with $f$ new edges, we get

    \begin{equation}
    n_i = A_r + 0 = A_r \leq n_r < \tau_l (n_r + f).
     \end{equation}
     
   The inequity means the node tends to choose strategy $A$, indicating that this is not an equilibrium state, which contradicts the assumption that the system has reached an equilibrium state.

    \item If $A_l \neq n_l$, then there are only two possible scenarios: $A_l = 0, A_r = n_r$ or $A_l \neq 0, A_r = 0$.

    In the first scenario, according to best response dynamics, it is clear that this is not an equilibrium, a contradiction.
    
    In the second scenario, there exists a node with $g_i$ new edges and the best response dynamics lead it to choose strategy $B$, i.e.
    
    \begin{equation}
    \tau_l (n_r + g_i) \leq A_r + a = a,
    \end{equation}
    where $a \leq g_i$ and $\tau_l (n_r + k) > \frac{n_r (n_r + k)}{n_r + f} \geq \frac{n_r^2}{n_r + f} \geq \frac{n_r^2}{n_r + n_r} \geq n_r$.
    Therefore, even if $a$ reaches its maximum, it must hold that
    
    \begin{equation}
    n_r < k \leq f,
    \end{equation}
    which contradicts the condition.

\end{itemize}
Overall, the assumption does not hold, thus if $n_r \geq f$, starting from $(0, n_r)$, the system definitely reaches $(n_l, 0)$.

\item Necessity

Again, we prove it by contradiction. 
Assume that starting from the state $(0, n_r)$, the system can reach $(n_l, 0)$, but $n_r < f$.

According to best response dynamics, when the system is at equilibrium in the state $(n_l, 0)$, for an individual on the left with $f$ new edges, it must hold

\begin{equation}
n_i = f \leq \tau_l (n_r + f).
\end{equation}

Given the range for $\tau_l$, we know

\begin{equation}
n_r + 1 \geq \tau_l (n_r + f) > n_r.
\end{equation}

Also, since $f$ is an integer, under the assumption, we obtain

\begin{equation}
f \geq n_r + 1,
\end{equation}
which makes

\begin{equation}
f \geq \tau_l (n_r + f).
\end{equation}

Under these circumstances, only if $f = \tau_l (n_r + f)$, when exactly $f$ edges are added to the left, the system remains in the state $(0, n_r)$ with no change in strategy.
This contradicts that at least one individual tends to switch his strategy, thus the assumption does not hold.

Therefore, if starting from $(0, n_r)$, the system can reach $(n_l, 0)$, then it must be that $n_r \geq f$.

\end{enumerate}

In conclusion, the necessary and sufficient condition is proven.

\end{proof}

The case when the initial state is $(n_l,0)$ follows a similar analysis.

In addition, given the initial state $(\tau_l n_r, \tau_r n_l)$, the analysis is similar to that in the coordination game. 
That is, randomly selecting two nodes to form a new edge, there are six cases (left side $A-A$, left side $B-B$, left side $A-B$, right side $A-A$, right side $B-B$, right side $A-B$). 
We mainly focus on the first three cases since the analysis for the next three cases is analogous.
\begin{enumerate}[I.]
    \item \textbf{Left side A-A}
    For the two endpoints of the new edge, $n_i = \tau_l n_r + 1 > \tau_l (n_r + 1)$. The initially activated node switches to strategy $B$. And the system eventually stabilizes at the state $(0, n_r)$.
    \item \textbf{Left side B-B}
    For the two endpoints of the new edge, $n_i = \tau_l n_r < \tau_l (n_r + 1)$. The initially activated node switches to strategy $A$. The system stabilizes at the state $(n_l, 0)$.
    \item \textbf{Left side A-B}
    For the $A$ node of the new edge, $n_i = \tau_l n_r < \tau_l (n_r + 1)$, leading to no strategy change. For the $B$ node, $n_i = \tau_l n_r + 1 > \tau_l (n_r + 1)$, its strategy remains unchanged. Thus, the system state does not change.
\end{enumerate}

\section{Rich-club}
In rich club networks, some nodes exhibit exceptionally high degrees while others display relatively low degrees.
Focusing on a specific type of rich-club network where $n_r$ rich nodes are fully connected and $n_p$ poor nodes link to all rich nodes ($n_r +n_p =N$), the state (strategy composition) of the network can be described by a $2$-tuple $(A_r, A_p)$, where $A_r$ and $A_p$ denote the number of $A$-individuals among the rich nodes and the poor nodes respectively. 
Let $\tau_p$ and $\tau_r$ denote the behavioral switching threshold, respectively.
\subsection{Analysis of the number of equilibrium states}
\subsubsection{Coordination game}
There are only two equilibrium states $(0,0)$ and $(n_r, n_p)$. 
The analysis is similar to that in Section 5.1.2.
\subsubsection{Anti-coordination game}
For poor nodes: $n_i = A_r, k_i = n_r$. 
Therefore, we obtain
\begin{equation}
s_i(t+1) = \left\{
\begin{aligned}
A &, \quad A_r < \tau_p n_r \\
s_i(t) &, \quad A_r = \tau_p n_r \\
B &, \quad A_r > \tau_p n_r
\end{aligned}
\right.
\end{equation}

Consider the rich nodes divided into two groups, $A$-rich nodes and $B$-rich nodes, with counts $A_r$ and $n_r - A_r$ respectively.
For $A$-rich nodes, we know $n_j= A_p + A_r - 1, k_j = n_r + n_p - 1$, thus
\begin{equation}
s_i(t+1) = \left\{
\begin{aligned}
A &, \quad A_p + A_r - 1 < \tau_r (n_r + n_p - 1) \\
s_i(t) &, \quad A_p + A_r - 1 = \tau_r (n_r + n_p - 1) \\
B &, \quad A_p + A_r - 1 > \tau_r (n_r + n_p - 1)
\end{aligned}
\right.
\end{equation}

For $B$-rich nodes, we have $n_i = A_p + A_r, k_i = n_r + n_p - 1$, therefore
\begin{equation}
s_i(t+1) = \left\{
\begin{aligned}
A &, \quad A_p + A_r < \tau_r (n_r + n_p - 1) \\
s_i(t) &, \quad A_p + A_r = \tau_r (n_r + n_p - 1) \\
B &, \quad A_p + A_r > \tau_r (n_r + n_p - 1)
\end{aligned}
\right.
\end{equation}

Based on the above equations, fifteen different scenarios can be discussed, and we find there are 11 types of equilibrium states (the number may be fewer because some equilibrium states could be merged).

\begin{enumerate} [I.]
\item $A_r < \tau_p n_r$, $A_p + A_r - 1 < \tau_r (n_p + n_r - 1)$, and $A_p + A_r > \tau_r (n_p + n_r - 1)$

poor nodes choose strategy $A$, $A$-rich nodes choose strategy $A$, and $B$-rich nodes choose strategy $B$.

To ensure network equilibrium, the number of $A$-poor nodes must be $n_p$, i.e., $A_p = n_p$.

For rich nodes, it must satisfy:
\begin{equation}
\tau_r (n_p + n_r - 1) - n_p < A_r < \tau_r (n_p + n_r - 1) - n_p + 1
\end{equation}
To ensure $0 \leq A_r \leq n_r$ and $A_r$ is an integer, $\tau_r(n_p + n_r - 1)$ must not be an integer and $\tau_r > \frac{n_p - 1}{n_p + n_r - 1}$.

Thus, $A_r = \lceil \tau_r (n_p + n_r - 1) - n_p \rceil = \lfloor \tau_r (n_p + n_r - 1) - n_p + 1 \rfloor$.

To ensure $A_r < \tau_p n_r$, it must satisfy $\tau_p > \frac{\lceil \tau_r (n_p + n_r - 1) - n_p \rceil}{n_r}$.

Hence, when these conditions are met, the system can balance with $n_p$ $A$-poor nodes and $\lceil \tau_r (n_p + n_r - 1) - n_p \rceil$ $A$-rich nodes, establishing an equilibrium state of $(\lceil \tau_r (n_p + n_r - 1) - n_p \rceil, n_p)$.

\item $A_r < \tau_p n_r$, $A_p + A_r - 1 = \tau_r (n_p + n_r - 1)$, and $A_p + A_r > \tau_r (n_p + n_r - 1)$

poor nodes choose strategy $A$, $A$-rich nodes maintain their strategy, and $B$-rich nodes choose strategy $B$.

To make these conditions valid while ensuring $0 \leq A_r \leq n_r$ and $A_r$ is an integer, it must be ensured that:
\begin{itemize}
    \item $\tau_r(n_p + n_r - 1)$ is an integer.
    \item $\tau_r \geq \frac{n_p - 1}{n_p + n_r - 1}$.
    \item $\tau_p > \frac{\tau_r (n_p + n_r - 1) - n_p + 1}{n_r}$.
\end{itemize}
Under these conditions, the system can balance with $n_p$ $A$-poor nodes and $\tau_r (n_p + n_r - 1) - n_p + 1$ $A$-rich nodes, establishing an equilibrium state of $(\tau_r (n_p + n_r - 1) - n_p + 1, n_p)$.

\item $A_r < \tau_p n_r$, $A_p + A_r - 1 < \tau_r (n_p + n_r - 1)$, and $A_p + A_r = \tau_r (n_p + n_r - 1)$

poor nodes choose strategy $A$, $A$-rich nodes choose strategy $A$, and $B$-rich nodes maintain their strategy.

To make these conditions valid while ensuring $0 \leq A_r \leq n_r$ and $A_r$ is an integer, it must be ensured that

\begin{itemize}
    \item $\tau_r(n_p + n_r - 1)$ is an integer.
    \item $\tau_r \geq \frac{n_p}{n_p + n_r - 1}$.
    \item $\tau_p > \frac{\tau_r (n_p + n_r - 1) - n_p}{n_r}$.
\end{itemize}
Under these conditions, the system can balance with $n_p$ $A$-poor nodes and $\tau_r (n_p + n_r - 1) - n_p$ $A$-rich nodes, establishing an equilibrium state of $(\tau_r (n_p + n_r - 1) - n_p, n_p)$.

\item $A_r < \tau_p n_r$, $A_p+A_r-1> \tau_r (n_p+n_r-1)$, $A_p+A_r> \tau_r (n_p+n_r-1)$

poor nodes choose strategy $A$, $A$-rich nodes choose strategy $B$, and $B$-rich nodes also choose strategy $B$.

To achieve system equilibrium under these conditions, all poor nodes must adopt strategy $A$, and all rich nodes must adopt strategy $B$, i.e., $A_p=n_p$, $A_r=0$. Substituting these values into the inequalities yields:
\begin{itemize}
    \item $\tau_r < \frac{n_p-1}{n_p+n_r-1}$
    \item $\tau_p > 0$ (obviously)
\end{itemize}
With these conditions met, the system can reach equilibrium with $n_p$ poor nodes choosing $A$ and $0$ rich nodes choosing $A$, resulting in an equilibrium state of $(0,n_p)$.

\item $A_r < \tau_p n_r$, $A_p+A_r-1< \tau_r (n_p+n_r-1)$, $A_p+A_r< \tau_r (n_p+n_r-1)$

poor nodes choose strategy $A$, $A$-rich nodes choose strategy $A$, and $B$-rich nodes also choose strategy $A$. Therefore, the system will definitely not be in equilibrium.

\item $A_r > \tau_p n_r$, $A_p+A_r-1< \tau_r (n_p+n_r-1)$, $A_p+A_r> \tau_r (n_p+n_r-1)$

poor nodes choose strategy $B$, $A$-rich nodes choose strategy $A$, and $B$-rich nodes choose strategy $B$. In order for the system to balance, it must be that $A_p=0$ and
$$
\tau_r (n_p+n_r-1) < A_r < \tau_r (n_p+n_r-1)+1
$$
To ensure $0 \leq A_r \leq n_r$ and that $A_r$ is an integer, and $A_r > \tau_p n_r$:
\begin{itemize}
    \item $\tau_r(n_p+n_r-1)$ is not an integer
    \item $\tau_r < \frac{n_r}{n_p+n_r-1}$
    \item $\tau_p < \frac{\lceil \tau_r (n_p+n_r-1) \rceil}{n_r}$
\end{itemize}
Under these conditions, the system can reach equilibrium with $0$ poor nodes choosing $A$ and $\lceil \tau_r (n_p+n_r-1) \rceil$ rich nodes choosing $A$, resulting in an equilibrium state of $(\lceil \tau_r (n_p+n_r-1) \rceil,0)$.

\item $A_r > \tau_p n_r$, $A_p+A_r-1= \tau_r (n_p+n_r-1)$, $A_p+A_r> \tau_r (n_p+n_r-1)$

poor nodes choose strategy $B$, $A$-rich nodes remain unchanged, and $B$-rich nodes choose strategy $B$.

To ensure that the above conditions hold, while also ensuring that $0 \leq A_r \leq n_r$ and that $A_r$ is an integer, the following must be satisfied:
\begin{itemize}
    \item $\tau_r(n_p+n_r-1)$ is an integer
    \item $\tau_r \leq \frac{n_r-1}{n_p+n_r-1}$
    \item $\tau_p < \frac{\tau_r (n_p+n_r-1) +1}{n_r}$
\end{itemize}
With these conditions, the system can reach equilibrium with $0$ poor nodes choosing $A$ and $\tau_r (n_p+n_r-1) +1$ rich nodes choosing $A$, resulting in an equilibrium state of $(\tau_r (n_p+n_r-1) +1,0)$.

\item $A_r > \tau_p n_r$, $A_p+A_r-1< \tau_r (n_p+n_r-1)$, $A_p+A_r= \tau_r (n_p+n_r-1)$

poor nodes choose strategy $B$, $A$-rich nodes choose strategy $A$, and $B$-rich nodes remain unchanged.

To ensure these conditions hold, while also ensuring that $0 \leq A_r \leq n_r$ and that $A_r$ is an integer, the following must be satisfied:
\begin{itemize}
    \item $\tau_r(n_p+n_r-1)$ is an integer
    \item $\tau_r \leq \frac{n_r}{n_p+n_r-1}$
    \item $\tau_p < \frac{\tau_r (n_p+n_r-1)}{n_r}$
\end{itemize}
With these conditions, the system can achieve equilibrium with $0$ poor nodes choosing $A$ and $\tau_r (n_p+n_r-1)$ rich nodes choosing A, resulting in an equilibrium state of $(\tau_r (n_p+n_r-1), 0)$.

\item $A_r > \tau_p n_r$, $A_p+A_r-1> \tau_r (n_p+n_r-1)$, $A_p+A_r> \tau_r (n_p+n_r-1)$

poor nodes choose strategy $B$, $A$-rich nodes choose strategy $B$, and $B$-rich nodes also choose strategy $B$. Thus, the system will never be in equilibrium.

\item For the scenario where $A_r > \tau_p n_r$, $A_p+A_r-1< \tau_r (n_p+n_r-1)$, $A_p+A_r< \tau_r (n_p+n_r-1)$:
poor nodes choose strategy $B$, $A$-rich nodes choose strategy $A$, and $B$-rich nodes choose strategy $A$. 

To achieve system equilibrium under these conditions, all poor nodes must adopt strategy $B$, and all rich nodes must adopt strategy $A$, i.e., $A_p=0$, $A_r=n_r$. Substituting these values into the inequalities yields:
\begin{itemize}
    \item $\tau_r > \frac{n_r}{n_p+n_r-1}$
    \item $\tau_p < 1$ (obviously)
\end{itemize}
With these conditions met, the system can reach equilibrium with $0$ poor nodes choosing $A$ and $n_r$ rich nodes choosing $A$, resulting in an equilibrium state of $(n_r,0)$.

\item $A_r = \tau_p n_r$, $A_p+A_r-1< \tau_r (n_p+n_r-1)$, $A_p+A_r> \tau_r (n_p+n_r-1)$

poor nodes maintain their current strategy, $A$-rich nodes choose strategy $A$, and $B$-rich nodes choose strategy $B$.

At the same time, $A_r= \tau_p n_r$ and $\tau_r(n_p+n_r-1)-\tau_p n_r < A_p < \tau_r(n_p+n_r-1)-\tau_p n_r +1$

To ensure $0 \leq A_p \leq n_p$ and that all conditions are satisfied, it must be ensured that:
\begin{itemize}
    \item $\tau_r(n_p+n_r-1)$ is not an integer
    \item $\tau_p n_r$ is an integer
    \item $\frac{\tau_p n_r -1}{n_p+ n_r -1} <\tau_r < \frac{n_p+ \tau_p n_r}{n_p +n_r -1}$ or $\frac{\tau_r(n_p+ n_r -1)-n_p}{ n_r } <\tau_p < \frac{\tau_r( n_r +n_p-1)+1}{n_r}$
\end{itemize}
Under these conditions, the system can achieve equilibrium with $\lceil \tau_r (n_p+n_r-1)-\tau_p n_r \rceil$ poor nodes choosing $A$ and $\tau_p n_r$ rich nodes choosing $A$, resulting in an equilibrium state of $(\tau_p n_r,\lceil \tau_r (n_p+n_r-1)-\tau_p n_r \rceil)$.

\item $A_r = \tau_p n_r$, $A_p+A_r-1= \tau_r (n_p+n_r-1)$, $A_p+A_r> \tau_r (n_p+n_r-1)$

poor nodes maintain their current strategy, $A$-rich nodes remain unchanged, and $B$-rich nodes choose strategy $B$.

To ensure $0 \leq A_p \leq n_p$ and that all conditions are satisfied, the following must be ensured:
\begin{itemize}
    \item $\tau_r(n_p+n_r-1)$ is an integer
    \item $\tau_p n_r$ is an integer
    \item $\frac{\tau_p n_r -1}{n_p+ n_r -1} \leq \tau_r \leq \frac{n_p+ \tau_p n_r -1}{n_p +n_r -1}$ or $\frac{\tau_r(n_p+ n_r -1)-n_p +1}{ n_r } \leq \tau_p \leq \frac{\tau_r( n_r +n_p-1)+1}{n_r}$
\end{itemize}
With these conditions met, the system can achieve equilibrium with $ \tau_r (n_p+n_r-1)-\tau_p n_r +1$ poor nodes choosing $A$ and $\tau_p n_r$ rich nodes choosing $A$, resulting in an equilibrium state of $(\tau_p n_r,\tau_r (n_p+n_r-1)-\tau_p n_r +1)$.

\item $A_r = \tau_p n_r$, $A_p+A_r-1< \tau_r (n_p+n_r-1)$, $A_p+A_r= \tau_r (n_p+n_r-1)$

poor nodes maintain their current strategy, $A$-rich nodes choose strategy $A$, and $B$-rich nodes remain unchanged.

To ensure $0 \leq A_p \leq n_p$ and that all conditions are satisfied, the following must be ensured:
\begin{itemize}
    \item $\tau_r(n_p+n_r-1)$ is an integer
    \item $\tau_p n_r$ is an integer
    \item $\frac{\tau_p n_r }{n_p+ n_r -1} \leq \tau_r \leq \frac{n_p+ \tau_p n_r }{n_p +n_r -1}$ or $\frac{\tau_r(n_p+ n_r -1)-n_p }{ n_r } \leq \tau_p \leq \frac{\tau_r( n_r +n_p-1)}{n_r}$
\end{itemize}
With these conditions met, the system can achieve equilibrium with $ \tau_r (n_p+n_r-1)-\tau_p n_r $ poor nodes choosing $A$ and $\tau_p n_r$ rich nodes choosing $A$, resulting in an equilibrium state of $(\tau_p n_r,\tau_r (n_p+n_r-1)-\tau_p n_r)$.

\item $A_r = \tau_p n_r$, $A_p+A_r-1> \tau_r (n_p+n_r-1)$, $A_p+A_r> \tau_r (n_p+n_r-1)$

The system is not in equilibrium.

\item $A_r = \tau_p n_r$, $A_p+A_r-1< \tau_r (n_p+n_r-1)$, $A_p+A_r< \tau_r (n_p+n_r-1)$

The system is not in equilibrium.

\end{enumerate}

\subsection{Analysis of the robustness}
\subsubsection{Coordination game}
There are only two equilibrium states $(0,0)$ and $(n_r,n_p)$.
Obviously, adding an edge does not have any effect.
\subsubsection{Anti-coordination game}
For all the conditions that can make the system in equilibrium, where I, II, III, and IV are similar to VI, VII, VIII, and X, only I, II, III, and IV are discussed, and the analysis last four are similar to them; XI, XII, XIII is the same case, we just consider XI.

\begin{itemize}
    \item For Scenario I, where the equilibrium state is $A_r=\lceil \tau_r(n_p+n_r-1)-n_p \rceil$, $A_p=n_p$, the conditions $\tau_r > \frac{n_p -1}{n_p+n_r-1}$ and $\tau_p > \frac{\lceil \tau_r(n_p+n_r-1)-n_p \rceil}{n_r}$ must hold. Considering the rich-club property, where all rich nodes are already connected to every other node, new edges can only be added between two poor nodes.

    In this situation, after adding an edge, the system's state change and the poor nodes' threshold $\tau_p$ are crucial. Examining the addition of a single edge, if an edge is added between two poor nodes, the system remains unchanged if $\tau_p \geq \frac{\lceil \tau_r(n_p+n_r-1)-n_p+1 \rceil}{n_r+1}$. If $\tau_p < \frac{\lceil \tau_r(n_p+n_r-1)-n_p+1 \rceil}{n_r+1}$, according to best response dynamics, one endpoint of the new edge switches to strategy $B$, causing a $B$- rich node to switch to strategy $A$, i.e., $A_r=\lceil \tau_r(n_p+n_r-1)-n_p \rceil+1, A_p=n_p-1$. Subsequently, poor nodes not connected by the new edge change their strategies according to best response dynamics, and for every poor node that switches to strategy $B$, a poor node switches to strategy $A$ until equilibrium is reached. Eventually, the system evolves into a new equilibrium state, where:

    \begin{equation}
    A_r^*=\min \{ n_r, \lceil \tau_r (n_p+n_r-1) \rceil \}  
    \end{equation}
    \begin{equation}
    A_p^*=0
    \end{equation}

    In more complex scenarios, the following conclusion can be drawn: if $\frac{\lceil \tau_r(n_p+n_r-1)-n_p \rceil+f-1}{n_r+f-1} < \tau_p <\frac{\lceil \tau_r(n_p+n_r-1)-n_p \rceil+f}{n_r+f}$, and if at least $f$ poor nodes (if fewer, then only those nodes and a corresponding number of rich nodes change) have been added at least $f$ edges each (these nodes do not connect to each other), the final equilibrium state of the system will be:
    \begin{equation}
    A_r^*=\min \{ n_r, \lceil \tau_r (n_p+n_r-1) \rceil \} 
    \end{equation}
    \begin{equation}
    A_p^*=0
    \end{equation}

    \item For Scenario II, where the equilibrium state is $A_r= \tau_r(n_p+n_r-1)-n_p +1$, $A_p=n_p)$, it follows that:
    \begin{equation}
    \tau_r (n_p+n_r-1)= n_p,\cdots,n_p+n_r-2
    \end{equation}
    \begin{equation}
    \tau_p>\frac{\tau_r(n_p+n_r-1)-n_p+1}{n_r}=\frac{k}{n_r}
    \end{equation}

    Similar conclusions can be drawn, indicating when $\frac{k+f-1}{n_r+f-1} \leq \tau_p <\frac{k+f}{n_r+f}$, if $f+1$ poor nodes (if fewer, then only those nodes and a corresponding number minus one of rich nodes change) have been added at least $f$ (more than 1) edges each (these nodes do not connect to each other), then the final equilibrium state of the system will be:
    \begin{equation}
    A_r^*=\min \{ n_r, \lceil \tau_r (n_p+n_r-1) -1\rceil \} 
    \end{equation}
    \begin{equation}
    A_p= 0
    \end{equation}

    \item Scenario III is similar to I.

    \item Scenarios VI, VII, VIII, and X follow similar analysis methods.

    \item For Scenario IV, when the equilibrium state is $A_r^*=0, A_p^*=n_p$, and it satisfies $0<\tau_r<\frac{n_p-1}{n_p+n_r-1}$, the following holds:

    If $m$ poor nodes have been added $f$ edges (these nodes do not connect to each other), the dynamics are influenced by:
    \begin{itemize}
        \item If $\tau_p>\frac{f}{n_r+f}$, there is no change.
        \item If $\frac{f-1}{n_r+f-1}\leq \tau_p<\frac{f}{n_r+f}$ and if $\frac{n_p-k}{n_p+n_r-1}<\tau_r \leq \frac{n_p-k+1}{n_p+n_r-1}$, and provided $m-k+1 \geq f, n_r \geq f$, then the final equilibrium state of the system is:
        \begin{equation}
        A_p=0
        \end{equation}
        \begin{equation}
        A_r=\min \{ n_r, n_p-k+1\}
        \end{equation}
    \end{itemize}

    \item For Scenario XI, the equilibrium state is characterized by $\tau_p n_r=k, k=1,2,\cdots,n_r-1$, $\frac{k-1}{n_p+n_r-1}<\tau_r<\frac{k+n_p}{n_p+n_r-1}$. Given this equilibrium state, both $A$ and strategy $B$ agents exist externally, and since this equilibrium is inherently unstable, the following conclusions can be drawn:

    \begin{itemize}
        \item If the endpoints of a new edge have different strategies, the system remains unchanged.
        \item If both endpoints of a new edge are $A$-individuals, the final equilibrium state will be:
        \begin{equation}
        A_p=0
        \end{equation}
        \begin{equation}
        A_r=\min \{ n_r, \tau_r \lceil n_p+n_r-1\rceil\}
        \end{equation}
        \item If both endpoints of a new edge are $B$-individuals, the final equilibrium state will be:
        \begin{equation}
        A_p=n_p
        \end{equation}
        \begin{equation}
        A_r=\max \{ 0, \tau_r \lfloor n_p+n_r-1\rfloor-n_p+1\}
        \end{equation}
    \end{itemize}
\end{itemize}

\section{Supplementary Discussion}
\subsection{The relationship between behavioral switching threshold $\tau$ and the average frequency of strategies in equilibrium states}
We analyze the average frequency of strategy $A$, denoted by
\begin{equation}
    \Bar{s} = \sum_{\mathbf{s} \in \mathbb{S}_{G}^{\tau+}} p(\mathbf{s}) \frac{||\mathbf{s}||}{N},
\end{equation}
or
\begin{equation}
    \Bar{s} = \sum_{\mathbf{s} \in \mathbb{S}_{G}^{\tau-}} p(\mathbf{s}) \frac{||\mathbf{s}||}{N},
\end{equation}
where $\mathbb{S}_{G}^{\tau+}$ (or $\mathbb{S}_{G}^{\tau-}$) is the set of network $G$'s equilibrium states in a coordinating (or anti-coordinating) game when the value of behavioral switching threshold equals $\tau$, and $p(\mathbf{s})$ is the probability that the system goes into the absorbing state $\mathbf{s}$ when the initial state is chosen from a uniform distribution.

Figure S19\textbf{ab} shows that the frequency of strategy is quite close to 1 when $\tau$ is less than $0.5$, which means the favored strategy dominates networks in coordination games.
Figure S19\textbf{cd} illustrates that the strategy frequency is quite closer to $0.5$ when $\tau$ is less than $0.5$, meaning two strategies coexist in anti-coordination games.
The above conclusion conforms to what we discussed in the main contents.

\subsection{The Role of Noise in Decision-Making}
Now, consider the decision-making process incorporates a stochastic component, which is fundamental to understanding the nature of the system's equilibria.
This stochasticity could be formally modeled through the concept of bounded rationality. 
The probability that an individual $i$ adopts strategy A is governed by a logistic choice function:
\begin{equation}
    p_i^A = \frac{\exp(\beta u_i^A)}{\exp(\beta u_i^A) + \exp(\beta u_i^B)} = \frac{1}{1+\exp[\beta(u_i^B-u_i^A)]},
    \label{eq:logit_choice}
\end{equation}
where $u_i^A$ and $u_i^B$ represent the payoffs for choosing strategies A and B, respectively, and the parameter $\beta \in [0, \infty)$ quantifies the level of individual rationality.
The parameter $\beta$ determines the sensitivity of an individual to payoff differences. 
A high value of $\beta$ signifies high rationality, where even small payoff advantages are detected and acted upon.
Conversely, as $\beta \to 0$, choices become nearly random.

From a stochastic process perspective, now the equilibrium states are not strictly absorbing. 
Once the system reaches such a state, there is always a non-zero probability of transitioning away from it.
This behavior contrasts with many standard imitation-based learning models. 
In imitation dynamics, noise often does not alter the fundamental properties of the corresponding Markov chains, which typically remain non-recurrent. 
In such cases, the system will inevitably become trapped in an absorbing state unless an explicit exploration mechanism (e.g., mutation) is introduced.

This formulation allows for a direct interpretation in terms of noise intensity by defining a noise parameter $\eta = 1/\beta$. 
Under this interpretation, low noise ($\eta \to 0$) corresponds to high rationality, enabling individuals to reliably choose the optimal strategy, while high noise ($\eta \to \infty$) increases the likelihood of erroneous choices.
A key consequence of this formulation is that for any finite level of noise, the system's equilibria are best described as metastable states. 
Although no state is strictly absorbing, the escape time required for the system to transition from one equilibrium's basin of attraction to another can be very large (Fig.~S\ref{rationality_noise}). 
Therefore, once the system settles into an equilibrium, it will primarily fluctuate around that state for extended periods. 

\subsection{ Comparison to other updating rules}
Consider the DB process when the selection intensity is infinite.
The probability that node $j$ is occupied by the offspring of $i$ is
\begin{equation}
    p_{ij} = \left\{
    \begin{aligned}
        &\frac{k_{ij}}{\sum_{\ell \in M_j} k_{\ell j}}, \quad i \in M_j,\\
        &0, \quad i \notin M_j,
    \end{aligned}
    \right.
\end{equation}
where $M_j = \{ \ell | u_{\ell} = \max_{n \in \mathcal{N}_j} u_n\}$.
Therefore, $\mathbf{s} = (s_1,s_2,\cdots,s_N)^{T}$ is an equilibrium state if and only if 
\begin{itemize}
    \item $\forall i \in \mathcal{N}, \forall j,\ell \in M_i, \quad s_j = s_{\ell}$,
    \item $\forall i \in \mathcal{N}, \forall j \in M_i, \quad s_j = s_i$.
\end{itemize}
Here we consider two types of regular graphs: ring lattice graphs and cyclic grid graphs.
For the coordination game
\begin{equation}
        \bordermatrix{
   & A & B \cr
A & 1,1  & 0,0 \cr
B & 0,0  & d,d \cr},
\end{equation}
in ring lattice graphs ($N = 16, 20$), there are only two equilibrium states: $\mathbf{s} = \mathbf{0}$ or $\mathbf{s} = \mathbf{1}$.
In cyclic grid graphs, the number of equilibrium states is determined by the value of $d$, as shown below
\begin{table}[h]
    \centering
\begin{tabular}{c c }
\toprule 
    $d$ & Equilibrium number  \\
    \midrule
    0.1& $2$\\
    0.3& $98$\\
    0.5& $18$\\
    0.8& $114$\\
    1.1& $114$\\
    1.3& $114$\\
    1.5& $114$\\
    1.6& $650$\\
    1.8& $650$\\
    2.0& $18$\\
    2.5& $18$\\
    2.8& $18$\\
    3.0& $18$\\
    3.2& $98$\\
    3.8& $98$\\
    \bottomrule
\end{tabular}
    \caption{\textbf{Non-linear relationship between payoff and equilibrium state number under DB updating.} Parameters: $N = 16$, $\Bar{k} =4.$}
    \label{tab:DB}
\end{table}

For the anti-coordination game
\begin{equation}
        \bordermatrix{
   & A & B \cr
A & 0,0  & 1,c \cr
B & c,1  & 0,0 \cr},
\end{equation}
both in ring lattice and cyclic grid graphs ($N = 9, 16$), there are only two equilibrium states: $\mathbf{s} = \mathbf{0}$ or $\mathbf{s} = \mathbf{1}$.
The results are quite different from coordinating or anti-coordinating systems.

Consider the IM process when the selection intensity is infinite.
The probability that node $j$ is occupied by the offspring of $i$ is
\begin{equation}
    p_{ij} = \left\{
    \begin{aligned}
        &\frac{k_{ij}}{\sum_{\ell \in M_j} k_{\ell j}}, \quad i \in M_j,\\
        &0, \quad i \notin M_j,
    \end{aligned}
    \right.
\end{equation}
where $M_j = \{ \ell | u_{\ell} = \max_{n \in \mathcal{N}_j \cup\{j\}} u_n\}$.
Therefore, $\mathbf{s} = (s_1,s_2,\cdots,s_N)^{T}$ is an equilibrium state if and only if 
$\forall i \in \mathcal{N}, \forall j \in M_i, s_j = s_i$ holds.

Meanwhile, consider the BD process when the selection intensity is infinite.
The probability that node $j$ is occupied by the offspring of $i$ is
\begin{equation}
    p_{ij} = \left\{
    \begin{aligned}
        &\frac{k_{ij}}{\sum_{\ell \in \mathcal{N}} k_{i\ell}}, \quad i \in M,\\
        &0, \quad i \notin M,
    \end{aligned}
    \right.
\end{equation}
where $M = \{ \ell | u_{\ell} = \max_{n \in \mathcal{N}} u_n\}$.
Therefore, $\mathbf{s} = (s_1,s_2,\cdots,s_N)^{T}$ is an equilibrium state if and only if 
$\forall i \in M, \forall j \in \mathcal{N}_i, s_j = s_i$ holds.

Different imitation rules makes the system dynamics diverse.
However, all imitation rules are sensitive to the perturbation (like the payoff structure), as shown in Fig.~\ref{strongSelection}, while systems under coordination/anti-coordination dynamics are robust to those perturbation.
This is one of the advantages of coordination/anti-coordination dynamics.
The above analysis reflects the differences between systems based on imitation and those based on coordination or anti-coordination.

\begin{table}[htbp]
\centering
\begin{tabular}{lcccl}
\toprule
Abbreviation & $N$ & $E$ & $\bar{k}$ & Description \\
\midrule
7G   & 29 & 240 & 16.55 & Friendship among 7th grade students \cite{vickers1981classroom_7thGrader}\\
BE   & 28 & 131 & 9.36 & Interaction among beetles \cite{nr-aaai15}\\
BC   & 25 & 90 & 7.20 & CEO club membership network \cite{konect:faust}\\
FE   & 32 & 266 & 16.63 & Personal message exchange \cite{freeman1979networkers}\\
KC   & 34 & 78 & 4.59 & Zachary's karate club friendship \cite{Zachary1977KarateClub}\\
MX   & 31 & 105 & 6.77 & Mexican political elites network \cite{1996MexicanPoliticalElites}\\
MZ   & 23 & 105 & 9.13 & Interaction among zebra \cite{Sundaresan2007ZebraNetwork}\\
MB   & 26 & 222 & 17.08 & Interaction among bison \cite{konect:lott}\\
MC   & 28 & 205 & 14.64 & Interaction among cattle \cite{konect:schein}\\
MK   & 17 & 91 & 10.71 & Kangaroo social network \cite{konect:grant}\\
MSa  & 18 & 126 & 14.00 & Relationships among monks \cite{konect:breiger}\\
MSh  & 28 & 235 & 16.79 & Relationships between bighorn sheep \cite{konect:hass}\\
PV   & 22 & 39 & 3.55 & Social network in a Papuan village \cite{schwimmer1970PapuanVillage}\\
RA   & 16 & 42 & 5.25 & Raccoon social network \cite{nr-aaai15}\\
RM   & 16 & 69 & 8.63 & Macaque social network \cite{konect:sade}\\
SM   & 17 & 70 & 8.24 & Spider monkey interaction network \cite{nr-aaai15}\\
TSF  & 39 & 158 & 8.10 & Tailor shop employee relationships \cite{Kapferer1972TailorShopEmployee_Zambia}\\
TR   & 16 & 58 & 7.25 & Relationships among New Guinea tribes \cite{Read1954Tribe}\\
VBF  & 32 & 356 & 22.25 & University freshman friendship network \cite{vandebunt1999friendship}\\
FB   & 32 & 124 & 7.75 & Online friendship network (Facebook) \cite{magnani2013combinatorial}\\
\bottomrule
\end{tabular}
\caption{Summary of the empirical networks used in this study.}
\label{tab:empirical_networks}
\end{table}

\makeatletter
\@fpsep\textheight
\makeatother

\newpage

\begin{figure}
    \centering
    \includegraphics[width=\linewidth]{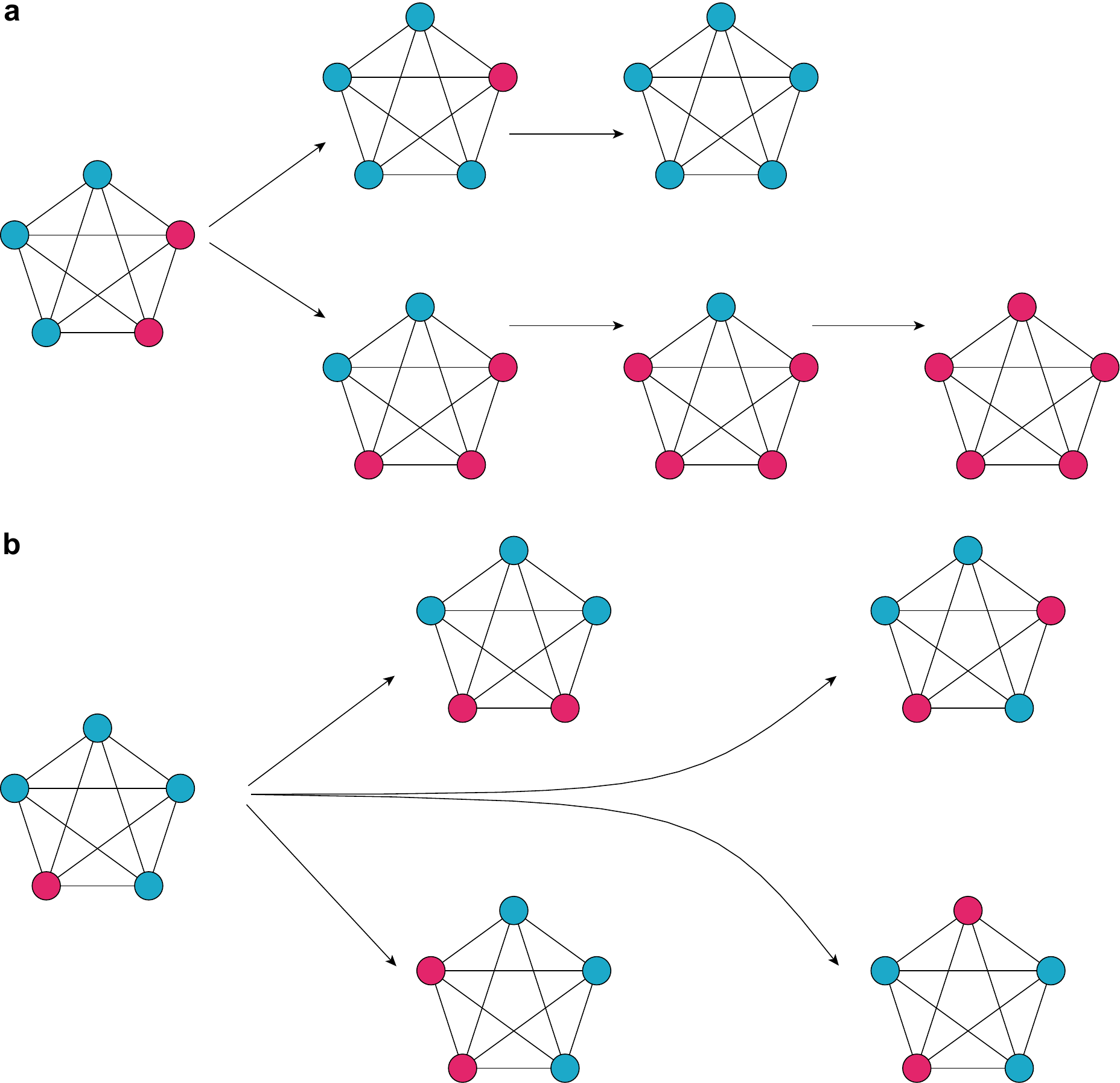}
    \caption{\textbf{Comparison between coordination game and anti-coordination game.} Panel \textbf{a} and \textbf{b} correspond to coordinating and anti-coordination game respectively. Parameter values: behavioral switching threshold $\tau=0.37$. }
\end{figure}

\newpage

\begin{figure}
    \centering
    \includegraphics[width=\linewidth]{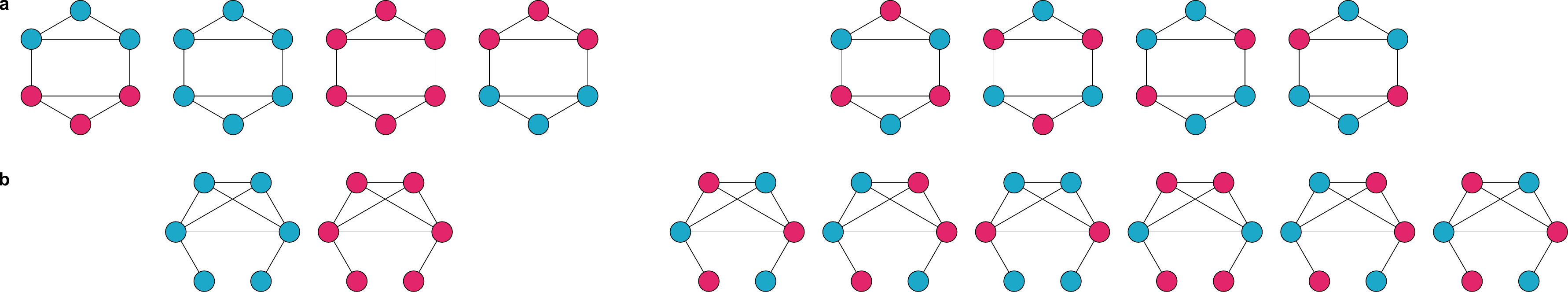}
    \caption{\textbf{Example of network with high clustering coefficient.} Panel \textbf{a} and \textbf{b} correspond to networks with low and high clustering coefficients respectively. Parameter values: behavioral switching threshold $\tau=0.37$. }
\end{figure}

\newpage

\begin{figure}
    \centering
    \includegraphics[width=\linewidth]{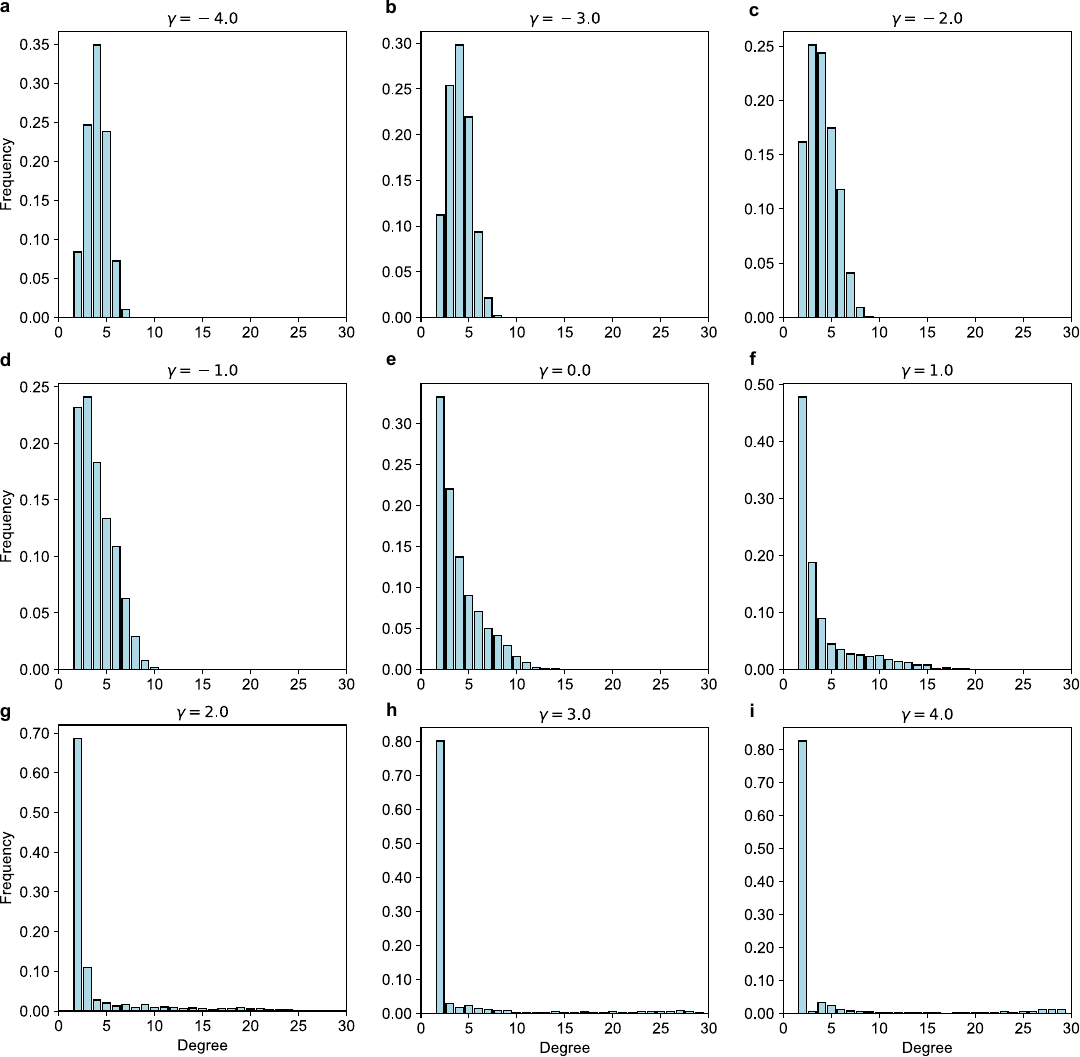}
    \caption{\textbf{Degree distribution of networks with different connection kernel.} In each panel, we take 100 networks to calculate the average degree distribution. Parameters: network size $N=30$, average degree $\Bar{k}=4.0$}
    \label{fig:si_dd}
\end{figure}

\newpage

\begin{figure}
    \centering
    \includegraphics[width=\linewidth]{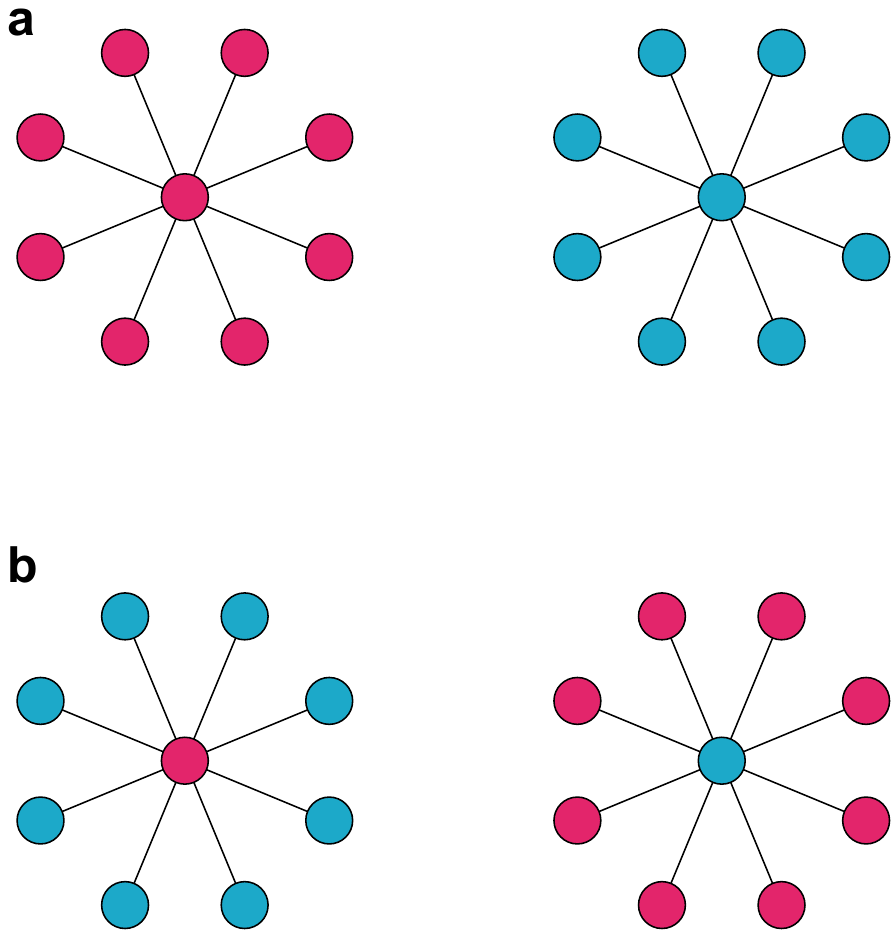}
    \caption{\textbf{Example of heterogeneous network.} Panel \textbf{a} and \textbf{b} correspond to coordinating and anti-coordinating game respectively. Parameter values: behavioral switching threshold $\tau=0.37$. }
\end{figure}

\newpage

\begin{figure}
    \centering
    \includegraphics[width=\linewidth]{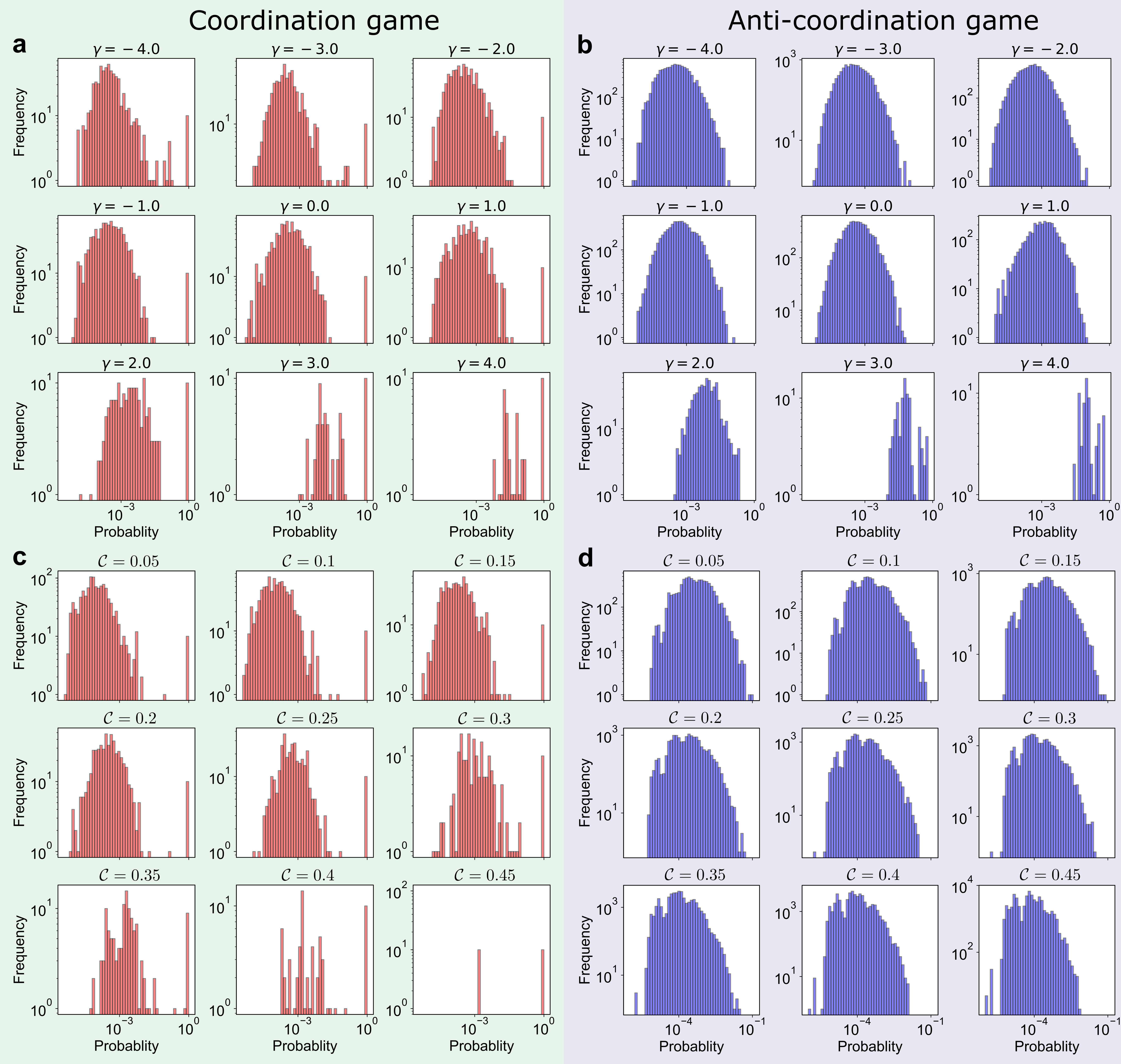}
    \caption{\textbf{There exists equilibrium states with high probabilities of being reached in any coordination game and anti-coordination game with significant node degree distribution heterogeneity.} We analyzed the arrival probability of each equilibrium state. The coordinating(\textbf{ac}) and anti-coordinating(\textbf{bd}) games are both considered. In each subfigure of panel \textbf{a-d}, the horizontal axis represents the arrival probability, while the vertical axis indicates the number of equilibrium states whose reaching probabilities fall within the intervals defined on the horizontal axis. \textbf{ab} and \textbf{cd} indicate the influence of the clustering coefficient($\mathcal{C}$) and the node degree distribution on the probability. Each panel is the result of 30000000 simulations. Parameter values: network size $|\mathcal{N}|=30,$ average degree $\Bar{k}=4$. }
    \label{fig5}
\end{figure}

\newpage

\begin{figure}
    \centering
    \includegraphics[width=\linewidth]{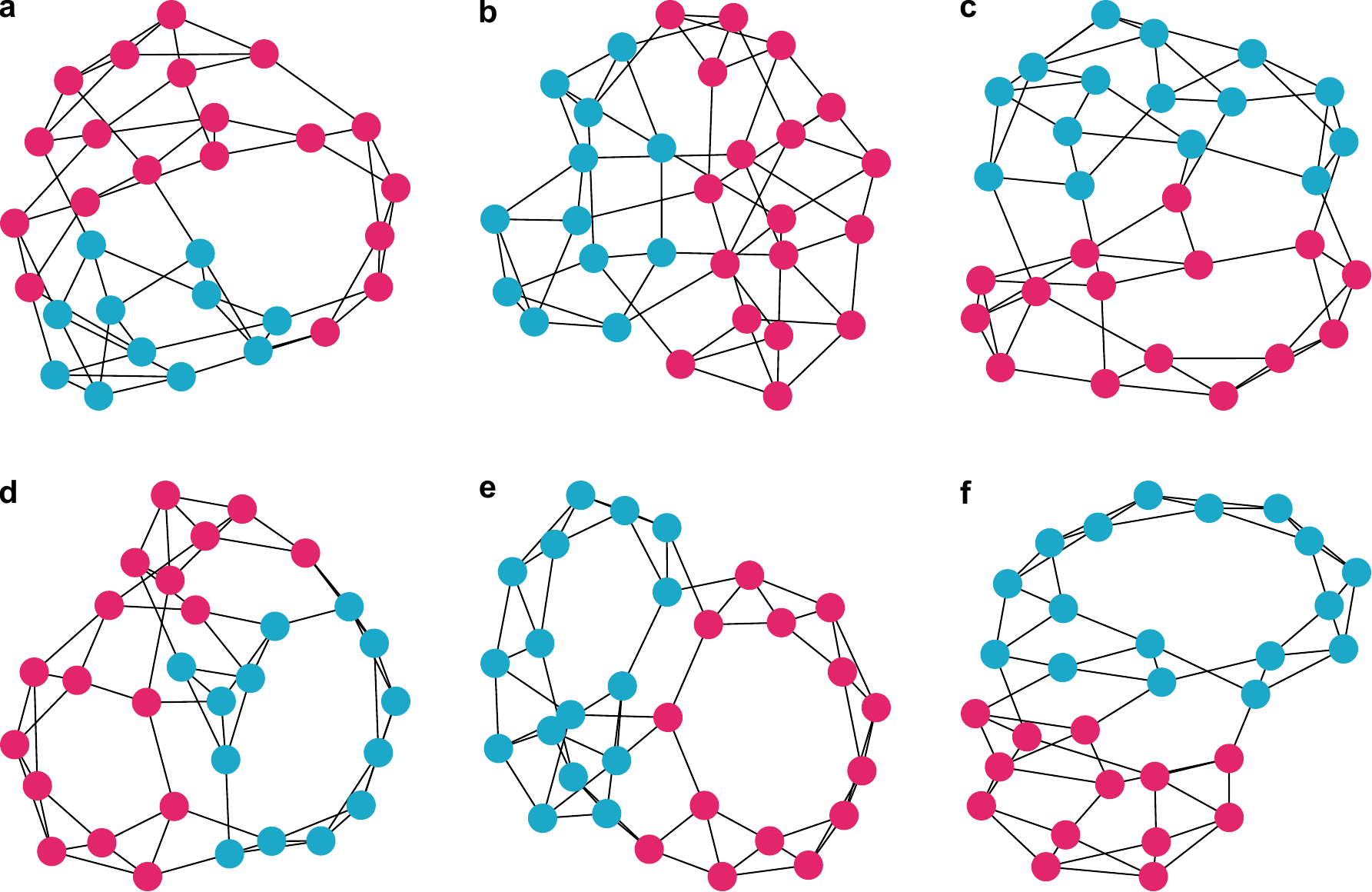}
    \caption{\textbf{Strategies coexist in networks with local clusters.} The clustering coefficient of networks in panel \textbf{a-c} and \textbf{e-f} are 0.25 and 0.35 respectively. Parameters: network size $N = 30$, behavioral switching threshold $\tau = 0.37$, average degree $\Bar{k}=4$}
    \label{fig:placeholder1}
\end{figure}

\newpage

\begin{figure}
    \centering
    \includegraphics[width=\linewidth]{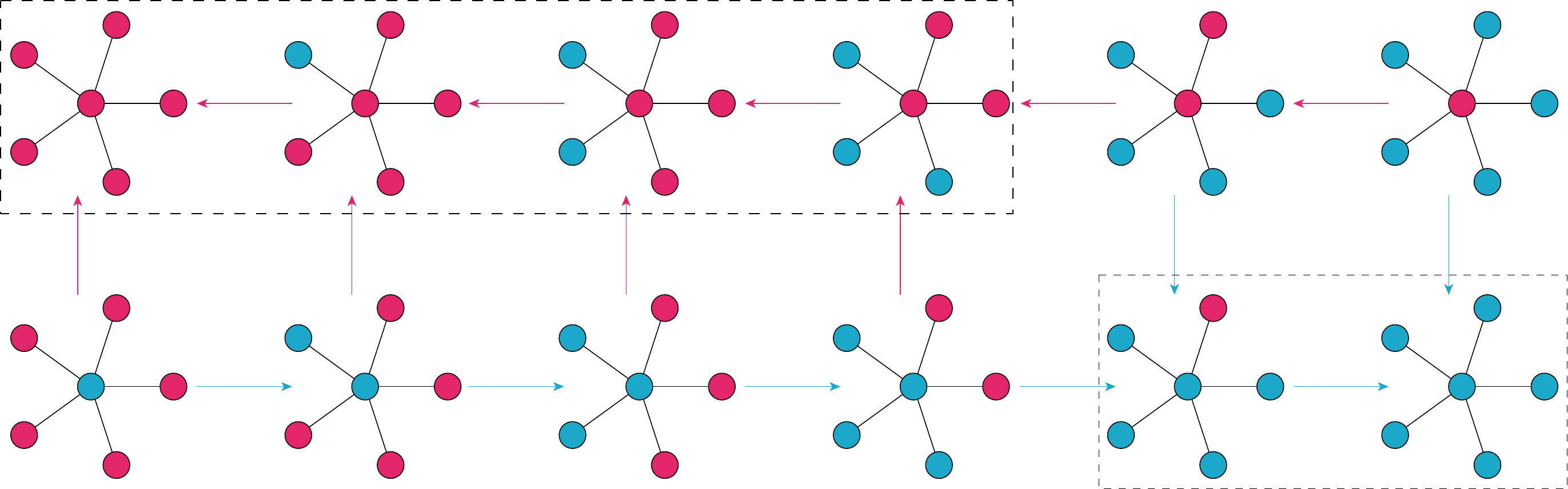}
    \caption{\textbf{State transition graph of star network in the coordination game.} The average equilibrium time of this system is $2.625$. Parameter values: network size $N=6,$ behavioral switching threshold $\tau=0.37$. }
\end{figure}

\newpage

\begin{figure}
    \centering
    \includegraphics[width=\linewidth]{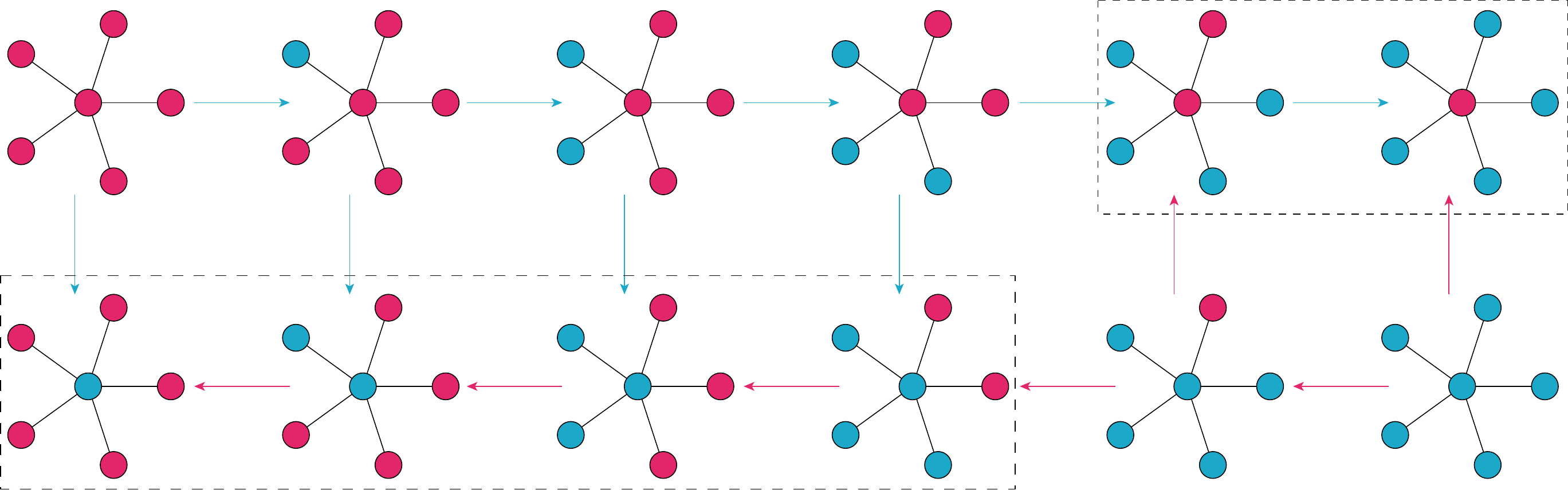}
    \caption{\textbf{State transition graph of star network in the anti-coordination game.} The average equilibrium time of this system is $2.625$. Parameter values: network size $N=6,$ behavioral switching threshold $\tau=0.37$. }
\end{figure}

\newpage

\begin{figure}
    \centering
    \includegraphics[width=\linewidth]{Figure_chain_coor_origin_rere.pdf}
    \caption{\textbf{State transition graph of chain network in the coordination game.} The average equilibrium time of this system is $3.325$. Parameter values: network size $N=6,$ behavioral switching threshold $\tau=0.37$. }
\end{figure}

\newpage

\begin{figure}
    \centering
    \includegraphics[width=\linewidth]{Figure_chain_coor_example_rere.pdf}
    \caption{\textbf{Example of the evolution process of chain network in the coordination game.} Parameter values: network size $N=6$, behavioral switching threshold $\tau=0.37$. }
\end{figure}

\newpage

\begin{figure}
    \centering
    \includegraphics[width=\linewidth]{Figure_chain_anti_origin_rere.pdf}
    \caption{\textbf{State transition graph of chain network in the anti-coordination game.}The average equilibrium time of this system is $2.050$. Parameter values: network size $N=6,$ behavioral switching threshold $\tau=0.37$. }
\end{figure}

\newpage

\begin{figure}
    \centering
    \includegraphics[width=\linewidth]{Figure_chain_anti_example_rere.pdf}
    \caption{\textbf{Example of the evolution process of chain network in the anti-coordination game.} Parameter values: network size $N=6,$ behavioral switching threshold $\tau=0.37$. }
\end{figure}

\newpage

\begin{figure}
    \centering
    \includegraphics[width=\linewidth]{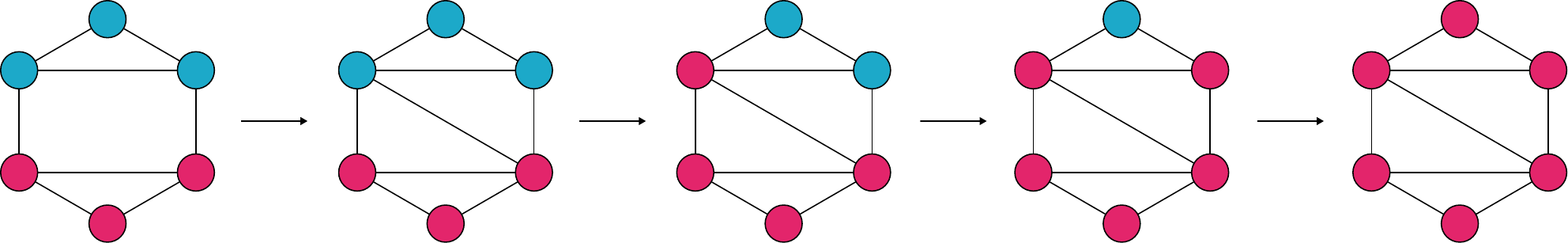}
    \caption{\textbf{Adding one new edge can make the favored strategy dominate the whole system.} Parameters: behavioral switching threshold $\tau=0.37$}
    \label{fig:si_add}
\end{figure}

\newpage

\begin{figure}
    \centering
    \includegraphics[width=.8\linewidth]{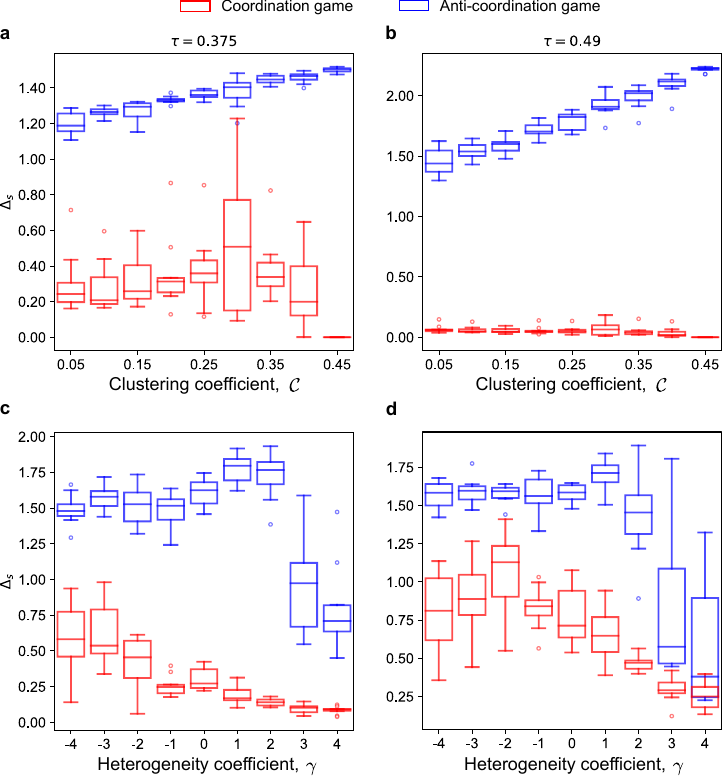}
    \caption{\textbf{Robustness analysis with adding 5 new edges.} Each red (respectively blue) box represents the average strategy changing under the coordination game (respectively under the anti-coordination game) in 50 networks, showing the median, quartiles, and outliers. We simulate each network of $2 \times 10^8$ times, where five edges are randomly added to each simulation. Panel \textbf{ab} displays the variations in the average number of strategy switches as the clustering coefficients change, and panel \textbf{cd} depicts how the average number of strategy switches evolves with the increase in degree heterogeneity.}
    \label{fig6}
\end{figure}

\newpage

\begin{figure}
    \centering
    \includegraphics[width=\linewidth]{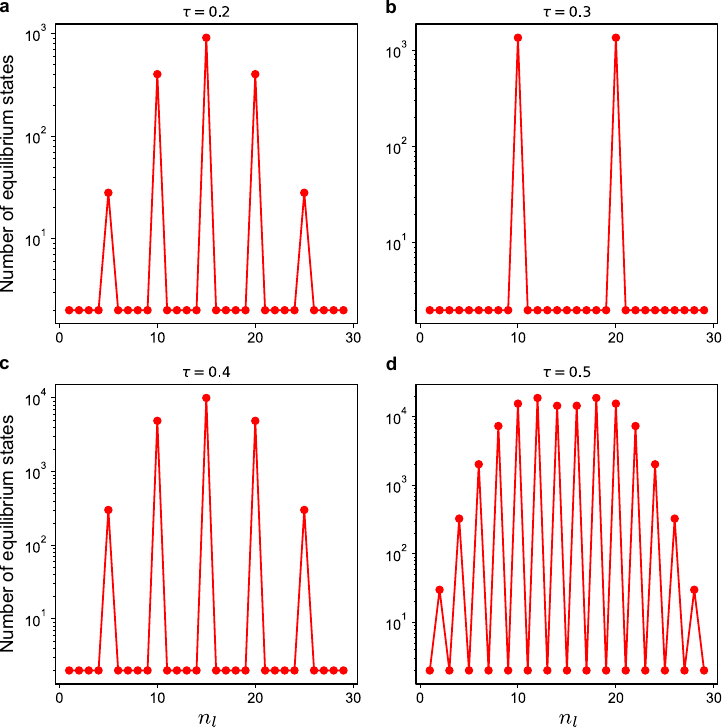}
    \caption{\textbf{Number of equilibrium states of the bipartite graph could be extremely high.} Presented is how the number of equilibrium states changes with the increase of left-side nodes. The total number of nodes remains unchanged. Parameters: network size $N = 30$}
    \label{fig:placeholder2}
\end{figure}

\newpage

\begin{figure}
    \centering
    \includegraphics[width=\linewidth]{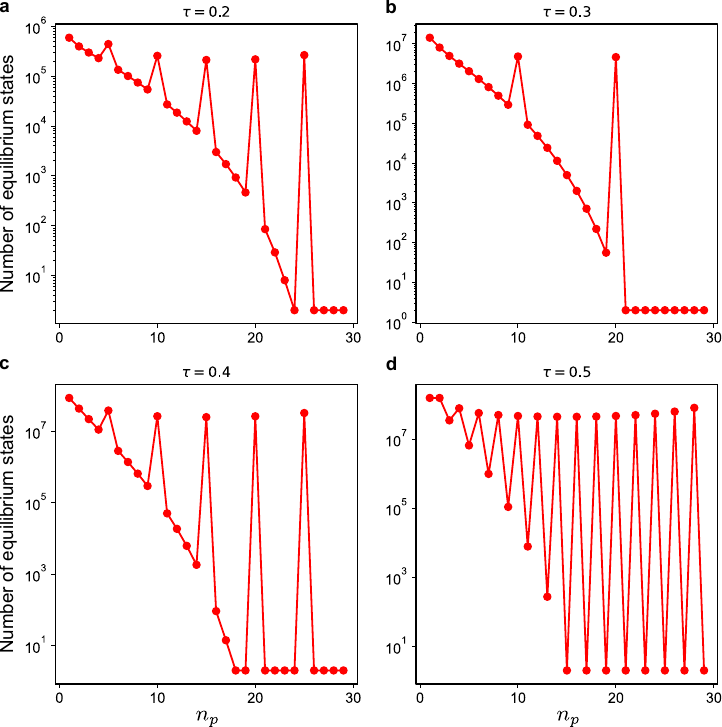}
    \caption{\textbf{Number of equilibrium states of the rich-club network could be extremely high.} Presented is how the number of equilibrium states changes with the increase of poor nodes. The total number of nodes remains unchanged. Parameters: network size $N = 30$}
    \label{fig:rich_special}
\end{figure}

\newpage

\begin{figure}
    \centering
    \includegraphics[width=\linewidth]{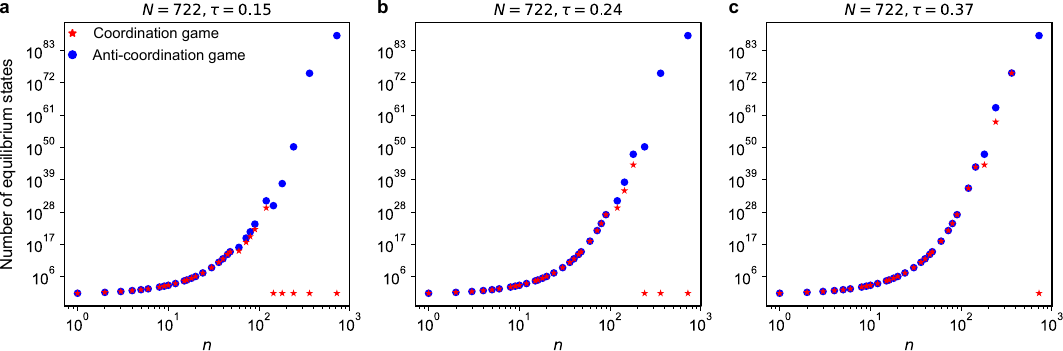}
    \caption{\textbf{The number of equilibrium states rises to a peak before decreasing in coordination games and increases with the increase of the average eccentricities of the network in anti-coordination games.}}
    \label{si:fig2}
\end{figure}

\newpage

\begin{figure}
    \centering
    \includegraphics[width=.6\linewidth]{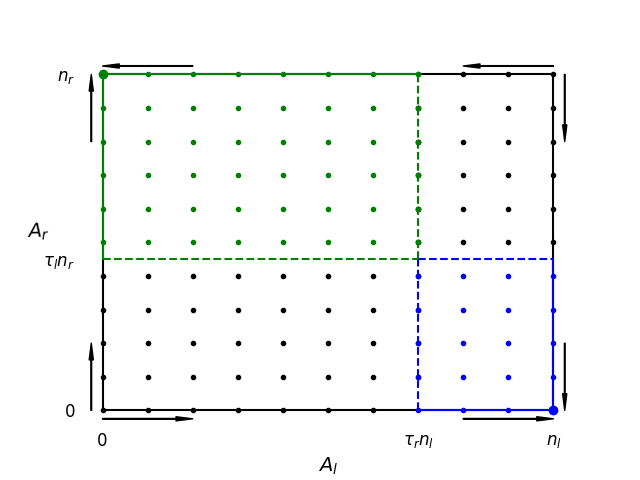}
    \caption{\textbf{The influence of $A$-individuals on the equilibrium states}}
    \label{ALAR}
\end{figure}

\newpage

\begin{figure}
    \centering
    \includegraphics[width=\linewidth]{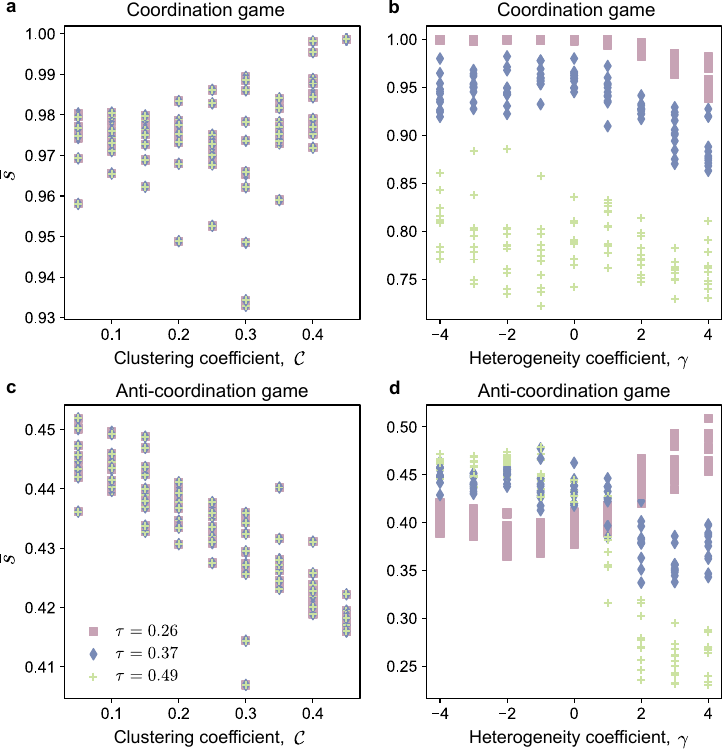}
    \caption{\textbf{The average frequency of strategy $A$ at different $\tau$ values.} Here each marker is the average frequency of strategy $A$ in a network. Panel \textbf{ac} illustrates the results in 4-regular networks with different clustering coefficients, while panel \textbf{bd} shows the result in networks with different degree distribution heterogeneity. The results in 4-regular networks remain unchanged at different $\tau$ values because the three values are in the interval of $(0.25,0.5)$ and any value of $\tau$ in this interval has the same effect on the dynamics of the 4-regular network according to Eq.~\ref{eq:coor} and Eq.~\ref{eq:anti}.}
\end{figure}

\newpage

\begin{figure}
    \centering
    \includegraphics[width=\linewidth]{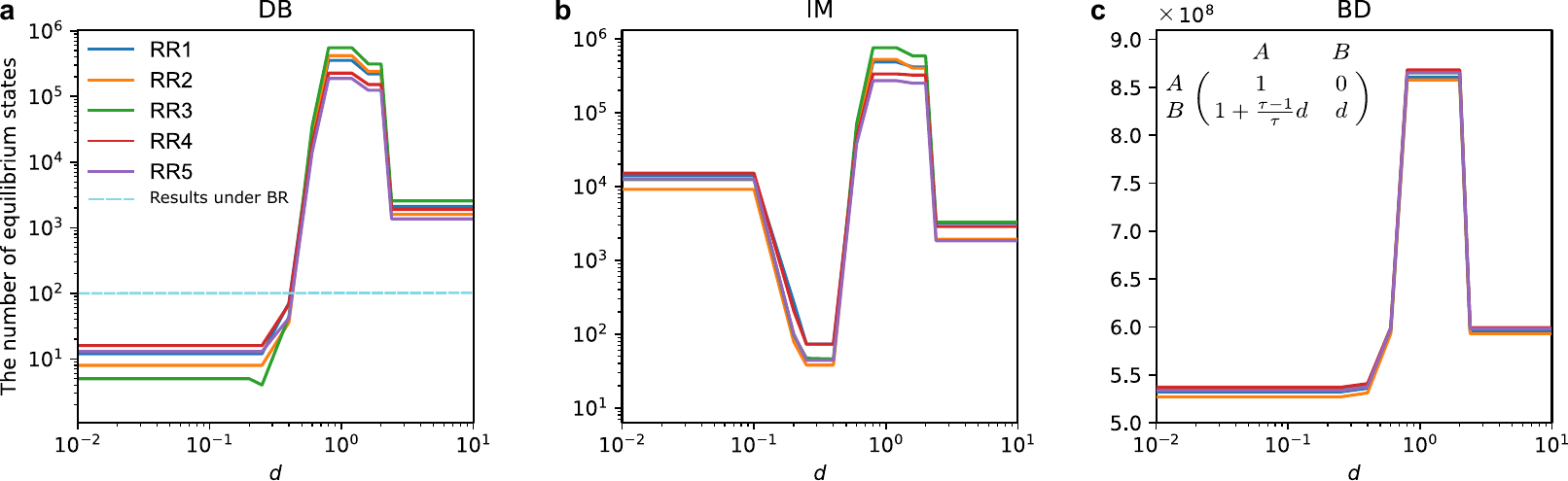}
    \caption{\textbf{Results under DB/IM/BD updating are sensitive to the payoff value.} 
    Here we compare the death-birth, imitation and death-birth learning rules in \textbf{a-c}.
    Different from the results under best-response dynamics, the number of equilibrium states under DB/IM/BD updating changes with the variation of $d$ even if the value of $\tau$ is constant.
    Parameter values: network size $N=30$ and average degree $\bar{k}=4$.}
    \label{strongSelection}
\end{figure}

\newpage

\begin{figure}
    \centering
    \includegraphics[width=.7\linewidth]{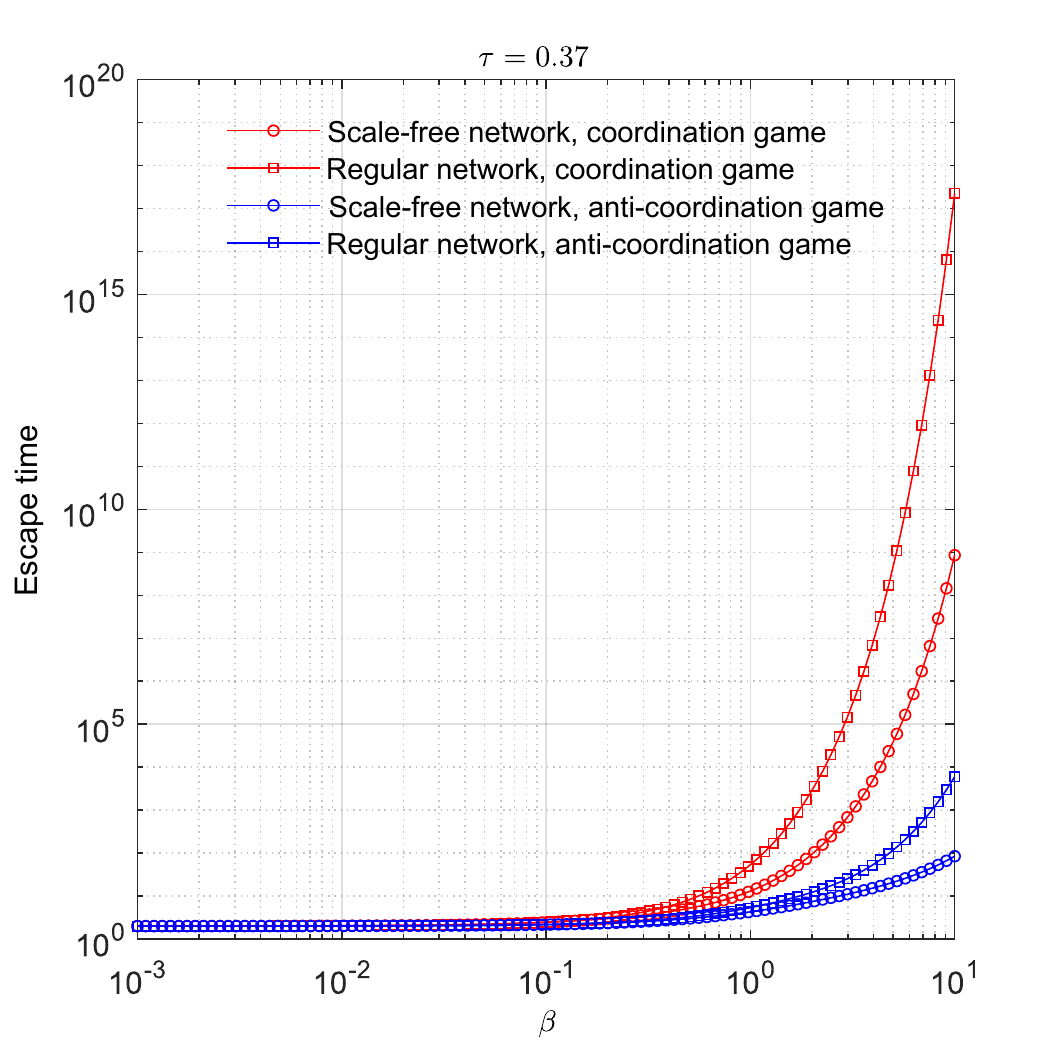}
    \caption{\textbf{Escape time as the players' rationality level increases (noise level decreases).} }
    \label{rationality_noise}
\end{figure}

\newpage

\begin{figure}
    \centering
    \includegraphics[width=.7\linewidth]{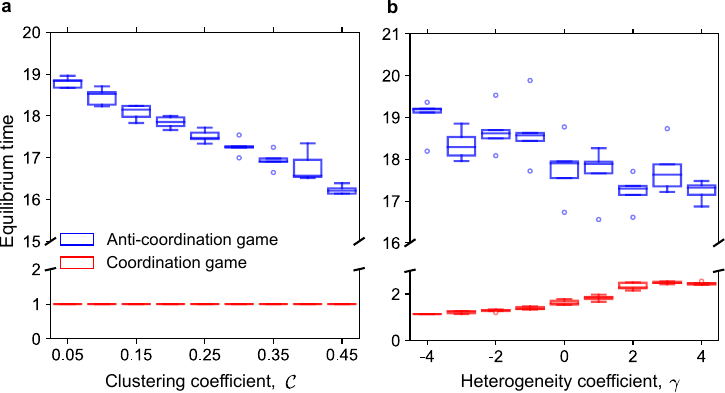}
    \caption{\textbf{Equilibrium time when only one $A$-player at the initial state}. Parameters: network size $N = 30$, behavioral switching threshold $\tau=0.37$}
\end{figure}

\begin{figure}
    \centering
    \includegraphics[width=\linewidth]{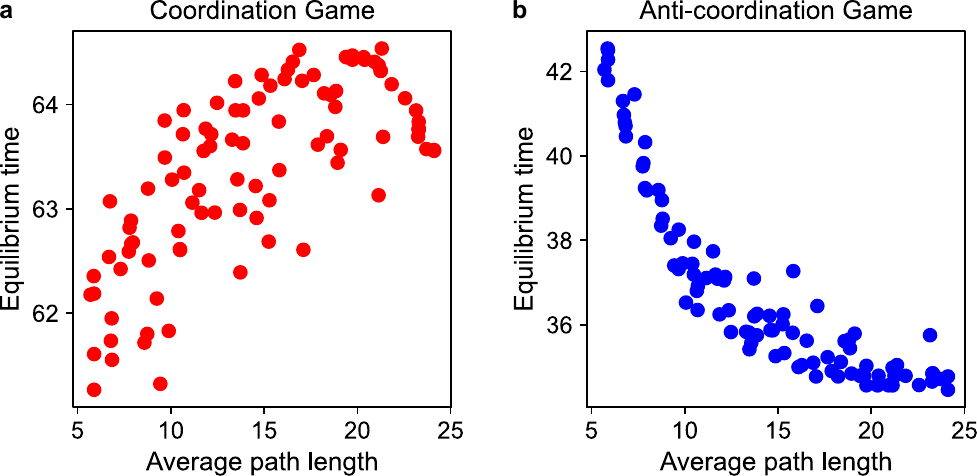}
    \caption{\textbf{Average equilibrium time grows (shrinks) as the average path length increases under coordination (anti-coordination) games in larger networks.}
    The average equilibrium time of different networks is shown as a function of the average path length.
    Each dot in panels \textbf{a} and \textbf{b} represents the average equilibrium time for one network starting from random initial states.
    The simulation is repeated $3\times10^7$ times for each network.
    Parameter values: network size $N=100$ and average degree $\bar{k}=4$.
    }   
\end{figure}

\begin{figure}[htb]
\centering
\includegraphics[width=\textwidth]{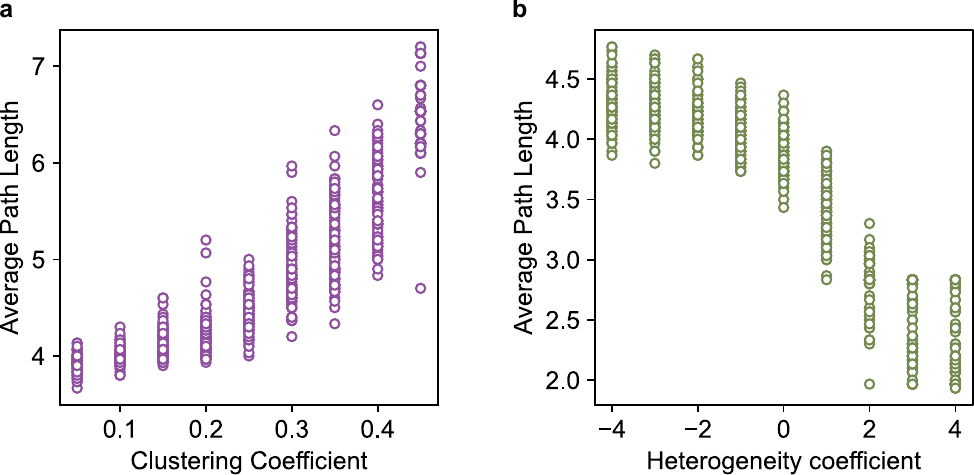}
\captionsetup{font={small,}}
\caption{
{\textbf{Average path length increases as the clustering coefficient increases, while decreases as the heterogeneity coefficient increases.}
}
}
\label{fig:R4}
\end{figure}

\end{document}